\documentclass[a4paper,11pt]{article}
\pdfoutput=1
\usepackage[utf8]{inputenc}
\usepackage[T1]{fontenc}
\usepackage[english]{babel}
\usepackage{amsmath,amsthm,amsbsy,amssymb,amsfonts,graphicx,mathrsfs,bm,cite,color,accents}
\usepackage{marvosym}
\usepackage{tablefootnote}
\usepackage{enumitem}
\usepackage{tikz}
\usepackage[colorlinks=true
,urlcolor=blue
,anchorcolor=blue
,citecolor=blue
,filecolor=blue
,linkcolor=blue
,menucolor=blue
,linktocpage=true
,pdfproducer=medialab
,pdfa=true
]{hyperref}
\newcommand{\nn}{\nonumber}
\newcommand{\dd}{\mathrm{d}}

\newcommand{\Ah}{\hat{A}}
\newcommand{\Bh}{\hat{B}}
\newcommand{\Ch}{\hat{C}}
\newcommand{\Dh}{\hat{D}}
\newcommand{\Eh}{\hat{E}}
\newcommand{\Fh}{\hat{F}}

\newcommand{\Ac}{\cA}
\newcommand{\Bc}{\mathcal{B}}
\newcommand{\Cc}{\mathcal{C}}
\newcommand{\Dc}{\mathcal{D}}
\newcommand{\Ec}{\cE}
\newcommand{\Fc}{\cF}
\newcommand{\Mc}{\mathcal{M}}
\newcommand{\Nc}{\mathcal{N}}
\newcommand{\Ic}{\mathcal{I}}
\newcommand{\Jc}{\mathcal{J}}
\newcommand{\Kc}{\mathcal{K}}
\newcommand{\Lc}{\mathcal{L}}
\newcommand{\Tc}{\mathcal{T}}
\newcommand{\Rc}{\mathcal{R}}
\newcommand{\Acl}{\underline{\mathcal{A}}}
\newcommand{\Acr}{\overline{\mathcal{A}}}
\newcommand{\Bcl}{\underline{\mathcal{B}}}
\newcommand{\Bcr}{\overline{\mathcal{B}}}
\newcommand{\Ccl}{\underline{\mathcal{C}}}
\newcommand{\Ccr}{\overline{\mathcal{C}}}
\newcommand{\Dcl}{\underline{\mathcal{D}}}
\newcommand{\Dcr}{\overline{\mathcal{D}}}

\newcommand{\alphaL}{\underline{\alpha}}
\newcommand{\alphaR}{\overline{\alpha}}
\newcommand{\betaL}{\underline{\beta}}
\newcommand{\betaR}{\overline{\beta}}
\newcommand{\gammaL}{\underline{\gamma}}
\newcommand{\gammaR}{\overline{\gamma}}
\newcommand{\deltaL}{\underline{\delta}}
\newcommand{\deltaR}{\overline{\delta}}
\newcommand{\epsilonL}{\underline{\epsilon}}
\newcommand{\epsilonR}{\overline{\epsilon}}

\newcommand{\aL}{\underline{a}}
\newcommand{\aR}{\overline{a}}
\newcommand{\bL}{\underline{b}}
\newcommand{\bR}{\overline{b}}
\newcommand{\cL}{\underline{c}}
\newcommand{\cR}{\overline{c}}
\newcommand{\dL}{\underline{d}}
\newcommand{\dR}{\overline{d}}
\newcommand{\eL}{\underline{e}}
\newcommand{\eR}{\overline{e}}

\newcommand{\AR}{\overline{A}}
\newcommand{\AL}{\underline{A}}
\newcommand{\BR}{\overline{B}}
\newcommand{\CR}{\overline{C}}

\newcommand{\la}{\langle}
\newcommand{\ra}{\rangle}

\newcommand{\al}{\underline{a}}
\newcommand{\ar}{\overline{a}}
\newcommand{\bl}{\underline{b}}
\newcommand{\br}{\overline{b}}
\newcommand{\cl}{\underline{c}}
\newcommand{\Cr}{\overline{c}}

\newcommand{\cA}{\mathcal{A}}
\newcommand{\cE}{\mathcal{E}}
\newcommand{\cF}{\mathcal{F}}
\newcommand{\cH}{\mathcal{H}}
\newcommand{\cM}{\mathcal{M}}

\newcommand{\OO}{\mathrm{O}}

\newcommand{\GL}{\mathrm{GL}}

\newcommand{\GD}{\mathrm{G_D}}
\newcommand{\GM}{\mathrm{G}_\mathcal{M}}
\newcommand{\GPS}{\mathrm{G_{PS}}}
\newcommand{\GS}{\mathrm{G_{S}}}

\newcommand{\Omegat}{\text{\Large\Leo}}

\allowdisplaybreaks

\numberwithin{equation}{section}

\begin{document}
\hypersetup{pageanchor=false}
\begin{titlepage}
	\renewcommand{\thefootnote}{\fnsymbol{footnote}}

	\vspace*{1.0cm}

	\centerline{\LARGE\textbf{$\alpha'$-Bootstrap}}

	\vspace{1.0cm}

    \vspace{1.0cm}

	\centerline{
		{Achilleas Gitsis}%
		\footnote{E-mail address: achilleas.gitsis@uwr.edu.pl},
		{Falk Hassler}%
		\footnote{E-mail address: falk.hassler@uwr.edu.pl}
		and
			{Luca Scala}%
		\footnote{E-mail address: luca.scala@uwr.edu.pl}
	}

	\begin{center}
		${}^\ast {}^\dagger {}^\ddagger${\it University of Wroc\l{}aw, Faculty of Physics and Astronomy,}\\
		{\it Maksa Borna 9, 50-204 Wrocław, Poland}
	\end{center}

	\begin{abstract}
Due to the exponential growth in the number of terms, computing $\alpha'$-corrections to string theory's low-energy effective actions is a challenging matter. In order to fix all the couplings, one has usually to deal with a large number of string scattering amplitudes. This difficulty can be overcome by exploiting T-duality, which severely constrains the allowed structure of the effective action. It is then convenient to work in a formulation where T-duality is a manifest symmetry. Building on a series of previous works by some of the authors, \cite{Hassler:2024yis,Gitsis:2024gfb,Gitsis:2025clo}, we present a refined version of an elegant and effective procedure that allows to obtain all the higher-derivative corrections of the NS-NS sector of sting theories at order $\alpha'$ and $\alpha'^2$, up to an overall coefficient. We dub this approach \emph{$\alpha'$-bootstrap}, since it is based only on consistency conditions and avoids the direct computation of scattering amplitudes. The procedure relies on an infinite dimensional algebraic structure that we present in full detail, and it is conjectured to work at all orders. Although, at the moment, it still misses the $\zeta$-like corrections starting at order $\alpha'^3$, the ease with which it can be generalized is promising for future developments in this direction.
	\end{abstract}

	\thispagestyle{empty}
\end{titlepage}
\hypersetup{pageanchor=true}

\setcounter{footnote}{0}

\newpage

\hrule

\tableofcontents
\vspace{2em}
\hrule

\section{Introduction}\label{sec:intro}
Dealing with the infinite tower of massive modes in string theory is a challenging matter and for most applications unfeasible. This is the reason why one usually considers a low-energy effective theory of the massless modes and its perturbative corrections. These corrections take the form of an expansion in two parameters, the inverse string tension $\alpha'$ (corresponding to higher-derivative corrections) and the string coupling $g_s$ (corresponding to string loop corrections).

In this work we restrict our investigation to the $\alpha'$-corrections of the low-energy effective theory describing the bosonic NS-NS sector, common to bosonic string theory and all superstring theories. The action we consider, therefore, takes the following form in $d$ dimensions:
\begin{equation}\label{eq:stringaction}
    S=\int \dd^dx \sqrt{-g} e^{-2\phi} \Big( R+4(\partial \phi)^2 -\tfrac{1}{12} H^2 + \mathcal{O}(\alpha')\Big),
\end{equation}
where $g$ is the determinant of the $d$-dimensional metric, $R$ the Ricci scalar, $\phi$ the dilaton and $H=\dd B$ the curvature of the Kalb-Ramond field $B$.

Correction terms proportional to powers of $\alpha'$ correspond to higher-derivative corrections of this low-energy action: For each power of $\alpha'$, two more derivatives appear. They naturally incorporate modifications to gravity, containing powers of Ricci scalars and contractions of Ricci and Riemann tensors. These corrections play an important role, for instance, in string cosmology \cite{Gasperini:2002bn,Becker:2002nn}, in the context of the AdS/CFT correspondence \cite{Buchel:2004di, Gubser:1998nz,Banks:1998nr} and for black hole thermodynamics \cite{CALLAN1989673, Myers:1987yn, Sen:2005wa, Mohaupt:2005jd}.

The standard way of calculating them is by matching string scattering amplitudes with their field theory counterparts. Alternatively, one might also impose the vanishing of $\beta$-functions on the string's worldsheet to reconstruct the low-energy effective action. In principle, one might compute the full expansion order-by-order; However, it turns out that the calculations involved are way too complicated beyond the leading orders so that, in practice, they are fully known only up to $\alpha'^2$ without recurring to more advanced techniques.
A question that naturally arises is whether it is possible to simplify these computations and make them more accessible. A natural idea is to exploit symmetry arguments. In principle, one might write all the possible terms appearing order-by-order in the expansion as contractions of the fields that constitute the massless spectrum, and their covariant derivatives. These terms in the action should, of course, be invariant under the symmetries of string theory, namely target space diffeomorphisms, $B$-field gauge transformations $B\rightarrow \dd A$ and worldsheet parity under the inversion $B\rightarrow -B$ for bosonic and type II string theory. While the number of allowed terms is reduced after imposing these symmetries, there are still a lot of free coefficients left. Their total number is summarized in the following table:

\begin{table}[ht]
\centering
    \begin{tabular}{c|c|c|c|c}
        & \multicolumn{3}{|c|}{minimal number of NS-NS coefficients to match} \rule{0pt}{2.5ex}\\
        \hline
        order & bosonic \cite{Garousi:2019cdn,Garousi:2019mca,Garousi:2020lof,Gholian:2023kjj,Ameri:2025bei} & heterotic \cite{Garousi:2023kxw,Garousi:2024rzh}  & type II \cite{Garousi:2020mqn,Garousi:2020gio} \rule{0pt}{2.5ex}\\
        \hline
        $(\alpha')^0$ & 3 (1) & 3 (1) &  3 (1)  \rule{0pt}{2.5ex}\\
        $(\alpha')^1$ & 8 (1)& 8 (1) & 0 \rule{0pt}{2.5ex}\\
        $(\alpha')^2$ & 60 (0)\tablefootnote{Imposing T-duality, all the coefficients can be fixed in terms of the single coefficient at order $\alpha'$.} & 73\tablefootnote{Comparing with bosonic, the additional couplings here come from odd-parity $B$-field terms.} (0) &0 \rule{0pt}{2.5ex}\\
        $(\alpha')^3$ & 872 (1)\tablefootnote{At this order, everything can be fixed in terms of the coefficient at order $\alpha'$ and one proportional to $\zeta(3)$.}& 1349 (1) & 872 (1)\tablefootnote{Here, the only coefficient to fix is the one proportional to $\zeta(3)$.} \rule{0pt}{2.5ex}\\
        $\cdots$ & $\cdots$ & $\cdots$ & $\cdots$ \rule{0pt}{2.5ex}
    \end{tabular}
\end{table}

These numbers already take into account redundancies coming from Bianchi identities, boundary terms (which arise after integration by parts) and field redefinitions. More recent approaches exploit, on top of the aforementioned symmetries, constraints coming from T-duality. The number of new independent parameters is then further reduced and given in parentheses in the table above.

In contrast to target space diffeomorphisms and $B$-field transformations, T-duality acts non-linearly on the fields. Thus, it becomes quickly cumbersome to impose it at higher orders. It is therefore desirable to reformulate the low-energy effective action in a way that makes it manifestly covariant under T-duality transformations. Such a framework is provided by Double Field Theory (DFT)\cite{Hull:2009mi,Hohm:2010jy,Hohm:2010pp}, or the closely related generalized geometry\cite{2002math......9099H,2004math......1221G}. The main achievement of these theories is to unify T-duality with all the other symmetries of string theory into a split-orthogonal group $\OO(d,d)$. For the purposes of this article, DFT is always supplemented with the canonical realization of its section condition, making it essentially equivalent to generalized geometry. Therefore, from now on, we will only refer to DFT, while keeping in mind that all the discussions can be directly carried over to the generalized geometry language.

Currently the most powerful approach to understand $\alpha'$-corrections in relation to the $\OO(d,d)$ duality group is the DFT uplift of a procedure first introduced by Bergshoeff and de Roo in \cite{Bergshoeff:1989de}. Formulated in \cite{Baron:2018lve,Baron:2020xel}, this mechanism is called the \emph{generalized Bergshoeff-de Roo (gBdR) identification} and reduces the free coefficients needed up to order $\alpha'^2$ to two. Historically these coefficients were denoted $a$ and $b$, and are related to the two chiral sectors of the theories. In this way one is able to write general bi-parametric expressions subsuming higher-derivative corrections for different models. The relations between these parameters and some relevant theories is given by the following table:

\begin{center}
    \begin{tabular}{c|c|c}
        Theory & $a$ & $b$ \\
        \hline
        Bosonic & $-\alpha'$ & $-\alpha'$\\
        Heterotic & $-\alpha'$ & 0\\
        Type II & 0 & 0\\
        HSZ theory & $-\alpha'$ & $\alpha'$\,.
    \end{tabular}
\end{center}

Note that only the first three choices of $a$ and $b$ correspond to higher-derivative corrected string theories, while the last one results in another $\OO(d,d)$-covariant theory called Hohm-Siegel-Zwiebach (HSZ).

At order $\alpha'^3$, the situation becomes more complicated. Although DFT is assumed to be able to reproduce \emph{certain} higher-derivative corrections coupled in action through the coefficients $a^3$ and $b^3$, string scattering amplitudes and a direct application of T-duality constraints \cite{Garousi:2020lof,Wulff:2024mgu,Ameri:2025bei} require additional terms proportional to $\zeta(3)$ in the low-energy effective action of all perturbative string theories. How to obtain these corrections in the framework of DFT is still not clear. The reason is that even though the gBdR identification is postulated to work at all orders, it reproduces only one tower of corrections in $a$ and $b$, while it still leaves out the $\zeta$-like ones. Eventually, the goal is to extend this construction to incorporate all the other towers. To this end, some conceptual problems hinted in \cite{Hsia:2024kpi} have to be overcome. Still gBdR-inspired methods are currently among the most promising routes to better understand the structure of higher-derivative corrections in string theory.

Originally lacking an intuitive geometrical interpretation, the gBdR procedure was recently reformulated and geometrized in a series of papers \cite{Hassler:2024yis,Gitsis:2024gfb,Gitsis:2025clo}, building on a formalism proposed by Pol\'a\v{c}ek and Siegel in~\cite{Polacek:2013nla} which is closely related to the lift of Cartan geometry to generalized geometry \cite{Hassler:2024hgq,Hassler:2025rag}. In these papers, the space on which the theory is formulated is further extended beyond the usual doubling of DFT into a so-called \emph{megaspace} in order to facilitate a systematic construction of covariant curvatures and torsion tensors. Covariant is here understood with respect to the standard symmetries (target space diffeomorphisms, $B$-field transformations), and an additional gauge symmetry governed by the generalized structure group of the theory, $\GS$. Choosing the latter appropriately, the gBdR identification arises naturally as 1) torsion constraints, and 2) a partial gauge fixing on the megaspace \cite{Gitsis:2024gfb}. The resulting action was verified to match with the literature up to order $\alpha'$. The explicit distinction between 1) and 2) is not present in the original gBdR identification but it is crucial to make the computations at higher orders manageable. This was first shown in \cite{Gitsis:2025clo}, where an all-order result for generalized Green-Schwarz transformations was derived. While this work already fully exploited the partial gauge fixing, it still had some deficits in the explicit solution of the torsion constraints. We will resolve them here, and thereby initiate what we call the \emph{$\alpha'$-bootstrap} program. Its main goal is to derive the full structure of higher-derivative corrections via constraints, without resorting to string scattering amplitude calculations (except to fix a very few number of coefficients).

Compared to the original gBdR approach, we believe our method has three main advantages:
\begin{itemize}[topsep=2pt,itemsep=1.5pt,parsep=2pt]
    \item[a)] It better clarifies the geometrical structure behind the identification procedure;
    \item[b)] It relies on iterative formulas that can be easily implemented in computer algebra systems in order to compute order-by-order corrections;
    \item[c)] It is easy to generalize and thereby might provide new insights into the problem of capturing $\zeta(3)$-corrections.
\end{itemize}

The article is structured in the following way. In section~\ref{sec:reviewgBdR} we review the basic features of DFT and its heterotic extension and we briefly present a conceptual overview of the gBdR identification. The main part of this work then reflects its twofold goals: In section~\ref{sec:bootstrap} we revisit the results of \cite{Hassler:2024yis,Gitsis:2024gfb,Gitsis:2025clo}, solving the regularization issue referred in \cite{Gitsis:2024gfb} as \emph{towers collapse} and clarifying the underlying algebraic structure. Afterwards, in section~\ref{sec:actionandsym} we employ the formalism we developed to find and match with the literature the known actions up to order $\alpha'^2$. This will be done for the full $a$ and $b$ tower of corrections, and therefore will include both the heterotic and the bosonic cases. In the same section we also discuss the relevant symmetries of these actions, with particular emphasis on the Green-Schwarz transformations for chiral theories. At order $\alpha'^2$ there is already a considerable amount of freedom in performing field redefinitions, integration by parts and in using Bianchi identities. Hence, even comparing results is far from trivial. For example the action obtained through the original gBdR identification contains 747 terms, while we manage to express our bi-parametric action in the rather compact form
\begin{align}
    \Lc^{(6)}=&-\tfrac{a^2}{8}\left[3 H_{ij}{}^{k} R^{(-) ijln} \Omega^{(-)}_{kln}+ 3 R^{(-) ijkl} \nabla^{(-)}_{i} \Omega^{(-)}_{jkl}+ \tfrac{3}{2} \Omega^{(-)}_{i}{}^{jk} \Omega^{(-)i}{}_{jk}\right] \nn \\
    &+\tfrac{b^2}{8}\left[- 3 H_{ij}{}^{k} R^{(+) ijln} \Omega^{(+)}_{kln}+ 3 R^{(+) ijkl} \nabla^{(+)}_{i} \Omega^{(+)}_{jkl}- \tfrac{3}{2} \Omega^{(+)}_{i}{}^{jk} \Omega^{(+)i}{}_{jk} \right] \nn \\
    &+\tfrac{ab}{2} \bigg[ \tfrac{1}{3} R^{(-)}_{m}{}^{sku} R^{(-)j}{}_{s}{}^{l}{}_{u} R^{(-)m}{}_{jkl} - \tfrac{1}{8} R^{(-)rs}{}_{kl} R^{(-)tukl} R^{(+)}_{rstu} \nn \\
    &\phantom{+\tfrac{ab}{2} \bigg[} - \tfrac{1}{8} R^{(-)mnrs} R^{(+)}_{mn}{}^{tu} R^{(+)}_{rstu} + \tfrac{3}{4} H_{ij}{}^{q} R^{(-)rsij} \Omega^{(-)}_{qrs} \nn \\  
    &\phantom{+\tfrac{ab}{2} \bigg[}+ \tfrac{3}{4} H_{ij}{}^{q} R^{(+)rsij} \Omega^{(+)}_{qrs} + \tfrac{3}{4} R^{(-)mnpq} \nabla^{(-)}_{m} \Omega^{(+)}_{npq} \nn \\ 
    &\phantom{+\tfrac{ab}{2} \bigg[}- \tfrac{3}{4} R^{(+)mnpq} \nabla^{(+)}_{m} \Omega^{(-)}_{npq} + \tfrac{3}{4} \Omega^{(-)jln} \Omega^{(+)}_{jln} \nn \\
    &\phantom{+\tfrac{ab}{2} \bigg[}- \tfrac{1}{8} (\nabla^{(+)}_{i} R^{(+)mnpq}-2H_{ir}{}^{[m}R^{(+)n]rpq}) (\nabla^{(+)i} R^{(+)}_{mnpq}-2H^i{}_{r[m}R^{(+)}_{n]}{}^r{}_{pq})\bigg] \nn \\
    &+\Lc^{(6)}_{\text{redef}}\,.
\end{align}
As we will explain better in the next sections, this result is written in terms of Riemann tensors for two torsion-full connections with torsions $-H_{ijk}$ for $(+)$ and $+H_{ijk}$ for $(-)$. $\Omega^{(+)}_{ijk}$ and $\Omega^{(-)}_{ijk}$ are the Chern-Simons forms corresponding to these connections. More details on these quantities can be found in appendix~\ref{app:geometry}. Finally, $\Lc^{(6)}_{\text{redef}}$ is a term that can be removed through field redefinitions and integration by parts to match with other results in the literature. We conclude the article with a brief comment on further extensions of this procedure that could potentially incorporate all the other towers of corrections ($\zeta(3)$, $\zeta(5)$, \dots) and that we plan to investigate in an upcoming paper.

\section{A brief overview of DFT and of the gBdR identification}\label{sec:reviewgBdR}
In the past few decades, an effort was made to find alternative ways to understand the structure of higher-derivative corrections without relying completely on scattering amplitude computations. A broad class of said approaches tries to exploit the constraints coming from T-duality in order to restrict the plethora of possible terms allowed by the manifest symmetries of the theory. 
This can be done thanks to the fact that T-duality is an $\alpha'$-perturbative duality of string theories, and thus should be preserved order-by-order in the expansion. Therefore, the possible corrections are not only constrained by the symmetries of the theory, but also by compatibility with this duality. Naturally, this requires to understand how T-duality acts on all fields, which can be conveniently formulated in the framework of generalized geometry or DFT, where the theory becomes manifestly covariant under the T-duality group $\OO(d,d)$ at the two-derivative level. This, in turn, allows to consider compatibility under T-duality transformations as an additional symmetry that has to be respected by the higher-derivative terms.

\subsection{Fundamentals of DFT}\label{sec:introDFT}
In this section we will briefly review some core features of DFT. For a more pedagogical and comprehensive introduction, the reader might refer to the excellent exposition \cite{Aldazabal:2013sca}.
The core point of this theory is that invariance under diffeomorphisms and $B$-field gauge transformations (now unified under the name \emph{generalized diffeomorphisms}) can be embedded in the larger $\OO(d,d)$ group of T-duality. In order to capture the action of this new symmetry, one extends the tangent space of differential geometry to define tensors that are manifestly covariant under $\OO(d,d)$. In DFT this is done by doubling the number of coordinates on the spacetime, thus defining doubled indices that split as
\begin{equation}\label{eq:windsplit}
    x_I = \begin{pmatrix}\widetilde{x}_i & x^i\end{pmatrix},\qquad x^I = \begin{pmatrix} x^i & \widetilde{x}_i \end{pmatrix}.
\end{equation}
The metric $g_{ij}$ and the $B$-field $B_{ij}$ are unified into a single object, the $\emph{generalized metric}$~\cite{Hohm:2010pp}
\begin{equation}
    \cH_{IJ} =\begin{pmatrix}
        g_{ij}-B_{ik}g^{kl}B_{lj} & B_{ik}g^{kj}\\
        -g^{ik}B_{kj} & g^{ij}
    \end{pmatrix}\,.
\end{equation}
Similar to the metric, it is an $\OO(d,d)$-valued, symmetric tensor. The remaining degree of freedom of the NS-NS sector, namely the dilaton $\phi$, is encoded in the \emph{generalized dilaton}
\begin{equation}
    \Phi:=\phi -\tfrac{1}{2} \log(g)\,.
\end{equation}
In analogy with the generalized metric, it is constructed such that it transforms covariantly under the duality group $\OO(d,d)$.

Similarly to general relativity (GR), one can transition to a first-order formalism in terms of frames, or vielbeins. This possibility originates from the fact that the generalized metric is not an unconstrained element of $\OO(d,d)$, but is valued in the coset $\OO(d,d)/(\OO(1,d-1)\times \OO(d-1,1))$, like the metric which takes values in $\GL(d)/\OO(1,d-1)$. In analogy with GR, the group we are modding out is called the \emph{double Lorentz group}. Taking into account the additional local symmetries which are mediated by this group, it is possible to encode all the information carried by the generalized metric in a \emph{generalized frame}. However, in contrast to ordinary geometry, this time one needs to impose the two compatibility conditions
\begin{align}
    \cH_{IJ}&=E^A{}_I E^B{}_J \cH_{AB}\,, \qquad\text{and} \nn \\
    \eta_{IJ}&=E^A{}_I E^B{}_J \eta_{AB}\,,
\end{align}
on this frame. The second one arises because the generalized frame is an element of $\OO(d,d)$. Thus, it has to leave invariant the $\OO(d,d)$ metric
\begin{equation}
    \eta_{IJ}=\begin{pmatrix}
        0 & \delta_i{}^j\\
        \delta^i{}_j & 0
    \end{pmatrix}\,,
\end{equation}
often called the \emph{$\eta$-metric},

In contrast to the generalized metric, the generalized frame $E^A{}_I$ is an unconstrained element of $\OO(d,d)$. It relates the flat-indexed constant tensors $\cH_{AB}$ and $\eta_{AB}$ to their counterparts with curved indices. Moreover, the double Lorentz transformations are formally defined as all the transformations which leave $\cH_{AB}$ and $\eta_{AB}$ invariant.

In the following, we will often turn from the splitting \eqref{eq:windsplit} (usually called \emph{winding basis}) to the one that distinguishes between the two factors in the double Lorentz group (\emph{chiral basis}). This latter can be defined by introducing the projections
\begin{equation}\label{eqn:chiralbasis}
    V_{\overline{A}}=P_{AB} V^B\, \qquad \text{and} \qquad V_{\underline{A}}=\overline{P}_{AB} V^B
\end{equation}
on the first and second group in the product. The respective projectors are defined by
\begin{equation}\label{eq:chirproj}
    P_{AB}=\tfrac{1}{2} (\eta_{AB}+\cH_{AB}), \qquad \overline{P}_{AB}=\tfrac{1}{2} (\eta_{AB}-\cH_{AB}).
\end{equation}
For this reason, we will write the double Lorentz group as $\underline{\OO(1,d-1)}\times \overline{\OO(d-1,1)}$, by emphasizing that its two factors correspond to the two kinds of indices in the chiral projection \eqref{eqn:chiralbasis}.

As in differential geometry, a natural next step is to introduce a notion of Lie derivations. However, since in DFT we are dealing with an extension of the ordinary tangent bundle, the \emph{generalized Lie derivative} additionally needs to preserve the $\eta$-metric. After taking this constraint into account, one is left with
\begin{equation}
    \Lc_{\xi} E_A{}^I = \xi^J \partial_J E_A{}^I + (\partial^I \xi_J - \partial_J \xi^I)E_A{}^J\, ,
\end{equation}
for generalized vectors, and with
\begin{equation}
    \Lc_{\xi} e^{-2 \Phi} = \partial_I (\xi^I e^{-2\Phi})\, ,
\end{equation}
for generalized weight 1 densities. To guarantee closure of the Lie derivative, we need to impose the strong constraint/section condition
\begin{equation}
    \partial_I \cdot \partial^I \cdot = 0\,,
\end{equation}
where the $\cdot$ stands for any field, parameter of gauge transformation, or their combinations. A canonical solution to this constraint is requiring that
\begin{equation}
    \partial_I \, \cdot = \begin{pmatrix} \partial_i & \widetilde{\partial}^i \end{pmatrix} \,\cdot = \begin{pmatrix} \partial_i & 0 \end{pmatrix}\,\cdot
\end{equation}
holds, or equivalently imposing that nothing depends on the auxiliary doubled coordinates. This particular solution to the section condition, which we will adopt for the rest of this article, effectively renders DFT equivalent to generalized geometry, where only the tangent bundle is extended but not the spacetime manifold.

The introduction of a Lie derivative allows us to introduce a flux formulation for the theory \cite{Geissbuhler:2013uka} via $\Lc_{E_A}E_B{}^I=F_{AB}{}^C E_C{}^I$, and $\Lc_{E_A} e^{2 \Phi} = F_A e^{2 \Phi}$, where
\begin{align}
    F_{ABC} &= 3 D_{[A} E_B{}^I E_{C]I}\,, \qquad \text{and} \nn \\
    F_A &= 2 D_A \Phi - \partial_I E_A{}^I\,,
\end{align}
are called \emph{generalized fluxes} and play an analogous role to the anholonomy coefficients in GR. $D_A = E_A{}^I \partial_I$ is the generalized flat derivative.

One interesting feature of DFT is that, because $B$-field transformations and diffeomorphisms are embedded in the larger group $\OO(d,d)$, the theory is more restricted. In particular, it is possible to show that the additional constraints imposed by $\OO(d,d)$ can be used to fix the relative coefficients of the three terms in the action \eqref{eq:stringaction}. More precisely, the action can be recast as \cite{Hohm:2010pp}
\begin{equation}
    S=\int \dd^d x e^{-2\Phi}\Rc \,,
\end{equation}
where $\Rc$ is the \emph{generalized Ricci scalar} of the generalized Levi-Civita connection. 

There is a subtle point here, since this connection cannot be anymore fixed just in terms of metric compatibility (now with respect to $\cH$ and $\eta$) and (generalized) torsion constraints. However, the undetermined degrees of freedom of this connection will not appear in the action or in the equations of motion, and therefore can be neglected. Important consequences of this fact will be discussed in section \ref{sec:bootstrap}.

By using the generalized fluxes discussed above, the action can be expanded as \cite{Geissbuhler:2013uka}
\begin{align}
    S=\int \dd^d x e^{-2\Phi} \Big(&2\cH^{AB}D_A F_B -\cH^{AB} F_A F_B +\tfrac{1}{4}\cH^{AD}F_{ABC} F_D{}^{BC} \nn \\
    &-\tfrac{1}{12} \cH^{AB} \cH^{CD}\cH^{EF}F_{ACE}F_{BDF} \nn \\ \label{eq:DFTaction}
    &-\tfrac{1}{6}F^{ABC} F_{ABC}+ F^A F_A -2D_A F^A\Big),
\end{align}
where the last three terms drop out due to the section condition.
By means of the projectors \eqref{eq:chirproj}, one can alternatively rewrite it in the chiral basis as
\begin{equation}\label{eq:hetDFTaction}
    S = \int \dd^d x e^{-2\Phi} \left( 2 D_{\AR} F^{\AR} - F_{\AR} F^{\AR} + \tfrac12 F_{\AL\BR\CR} F^{\AL\BR\CR} + \tfrac16 F_{\AR\BR\CR} F^{\AR\BR\CR} - \text{c.c.} \right),
\end{equation}
where c.c stands for chiral conjugation and swaps under- and over-bars on the indices in each term.

In half-maximal theories, such as the low-energy limits of heterotic and type I string theories, the ordinary $\OO(d,d)$-covariant DFT has to be extended to the so-called \emph{heterotic DFT} \cite{Hohm:2011ex}, which is based on the larger duality group $\OO(d,d+n)$, where $n$ is the dimension of the gauge group $\mathcal{G}$ of the theory. In this case, the action \eqref{eq:DFTaction} has to be modified by replacing the three-, and one-index fluxes with
\begin{align}\label{eq:heteroticDFTfluxes}
    \cF_{\Ac\Bc\Cc} &= 3 \Dc_{[\Ac} \cE_{\Bc}{}^{\Ic} \cE_{\Cc]\Ic} + \cE_{\Ac}{}^{\Ic} \cE_{\Bc}{}^{\Jc} \cE_{\Cc}{}^{\Kc} f_{\Ic\Jc\Kc}\,, \quad \text{and} \\
    \cF_{\cA} &= 2 \Dc_{\cA} \Phi - \partial_{\Ic} \cE_{\Ac}{}^{\Ic}\,,
\end{align}
where $\cE_{\Ac}{}^{\Ic}\in\OO(d,d+n)$, while $f_{\Ic\Jc\Kc}$ are the structure constants of the Lie algebra associated to the gauge group $\mathcal{G}$. As before flat derivatives $\Dc_{\Ac} = \cE_{\Ac}{}^{\Ic} \partial_{\Ic}$ are defined by contraction with the generalized frame. All calligraphic capital Latin indices are evaluated now in $\OO(d,d+n)$. Moreover, the structure constants $f_{\Ic\Jc\Kc}$ have to be totally antisymmetric and satisfy the Bianchi identity
\begin{equation}
    f_{[\Ic\Jc}{}^{\Lc} f_{\Kc]\Lc}{}^{\Mc}=0\,.
\end{equation}

While the expression of the generalized Lie derivative of the density $e^{-2\Phi}$ is preserved, the one for the generalized frame,
\begin{equation}\label{eq:hetgenLie}
    \Lc_{\xi} \cE_{\Ac}{}^{\Ic} = \xi^{\Jc} \partial_{\Jc} \Ec_{\Ac}{}^{\Ic} + (\partial^{\Ic} \xi_{\Jc} - \partial_{\Jc} \xi^{\Ic})\Ec_{\Ac}{}^{\Jc} + f_{\Jc\Kc}{}^{\Ic} \xi^{\Jc} \Ec_{\Ac}{}^{\Kc}\,,
\end{equation}
picks up an additional torsion term. Consequentially, the strong constraint has to be supplemented with the requirement that
\begin{equation}
    f_{\Ic\Jc}{}^{\Kc} \partial_{\Kc} \, \cdot \, = 0\,.
\end{equation}
Its canonical solution then becomes
\begin{equation}
    \partial_{\Ic} \, \cdot \, = \begin{pmatrix} \partial_I & \partial^\alpha \end{pmatrix} \,\cdot= \begin{pmatrix} \partial_I & 0 \end{pmatrix}\,\cdot\,\,,
\end{equation}
where capital Latin indices are $\OO(d,d)$-indices and the Greek ones span the gauge group $\mathcal{G}$, after we perform the branching $\OO(d,d+n)\rightarrow \OO(d,d)\times\mathcal{G}$.

This will be relevant for the following discussion, since the starting point of both the gBdR identification as well as our $\alpha'$-bootstrap procedure is a heterotic-like extension of DFT.

\subsection{The generalized Bergshoeff-de Roo identification}
The gBdR identification relies on an extension of DFT's local symmetries by introducing new, auxiliary, degrees of freedom. The idea is to identify these additional degrees of freedom with existing combination of the physical fields and their derivatives. This way higher-derivative corrections arise.

More explicitly, the starting point is to consider a heterotic-like DFT extension of the standard duality group $\OO(d,d)$ to $\OO(d+k,d+k)$, with a mixed signature gauge group $\mathcal{G}$. Its dimension is $k=p+q$ where $p$ and $q$ keep track of the number of positive ($p$) and negative ($q$) eigenvalues of $\mathcal{G}$'s Killing metric. Performing an $\OO(d,d)$ decomposition, the theory looks like an ordinary $\OO(d,d)$-DFT coupled to $2k$ vectors and $k^2$ scalars. These additional fields appear as components of the generalized frame of the $\OO(d+k,d+k)$-DFT, and transform as generalized connections under the gauge group $\mathcal{G}$. Eventually, they are identified with the mixed-chiral components of the generalized fluxes $\Fc_{\Acl\Bcr\Ccr}, \Fc_{\Acr\Bcl\Ccl}$, treated as a generalized spin connections for the corresponding double Lorentz group. One is then led to the following identification between the corresponding groups:
\begin{equation}\
    \mathcal{G} \sim \underline{\OO(d+q-1,1+p)}\times \overline{\OO(1+p,d+q-1)}\,.
\end{equation}
At this point, one has to pay attention to an important caveat of the procedure. In order to identify these groups, their dimensions have to match. But instead we find
\begin{equation}
    k \ne ( d + k )( d + k - 1)\,.
\end{equation}
To overcome this problem, the procedure requires to take the limit $k\rightarrow\infty$ for the identification to make any sense. In this limit, where one naively encounters a matching between two infinities, it is possible to come up with an iterative procedure to obtain $\alpha'$-corrections order-by-order. Moreover, two parameters $a$ and $b$ (one for each chirality) appear naturally in the identification and control the higher derivative corrections in the effective action. Of extreme importance for us here are the two core ideas, namely
\begin{enumerate}[topsep=2pt,itemsep=1.5pt,parsep=2pt]
    \item the identification of auxiliary degrees of freedom with higher order physical ones, and
    \item the emergence of an infinite dimensional symmetry group.
\end{enumerate}
An important consequence of the procedure is a deformation of the double Lorentz symmetry called \emph{generalized Green-Schwarz transformations}. 

More details about the gBdR mechanism can be found in \cite{Baron:2020xel}, while a pedagogical introduction is provided in \cite{Lescano:2021lup}.

Two main issues with this procedure are its lack of a geometric interpretation and its computational complexity.
In the next section, we will review the approach developed in \cite{Hassler:2024yis,Gitsis:2024gfb}, aimed at solving these problems by transforming the gBdR identification to a geometrically meaningful setting on an extended space. Furthermore, we will present the main results of this article: The infinite dimensional structure group of the theory and how to use it to recovery $\alpha'$-corrections.

\section{The \texorpdfstring{$\alpha'$}{α'}-bootstrap programme}\label{sec:bootstrap}
It is well-known that lifting central concepts from differential geometry to DFT is not straightforward. At the heart of the difficulty lies the fact that in DFT the connection cannot be fully determined in terms of metric compatibility and torsion-freeness. A way to draw a more geometrically intuitive picture was introduced by Pol\'a\v{c}ek and Siegel in~\cite{Polacek:2013nla}. The idea there is to extend the physical space to a so-called \emph{megaspace}, where the presence of an extended generalized flat connection allows one to construct well-behaved geometrical objects (like torsions, Riemann curvature tensors, etc.), all covariant under generalized diffeomorphisms and double Lorentz transformations.

In section~\ref{sec:reviewgBdR} we stressed that our discussion is based on the invariance of the string's dynamic under T-duality. From the target space point of view, in $d$ dimensions the duality group governing this phenomena is $\GD=\OO(d,d)$. Suppose now that we extend the space of the theory by $n$ additional dimensions. This megaspace $\cM$ will be $(d+n)$-dimensional and its duality group will be naturally defined as $\GM=\OO(d+n,d+n)$.

\subsection{The generalized structure group of the megaspace}\label{sec:structgroup}
We now consider a flat connection for $\cM$. In this way all the geometrical information of the theory will be contained in its torsion. However, we are eventually interested only in components of this torsion that have a physical interpretation. We will now proceed to find them.

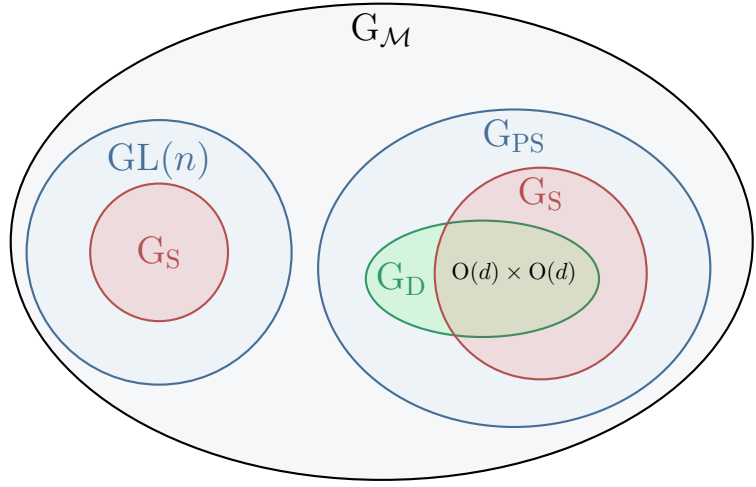
\begin{figure}
\begin{center}
\begin{tikzpicture}[
    scale=0.7,
    font=\sffamily\Large,
    every node/.style={text=black}
]
\definecolor{outerbg}{RGB}{245, 247, 248}
\definecolor{groupA}{RGB}{180, 80, 80}
\definecolor{groupB}{RGB}{70, 110, 150}
\definecolor{lightblue}{RGB}{173,216,230}
    \draw[thick, fill=outerbg] (0,0) ellipse (7cm and 4.5cm);
    \node[anchor=north] at (0, 4.5) {$\GM$};
    \begin{scope}[shift={(-4.2, -0.2)}]
        \draw[thick, groupB, fill=lightblue, fill opacity=0.1] (0,0) circle (2.5cm);
        \node[groupB, anchor=north] at (0, 2.3) {$\GL(n)$};
        \draw[thick, groupA, fill=red, fill opacity=0.1] (0,0) circle (1.3cm);
        \node[groupA, anchor=center] at (0, 0) {$\GS$};
    \end{scope}
    \begin{scope}[shift={(2.5, -0.5)}]
        \draw[thick, groupB, fill=lightblue, fill opacity=0.1] (0,0) ellipse (3.7cm and 3cm);
        \node[groupB, anchor=north] at (0, 3) {$\GPS$};
        \draw[thick, groupB!70!green, fill=green, fill opacity=0.1] (-0.6,-0.2) ellipse (2.2cm and 1.1cm);
        \node[groupB!70!green,anchor=west] at (-2.8, -0.2) {$\GD$};
        \draw[thick, groupA, fill=red, fill opacity=0.1] (0.5,-0.1) circle (2cm);
        \node[groupA,anchor=north] at (0.5, 1.9) {$\GS$};
        \node[scale=0.8,font=\sffamily\normalsize,anchor=center] at (0, -0.1) {$\OO(d)\times\OO(d)$};
    \end{scope}
\end{tikzpicture}
\end{center}
\caption{Venn diagram depicting the relations between the groups in our construction.}\label{fig:venn}
\end{figure}

The first observation one can make is that the \emph{generalized megaframe} $\widehat{E}\in \GM$ can always be parametrized as \cite{Hassler:2024yis}
\begin{equation}\label{eq:tildeEdecomp}
    \widehat{E}=\widetilde{M}\cE \widetilde{V}\,.
\end{equation}
Here, $\cE$ is valued in a subgroup $\GPS \subset \GM$, called the \emph{Pol\'a\v{c}ek-Siegel group}, and it only depends on the physical coordinates $x^i$. $\widetilde{M}$ and $\widetilde{V}$, on the other side, encode structural information and depend on the auxiliary coordinates $y^\mu$ of the additional $n$ dimensions. More precisely, $\widetilde{M}$ is a left-invariant Maurer-Cartan form satisfying
\begin{equation}
    \widetilde{M}^{-1} \dd \widetilde{M} = \hat{t}_\alpha \widetilde{V}_\mu{}^\alpha \dd y^\mu\,.
\end{equation}
From this equation, one sees that a well-defined $\widetilde{M}$ requires the generators $\hat{t}_\alpha$ to form a Lie algebra:
\begin{equation}\label{eq:thatcomm}
    [\hat{t}_\alpha,\hat{t}_\beta]=-f_{\alpha\beta}{}^\gamma \hat{t}_\gamma\,,
\end{equation}
and thus the corresponding structure constants $f_{\alpha\beta}{}^\gamma$ have to satisfy the Jacobi identity
\begin{equation*}\tag{c1}\label{eq:constr1}
    3 f_{[\alpha\beta}{}^\epsilon f_{\gamma]\epsilon}{}^\delta = 0\,.
\end{equation*}
The Lie group which arises from this algebra is the structure group of our theory, $\GS$. As detailed in \cite{Hassler:2024yis,Gitsis:2024gfb}, the generators $\hat{t}_\alpha$ can be decomposed into a part that parametrizes the embedding $\GS\subset\GL(n)$, denoted by $K_\alpha$, and the most general non-trivial action on $\GPS$, denoted as $t_\alpha$:
\begin{equation}\label{eqn:hattalpha}
    \hat{t}_\alpha = K_\alpha+t_\alpha\,.
\end{equation}
To visualize the mutual dependence of all the groups we encounter here, one may refer to the diagram in figure~\ref{fig:venn}. In particular, this picture emphasizes that we are dealing with two different representations of the structure group.

While important for the initial construction, the generators $\hat{t}_\alpha$ are not ideal for the rest of the discussion because they are not fully contained in $\GPS$. In fact, we will see in the following that all the physical information we are interested in is contained in this group. We will now show that there is a better choice of realization for $\GS$ that allows us to deal only with generators contained in $\GPS$. To make this statement clearer, let us take a look at the Lie algebra of $\GPS$. It is spanned by the generators
\begin{equation}
    \mathfrak{g}_{\text{PS}}=\{K_{AB},R_\alpha{}^A,R_{\alpha\beta}\}\,,
\end{equation}
where $K_{AB}$, with Latin indices running from $1$ to $2d$, generates the physical duality group $\GD$. Its commutation relations read
\begin{equation}
    [K_{AB}, K_{CD}]=2\eta_{[A|[C} K_{D]|B]}\,.
\end{equation}
For the other generators, with $\alpha=1,\dots,n$, the additional commutation relations are given by \cite{Hassler:2024yis}
\begin{align}
  \relax[R^A{}_{\alpha}, R^B{}_{\beta}] & = - 2 \kappa_{\alpha\beta}K^{AB} + \eta^{AB}R_{\alpha\beta}\,, \nn \\
  [K_{A B}, R^C{}_{\gamma}] & = -\delta_{[A}{}^C \eta_{B]D}R^D{}_{\gamma}\,, \nn \\
  [R^A{}_{\alpha}, R_{\beta\gamma}] & = -2\kappa_{\alpha[\beta}R^A{}_{\gamma]}\,, \nn \\ \label{eqn: coms} 
  [R_{\alpha\beta}, R_{\gamma\delta}] & = -4\kappa_{[a|[\gamma}R_{\delta]|\beta]}\, .
\end{align}
At this point, we should note an important difference with respect to the original work of Pol\'a\v{c}ek and Siegel: We are twisting the original commutation relations with a symmetric tensor $\kappa_{\alpha\beta}$. We will see later that this tensor plays a fundamental role in the description of higher-derivative corrections.
To these relations, one should then add the commutators of the $\GPS$ generators with the $\GL(n)$ generators needed to define $K_\alpha$ in \eqref{eqn:hattalpha}. They are reported in \cite{Gitsis:2024gfb}, but we will not need them for our discussion. In fact, as anticipated before, we can refine the embedding considering only generators that live in the Lie algebra $\mathfrak{g}_{\mathrm{PS}}$. Assuming that $\kappa_{\alpha\beta}$ can be inverted (at least formally), it is possible to define the new embedding
\begin{equation}\label{eqn:taualpha}
    \tau_\alpha = t_\alpha + R_\alpha\,,
\end{equation}
where
\begin{equation}\label{eq:Ralpha}
    R_\alpha = \tfrac12 f_{\alpha\beta}{}^\gamma \kappa^{\beta\delta} R_{\gamma\delta}\,.
\end{equation}
$R_\alpha$ is a new generator that acts on $\GPS$ generators in the same way as $K_\alpha$. Thus, we are free to choose this second, more convenient, embedding for the following discussion. In particular, we recover
\begin{equation}\label{eqn:structurealg}
    [\tau_\alpha,\tau_\beta] = -f_{\alpha\beta}{}^\gamma \tau_\gamma
\end{equation}
in analogy with \eqref{eq:thatcomm}.

Coming now back to the decomposition of the generators $\tau_\alpha$ in \eqref{eqn:taualpha}, we parametrize
\begin{equation}\label{eqn:paramtalpha}
    t_\alpha = (t_\alpha)_{BC} K^{BC}+(t_\alpha)_{B}{}^\gamma R_\gamma{}^{B}-\tfrac{1}{2} (t_\alpha)^{\beta\gamma} R_{\beta\gamma}
\end{equation}
as the most general embedding $\GS \subset \GPS$, with constants $(t_\alpha)_{BC}$, $(t_\alpha)_B{}^\gamma$, and $(t_\alpha)^{\beta\gamma}$. One can check then that
\begin{equation*}\tag{c2}\label{eqn:linearconst}
    [t_\alpha,t_\beta]=-[R_\alpha, t_\beta]-[t_\alpha, R_\beta]-f_{\alpha\beta}{}^\gamma t_\gamma
\end{equation*}
holds. Finally, as shown in \cite{Hassler:2024yis}, it is only possible to construct the full $\widetilde{V}$ in a consistent way if we impose the condition 
\begin{equation*}\tag{c3}\label{eq:constr3}
    f_{\alpha(\beta}{}^\delta \kappa_{\gamma)\delta}=0\,,
\end{equation*}
meaning that $\kappa_{\gamma\delta}$ has to be invariant under the adjoint action of $\hat{t}_\alpha$ (or equivalently $\tau_\alpha$).

\subsubsection{Characterization of \texorpdfstring{$\GS$}{G\_S}}\label{sec:charGS}

Now that we analyzed the structural properties of $\GS$, there is a central question left:
\begin{center}
    How can we characterize it?
\end{center}
In standard DFT the structure group is usually the double Lorentz group, the maximal compact subgroup of $\GD$.
For the case at hand however further investigation is necessary. First of all, we need to characterize the constants $(t_\alpha)_{BC}$, $(t_\alpha)_B{}^\gamma$, and $(t_\alpha)^{\beta\gamma}$ in \eqref{eqn:paramtalpha}. To this end, one has to specify the embedding $\GS \subset \GPS$ such that the constraints \eqref{eq:constr1}, \eqref{eqn:linearconst}, and \eqref{eq:constr3} are satisfied. In other words, we have to decide which generators of $\GPS$ should be contained in $\GS$. An approach proposed in \cite{Gitsis:2024gfb} is to choose \emph{all of them}. 

To do so, we first notice that the three constraints can be unified in terms of single constraint on the constant, intrinsic torsion\footnote{This part of the generalized torsion on the megaspace arises setting $\cE$ in the decomposition \eqref{eq:tildeEdecomp} as the identity. This situation corresponds to flat space and the physically relevant parts of the generalized torsion vanish. Still, certain components of the generalized torsion remain non-zero. Because they cannot be removed, we call them intrinsic.} $\hat{t}_{\Ah\Bh\Ch}$ of the megaspace as
\begin{equation}\label{eq:modifiedconstraint}
    \hat{t}_{[\Ah\Bh|\Eh} \hat{t}_{\Fh|\Ch\Dh]} \eta^{\Eh\Fh}-4 \hat{t}_{\alpha [\Bh |\Eh} \hat{t}_{\Fh|\Ch\Dh]} \eta^{\Eh\Fh}=0\,.
\end{equation}
In order to interpret this relation, we start decomposing hatted indices by following the convention
\begin{equation}
    V_{\Ah} = \begin{pmatrix} V_\alpha & V_A & V^\alpha \end{pmatrix}, \qquad
    V^{\Ah} = \begin{pmatrix} V^\alpha & V^A & V_\alpha \end{pmatrix}\,. 
\end{equation}
The index mismatch between the two terms in \eqref{eq:modifiedconstraint} indicates that the second term is projected from the full index $\Ah$ to its component $\alpha$. Since the torsion is totally antisymmetric, after writing it in components, we are left only with
\begin{equation}
    \hat{t}_{\Ah\Bh\Ch} = \begin{cases}
        \hat{t}_{\alpha\beta\gamma} &= - f_{\alpha\beta\gamma}= -f_{\alpha\beta}{}^\delta \kappa_{\delta\gamma}  \\
        \hat{t}_{\alpha\beta}{}^\gamma &= f_{\alpha\beta}{}^\gamma \\
        \hat{t}_{\alpha BC} &= (t_\alpha)_{BC} \\
        \hat{t}_{\alpha B}{}^\gamma &= (t_\alpha)_B{}^\gamma \\
        \hat{t}_{\alpha}{}^{\beta\gamma} &= (t_\alpha)^{\beta\gamma} \\
        \text{otherwise} &=0\,.
    \end{cases}
\end{equation}
The first equation stems from the antisymmetricity of $t_{\alpha\beta\gamma}$ and the $ad$-invariance of $\kappa_{\alpha\beta}$ required by \eqref{eq:constr3}. In order to analyze \eqref{eq:modifiedconstraint} further, we also need the components of the metric~\cite{Hassler:2024yis}
\begin{equation}
    \eta^{\Ah\Bh}=\begin{pmatrix}
        -\kappa_{\alpha\beta} & 0 & \delta_\alpha{}^\beta\\
        0 & \eta_{AB} & 0\\
        \delta^\alpha{}_\beta & 0 & 0
    \end{pmatrix}.
\end{equation}
Expanding \eqref{eq:modifiedconstraint}, one immediately gets the constraints \eqref{eq:constr1} and \eqref{eq:constr3}, while \eqref{eqn:linearconst} translates to
\begin{equation}\label{eqn:linearconstr2}
    \boxed{\Xi^{\Ac\Bc\gamma\delta}\left[\tfrac{1}{2} f_{\gamma\delta}{}^\mu (t_{\mu})_{\Ac\Bc} - (t_{\gamma})_{\Ac\Mc} (t_{\delta})_{\Bc\Nc} \eta^{\Mc \Nc}+2 (t_{\gamma})_{\Ac}{}^\mu f_{\delta\mu\Bc}\right]=0\,.}
\end{equation}
To avoid writing three independent equations, we introduced calligraphic indices that split according to
\begin{equation}
    V_{\Ac} = \begin{pmatrix} V_A & V^\alpha \end{pmatrix}\,,
    \qquad
    V^{\Ac} = \begin{pmatrix} V^A & V_\alpha \end{pmatrix}\,.
\end{equation}
They are raised/lowered with the metric
\begin{equation}\label{eq:modifiedeta}
    \eta^{\Ic\Jc} = \begin{pmatrix}
        \eta^{IJ} & 0 \\
        0 & \kappa_{\mu\nu}
    \end{pmatrix}\,,  \qquad
    \eta_{\Ic\Jc} = \begin{pmatrix}
        \eta_{IJ} & 0 \\
        0 & \kappa^{\mu\nu}
    \end{pmatrix}\,.
\end{equation}
Moreover, we introduced the symbols $\Xi^{\Ac\Bc \gamma \delta}$ that antisymmetrize quantities over $\Ac\Bc$ and $\gamma\delta$.

We now have to solve the constraint \eqref{eqn:linearconstr2}. Our approach is based on three observations:
\begin{enumerate}[topsep=2pt,itemsep=1.5pt,parsep=2pt]
    \item $f_{\alpha\beta}{}^\gamma$ appears linearly in the constraint, while
    \item $(t_{\alpha})_{\Bc\Cc}$ appears both linearly and quadratically,
    \item but we can choose $(t_{\alpha})_{\Bc\Cc}$ freely (provided it is antisymmetric in the last two indices), since no other constraints involve it.
\end{enumerate}
The strategy is therefore to choose the constants $(t_{\alpha})_{\Bc\Cc}$ in a convenient way to easily solve \eqref{eqn:linearconstr2} for the structure coefficients $f_{\alpha\beta}{}^\gamma$. This can be done introducing an operation that allows for a Greek index to be split into multiple calligraphic ones. To see how this works, let us consider a vector $V^\alpha$, and define
\begin{equation}\label{eq:splittingindices}
    V^\alpha (t_{\alpha})_{\Bc\Cc} = \begin{pmatrix}
        \phantom{-} V^\alpha (t_{\alpha})_{BC} & V^\alpha (t_{\alpha})_B{}^{\gamma} \\
        - V^\alpha (t_{\alpha})_C{}^{\beta} & V^\alpha (t_\alpha)^{\beta\gamma}
    \end{pmatrix} := V_{\la\Bc\Cc\ra}\,.
\end{equation}
Note that due to the antisymmetry in $\Bc$ and $\Cc$, the only independent components of $V_{\Bc\Cc}$ are $V_{BC}$, $V_B{}^{\gamma}$, and $V^{\beta\gamma}$. To have an explicit one-to-one correspondence between $V^\alpha$ and $V_{\Bc\Cc}$, we introduced the notation
\begin{equation}
    K: \quad V_{\la BC\ra } = V_{BC}\,, \qquad
    R_1: \quad V_{\la B}{}^{\gamma\ra } = V_B{}^{\gamma} = - V^{\gamma}{}_B\,, \qquad
    R_2: \quad V^{\la \beta\gamma\ra } = V^{\beta\gamma}\,,
\end{equation}
such that a contracted Greek index can be expanded as
\begin{equation}
    V^\alpha t_\alpha = V_{\la A_1 A_2\ra} t^{\la A_1 A_2\ra} + 2 V_{\la A_1}{}^{\alpha_2\ra} t^{\la A_1}{}_{\alpha_2\ra} + V^{\la \alpha_1 \alpha_2\ra} t_{\la \alpha_1 \alpha_2\ra}\,.
\end{equation}
Here we also label the respective components of $V_{\la\Bc\Cc\ra}$ according to the generators they are contracted with. $K$ stands for $K_{BC}$, $R_1$ for $R_\gamma^B$, and $R_2$ for $R_{\beta\gamma}$. As a result of this prescription, we realize the $(t_\alpha)_{\Bc_1\Bc_2}$ as
\begin{equation}
    (t^{\la\Ac_1\Ac_2\ra})_{\Bc_1\Bc_2} = \delta^{[\Ac_1}_{\Bc_1}\delta^{\Ac_2]}_{\Bc_2}\,,
\end{equation}
from which we obtain the components
\begin{alignat}{2}
    K: \qquad & (t^{\la A_1 A_2\ra })_{B_1 B_2}  &= \delta^{[A_1}_{B_1} \delta^{A_2]}_{B_2}\,, \nn \\
    R_1: \qquad &(t^{\la A}{}_{\alpha\ra })_{B}{}^{\beta} &= \tfrac12 \delta^A_B \delta_\alpha^\beta\,, \nn \\
    R_2: \qquad &(t_{\la \alpha_1 \alpha_2\ra })^{\beta_1 \beta_2} &= \delta_{[\alpha_1}^{\beta_1} \delta_{\alpha_2]}^{\beta_2}\,.
\end{alignat}
It is crucial to note that the generators $t_\alpha$ defined in this way \emph{do not form a Lie algebra} (as a result of the Jacobi identity violation of \eqref{eq:modifiedconstraint}). They rather give rise to an anticommutative algebra.

We can now solve the linear constraint \eqref{eqn:linearconstr2} to get
\begin{equation}\label{eq:expandedconstr}
    f^{\la \Ac_1 \Ac_2 \ra \la \Bc_1 \Bc_2 \ra}{}_{\la \Cc_1 \Cc_2 \ra}=  2 \delta_{[\Cc_1}^{[\Ac_1} \eta^{\Ac_2][\Bc_2} \delta_{\Cc_2]}^{\Bc_1]}-2\delta_{[\Cc_1}^{[\Ac_1}f^{\la \Bc_1 \Bc_2 \ra | \Ac_2]}{}_{\Cc_2]} +2 \delta_{[\Cc_1}^{[\Bc_1}f^{\la \Ac_1 \Ac_2 \ra | \Bc_2]}{}_{\Cc_2]}\,.
\end{equation}
The important feature of this relation is that it allows to obtain the structure constants of $\GS$ in a recursive way, where the only input needed is $\kappa_{\alpha\beta}$.

\subsubsection{The choice of \texorpdfstring{$\kappa_{\alpha\beta}$}{kappa\_alphabeta}}
But now we are left with a second question:
\begin{center}
    How do we choose $\kappa_{\alpha\beta}$?
\end{center}
An exhaustive answer to this question would amount to classify all the possible $ad$-invariant \eqref{eq:constr3} symmetric bilinears we can form. We delay it to an upcoming work, and here only look at the most obvious (and relevant) example. This is given by a regularized version of the Killing metric for $\GS$. Starting from 
\begin{equation}
    \kappa^{\la\Ac_1 \Ac_2\ra\la\Bc_1\Bc_2\ra} = f^{\la\Ac_1 \Ac_2\ra \la\Cc_1 \Cc_2\ra}{}_{\la\Dc_1\Dc_2\ra}f^{\la\Bc_1\Bc_2\ra\la\Dc_1\Dc_2\ra}{}_{\la\Cc_1\Cc_2\ra}\,,
\end{equation}
one finds that
\begin{align}
    \kappa^{\la\Ac_1 \Ac_2\ra\la\Bc_1\Bc_2\ra}=\mathcal{N}\Big(&\kappa^{\la\Ac_1 \Ac_2\ra\la\Bc_1\Bc_2\ra}-\eta^{\Ac_1[\Bc_1}\eta^{\Ac_2|\Bc_2]} \nn \\
    &- f^{\la\Ac_1\Ac_2\ra [\Bc_1}{}_{\Dc} \eta^{\Dc|\Bc_2]}- f^{\la\Bc_1\Bc_2\ra [\Ac_1}{}_{\Dc} \eta^{\Dc|\Ac_2]}\Big)+\mathcal{O}(1)\,,
\end{align}
where $\mathcal{N}$ is the dimension of the algebra, or equivalently, the range of the calligraphic indices $\Ac=1, \dots, \mathcal{N}$. The procedure of splitting indices explained above makes $\mathcal{N}$ infinite, thus we regularize the expression by considering the limit $\lim\limits_{\mathcal{N}\rightarrow\infty} \tfrac{1}{\mathcal{N}} \kappa^{\la\Ac_1 \Ac_2\ra\la\Bc_1\Bc_2\ra}=0$, such that
\begin{equation}\label{eq:splitkappa}
    \kappa^{\la\Ac_1 \Ac_2\ra\la\Bc_1\Bc_2\ra}=\eta^{\Ac_1[\Bc_1}\eta^{\Ac_2|\Bc_2]}+ f^{\la\Ac_1\Ac_2\ra [\Bc_1}{}_{\Dc} \eta^{\Dc|\Bc_2]}+ f^{\la\Bc_1\Bc_2\ra [\Ac_1}{}_{\Dc} \eta^{\Dc|\Ac_2]}\,.
\end{equation}
Again this procedure is fully recursive, and it only needs knowledge of $\eta_{AB}$.

Following \eqref{eqn:chiralbasis},  we perform a chiral splitting of indices
\begin{equation}\label{eq:chiralsplitting}
    V_{A} = \begin{pmatrix} V_{\al} & V_{\ar} \end{pmatrix}\,,
    \qquad\text{and}\qquad
    V^{A} = \begin{pmatrix} V^{\al} & V^{\ar} \end{pmatrix}\,,
\end{equation}
to bring the $\eta$-metric into the diagonal form
\begin{equation}
    \eta_{AB} = \begin{pmatrix} \eta_{\al\bl} & 0 \\
    0 & \eta_{\ar\br} \end{pmatrix} = \begin{pmatrix} + \eta_{a b} & 0 \\ 0 & -\eta_{a b} \end{pmatrix}\,.
\end{equation}
Here $\eta_{ab}$ is the flat Minkowski metric. Due to the diagonal form of $\eta_{AB}$, $\GS$ decomposes into a direct product of a left and a right chiral sector. The splitting \eqref{eq:chiralsplitting} naturally uplifts to calligraphic indices with
\begin{equation}
    V^{\Acl} = \begin{pmatrix} V^{\al} & V_{\alphaL} \end{pmatrix}\,,
    \qquad\text{and}\qquad
    V^{\Acr} = \begin{pmatrix} V^{\ar} & V_{\alphaR} \end{pmatrix}\,.
\end{equation}
In the end, the most general $\kappa_{\alpha\beta}$ which arises from the Killing metric carries two parameters: $a$ for the right-moving sector, and $b$ for the left-moving one. Hence, we define them as\footnote{Note that we choose the explicit prefactors to match with the conventions of \cite{Baron:2020xel}.}
\begin{align}
    \kappa^{\la\Acr_1 \Acr_2\ra\la\Bcr_1\Bcr_2\ra}&\rightarrow\phantom{-}\tfrac{a}{2}\kappa^{\la\Acr_1 \Acr_2\ra\la\Bcr_1\Bcr_2\ra}\,, \nn \\ \label{eq:kappasplitchiral}
    \kappa^{\la\Acl_1 \Acl_2\ra\la\Bcl_1\Bcl_2\ra}&\rightarrow-\tfrac{b}{2}\kappa^{\la\Acl_1 \Acl_2\ra\la\Bcl_1\Bcl_2\ra}\,.
\end{align}

The recursive nature of these equations encode a very interesting feature: Once we turn on the $a$ or $b$ deformation at the first order, it automatically propagates to higher ones. This is the core feature of our procedure, and the mechanism that allows to recover the full $a$ and $b$ towers of corrections.

From \eqref{eq:expandedconstr} and \eqref{eq:splitkappa} one also finds that $f_{\alpha\beta\gamma}=f_{\alpha\beta}{}^\delta \kappa_{\delta\gamma}$ is completely fixed by
\begin{align}
    f^{\la \Acr_1 \Acr_2 \ra \la \Bcr_1 \Bcr_2 \ra \la \Ccr_1 \Ccr_2 \ra}=a&\left( \eta^{[\Acr_1|[\Bcr_2}\eta^{\Bcr_1][\Ccr_1}\eta^{\Ccr_2]|\Acr_2]} + f^{\la \Acr_1\Acr_2\ra [\Bcr_1}{}_{\Dcr} f^{\la \Ccr_1\Ccr_2\ra \Dcr|\Bcr_2]}  \right. \nn \\
    &-f^{\la \Bcr_1\Bcr_2\ra [\Acr_1}{}_{\Dcr} f^{\la \Ccr_1\Ccr_2\ra \Dcr|\Acr_2]} +f^{\la \Acr_1\Acr_2\ra [\Bcr_2 |[\Ccr_2} \eta^{|\Bcr_1]|\Ccr_1]} \nn \\
    &\left.-f^{\la \Bcr_1\Bcr_2\ra [\Acr_2 |[\Ccr_2} \eta^{|\Acr_1]|\Ccr_1]}+f^{\la \Ccr_1\Ccr_2\ra [\Acr_2 |[\Bcr_2} \eta^{|\Acr_1]|\Bcr_1]}\right)\,, \nn \\
    f^{\la \Acl_1 \Acl_2 \ra \la \Bcl_1 \Bcl_2 \ra \la \Ccl_1 \Ccl_2 \ra}=-b &\left(\eta^{[\Acl_1|[\Bcl_2}\eta^{\Bcl_1][\Ccl_1}\eta^{\Ccl_2]|\Acl_2]} + f^{\la \Acl_1\Acl_2\ra [\Bcl_1}{}_{\Dcl} f^{\la \Ccl_1\Ccl_2\ra \Dcl|\Bcl_2]}  \right. \nn \\
    &-f^{\la \Bcl_1\Bcl_2\ra [\Acl_1}{}_{\Dcl} f^{\la \Ccl_1\Ccl_2\ra \Dcl|\Acl_2]} +f^{\la \Acl_1\Acl_2\ra [\Bcl_2 |[\Ccl_2} \eta^{|\Bcl_1]|\Ccl_1]} \nn \\ \label{eq:fdownsplitchiral}
    &\left.-f^{\la \Bcl_1\Bcl_2\ra [\Acl_2 |[\Ccl_2} \eta^{|\Acl_1]|\Ccl_1]}+f^{\la \Ccl_1\Ccl_2\ra [\Acl_2 |[\Bcl_2} \eta^{|\Acl_1]|\Bcl_1]}\right).
\end{align}

\subsubsection{The physical frame}
Now that we have fully characterized the structure algebra, and as a consequence $\widetilde{M}$ and $\widetilde{V}$, we can focus on the generalized frame $\cE$ living in the subgroup $\GPS$.

To extract the physical information from our construction, we have to further reduce the frame in $\GPS$ to the physical duality group $\GD$. Practically this is done by identifying a set of auxiliary fields $\cA$ in terms of the $\GD$-frame $E$ and its derivatives in the spirit of the original BdR identification \cite{Bergshoeff:1989de} and its generalizations \cite{Baron:2018lve,Baron:2020xel}. Consequentially, the frame $\cE$ is splits into
\begin{equation}\label{eqn:decompgenframe}
    \cE = E\cA\,,
\end{equation}
where $E\in \GD$ is the physical frame, and $\cA$ is a coset element in $ \GD\setminus \GPS $. Finally, in order to fix $\cA$ completely, we resort to two mechanisms:
\begin{enumerate}[topsep=2pt,itemsep=1.5pt,parsep=2pt]
    \item Partial gauge fixings of the structure group $\GS$\,, and
    \item torsion constraints on the megaspace.
\end{enumerate}

\subsection{Torsion constraints and gauge fixing}\label{sec:torsionconst}
The main idea of the Pol\'a\v{c}ek-Siegel approach to the gBdR mechanism is to reinterpret the identifications as constraints on the extrinsic\footnote{As the name suggests, this is everything which is left from the full generalized torsion on the megaspace after removing the intrinsic torsion which we encountered in section~\ref{sec:charGS}.} part of the megaspace's generalized torsion.

A key observation of~\cite{Hassler:2024yis, Gitsis:2024gfb}, is that the relevant torsion components can be restricted to $\GPS$, where they take the simple form
\begin{align}
    \Tc_{\Ac\Bc\Cc}&=\cF_{\Ac\Bc\Cc}+3\Ac_{[\Ac}{}^{\delta}(\tau_\delta)_{|\Bc\Cc]}\,,\label{eq:megaspacetorsion}\\
    \Tc_{\Ac}&=\cF_{\Ac}+3\Ac_{\Bc}{}^{\delta}(\tau_\delta)^{\Bc}{}_{\Ac}\,. 
\end{align}
Here $\cF_{\Ac\Bc\Cc}$ and $\cF_{\Ac}$ denote the heterotic generalized fluxes, which we already discussed in section~\ref{sec:introDFT}, and the additional terms are the components of the generalized higher connections $\Ac$ appearing in the decomposition~\eqref{eqn:decompgenframe}.

The gBdR-like identification will then correspond to setting certain components of the generalized torsions to zero, such that the corresponding fluxes completely determine the respective components of the higher connections in $\cA_{\Ac\Bc}$. As shown in~\cite{Hassler:2024yis, Gitsis:2024gfb}, in order to do this it is sufficient to impose constraints only on the three-index torsion, leaving the dilatonic one untouched. However, due to our choice for $\GS$, out of the four projections $\Tc_{\Acl\Bcl\Ccl}, \Tc_{\Acl\Bcl\Ccr}, \Tc_{\Acr\Bcr\Ccl}, \Tc_{\Acr\Bcr\Ccr}$, the fully-chiral ones pick up a shift term in their transformation. Hence, setting them to zero is not a covariant constraint, leaving us able to fix to zero only the mixed ones, that is
\begin{align}
    \Tc_{\Acl\Bcl\Ccr}&=0\,, \nn \\ \label{eq:gBdRi}
    \Tc_{\Acr\Bcr\Ccl}&=0\,.
\end{align}
Doing so completely specifies the mixed-chiral components of $\cA_{\Ac\Bc}$, but leaves undetermined $\Ac_{\Acl\Bcl}$ and $\Ac_{\Acr\Bcr}$. The idea is then to exploit the gauge freedom and to perform appropriate $\GS$-transformations to fix these remaining components too.

To better understand the resulting gauge fixing constraints, it is convenient to look at the explicit parametrization of the frame presented in \eqref{eqn:decompgenframe}, namely
\begin{equation}\label{eqn:mega-frame}
    \cE_{\Ac}{}^{\Jc} = E_{\Ac}{}^{\Ic} \cA_{\Ic}{}^{\Jc}\,,
\end{equation}
with
\begin{equation}\label{eqn:paramcE}
    E_{\Ac}{}^{\Ic} = \begin{pmatrix}
        E_A{}^I & 0 \\
        0 & \delta^\alpha{}_\nu
    \end{pmatrix}, \qquad
    \mathcal{A}_{\Ic}{}^{\Jc} = \begin{pmatrix}
        \delta_I{}^J - \tfrac12 A_I{}^\rho A^{J\lambda} \kappa_{\rho\lambda} & A_I{}^\rho \kappa_{\rho\nu} \\
        - \Pi^\mu{}_\rho A^{J\rho} & \Pi^\mu{}_\nu
    \end{pmatrix}.
\end{equation}
As we will see in the next sections, $A_I{}^\mu$ and $\Pi^\mu{}_\nu$ are the basic fields we have to identify with higher geometrical objects in order to recover the $\alpha'$-corrected action.

The partial gauge fixing we want to impose takes the form\footnote{We cannot thank enough Daniel Butter who suggested this gauge fixing. It simplifies computations considerably and thereby makes the rest of the discussion feasible.}
\begin{equation}\label{eqn:AAzero}
  A_I{}^\mu A^{I\nu} = A_i{}^\mu A^{i\nu} + A^{i\mu} A_i{}^\nu = 0\,.  
\end{equation}
We would like the one-form $A_i{}^{\alpha}$ to be fixed by the torsion constraints, like in the original BdR identification. Therefore, we pick the solution 
\begin{equation}\label{eq:gaugefixA}
    A^{i\alpha}=0
\end{equation}
to \eqref{eqn:AAzero}, and thereby complete our gauge fixing procedure. 

As en element of $\GPS$, $\cE_{\Ac}{}^{\Ic}$ has to leave the metric
\begin{equation}
    \eta_{\Ic\Jc} = \begin{pmatrix}
        \eta_{IJ} & 0 \\
        0 & \kappa^{\mu\nu}
    \end{pmatrix}
\end{equation}
invariant. Consequentially, $\Pi^\mu{}_\nu$ is a constrained field. A natural choice for it, motivated by the observation that $\cE \in \GPS$, would be $\Pi=\exp(\rho^{\alpha\beta}R_{\alpha\beta})$. But recalling that $\kappa_{\alpha\beta}$ plays the role of an invariant metric for the subgroup of $\GPS$ generated by $\{R_{\alpha\beta}\}$, we find the general constrain
\begin{equation}\label{eqn:constrPI}
    \Pi^\mu{}_{\rho} \Pi^\nu{}_\lambda \kappa^{\rho\lambda} = \kappa^{\mu\nu}\,.
\end{equation}
Together with \eqref{eqn:AAzero}, this is sufficient to check that 
\begin{equation}
    \eta_{\Ac\Bc} = \cE_{\Ac}{}^{\Ic} \eta_{\Ic\Jc} \cE_{\Bc}{}^{\Jc} = \begin{pmatrix}
        \eta_{AB} & 0 \\
        0 & \kappa^{\alpha\beta}
    \end{pmatrix}
\end{equation}
holds as required.

Now that we have analyzed the constraints required for the identification, we are ready to compute explicitly the $\GPS$ components of the generalized fluxes on the megaspace.

\subsection{Generalized fluxes}\label{sec:genflux}
In analogy with the ordinary heterotic-DFT (see for example \eqref{eq:heteroticDFTfluxes}), we define the generalized fluxes corresponding to the frame $\cE_{\Ac}{}^{\Ic}$ as
\begin{equation}
    \cF_{\Ac\Bc\Cc} = 3 \cE_{[\Ac|}{}^I \partial_I \cE_{|\Bc|}{}^{\Jc} \cE_{|\Cc]\Jc} + \cE_{\Ac}{}^{\mu} \cE_{\Bc}{}^{\nu} \cE_{\Cc}{}^{\rho} f_{\mu\nu\rho}\,.
\end{equation}
Employing the parametrization \eqref{eqn:mega-frame}, we get
\begin{equation}\label{eqn:ident}
    \cF_{\Ac\Bc\Cc} = F_{\Ac\Bc\Cc} + E_{\Ac}{}^{\Ic} E_{\Bc}{}^{\Jc} E_{\Cc}{}^{\Kc} \left( 3 \partial_{[\Ic|} \cA_{|\Jc|}{}^{\Lc} \cA_{|\Kc]\Lc} + \cA_{\Ic}{}^\rho \cA_{\Jc}{}^\lambda \cA_{\Kc}{}^\kappa f_{\rho\lambda\kappa} \right).
\end{equation}
In order to make further progress on the first term, namely the fluxes $F_{\Ac\Bc\Cc}$, it is convenient to parametrize the generalized frame in the standard way with
\begin{equation}
    E_A{}^I = \frac{1}{\sqrt{2}}\begin{pmatrix}
        e_{\al}{}^i && e_{\al}{}^j B_{j i} + e_{\al i}\\
        e_{\ar}{}^i && e_{\ar}{}^j B_{j i} - e_{\ar i}
    \end{pmatrix},
\end{equation}
and
\begin{equation}\label{eqn:frame}
    E^{AI} = \frac{1}{\sqrt{2}}\begin{pmatrix}
        \phantom{-}e^{\al i} && \phantom{-}e^{\al j} B_{j i} + e^{\al}{}_i\\
        -e^{\ar i} && -e^{\ar j} B_{j i} + e^{\ar}{}_i
    \end{pmatrix}.
\end{equation}
From this choice, we can immediately compute the non-vanishing contributions of $F_{\Ac\Bc\Cc}$ as
\begin{align}
    F_{\ar\br\Cr} &= \tfrac{1}{2\sqrt{2}}(H_{\ar\br\Cr} - 6 \omega_{[\ar\br\Cr]})\,, &  &&
    F_{\al\br\Cr} &= \tfrac{1}{2\sqrt{2}}(H_{\al\br\Cr} - 2 \omega_{\al\br\Cr})\,, \nn \\
    F_{\al\bl\Cr} &= \tfrac{1}{2\sqrt{2}}(H_{\al\bl\Cr} + 2 \omega_{\Cr\al\bl})\,, & &\text{and}&
    F_{\al\bl\cl} &= \tfrac{1}{2\sqrt{2}}(H_{\al\bl\cl} + 6 \omega_{[\al\bl\cl]})\,,
\end{align}
where $\omega_{ibc}$ is the spin connection associated with the frames $e_a{}^i$, and $\omega_{abc}=e_a{}^i \omega_{ibc}$. The second term in \eqref{eqn:ident},
\begin{equation}\label{eqn:gCSterm}
    \Omegat_{\Ic\Jc\Kc} = 3 \partial_{[\Ic|} \cA_{|\Jc|}{}^{\Lc} \cA_{|\Kc]\Lc} + \cA_{\Ic}{}^\rho \cA_{\Jc}{}^\lambda \cA_{\Kc}{}^\kappa f_{\rho\lambda\kappa}\,,
\end{equation}
requires more attention. It has the structure of a Chern-Simons form, and therefore we will call it a \emph{generalized Chern-Simons} term.
From \eqref{eq:gaugefixA}, one finds that the independent components of~\eqref{eqn:gCSterm} are
\begin{align}
    \Omegat^{\mu\nu\rho}   &= f_{\alpha\beta\gamma} \Pi^{\mu\alpha} \Pi^{\nu\beta} \Pi^{\rho\gamma}\,, \label{eq:omegamunurho}\\
    \Omegat_i{}^{\nu\rho}  &= \left(\partial_i \Pi^{\nu\alpha} + A_i{}^\beta \Pi^{\nu\gamma} f_{\beta\gamma}{}^\alpha \right) \Pi^\rho{}_\alpha\,, \label{eq:omegainurho}\\
    \Omegat_{ij}{}^{\rho}  &= \left( 2 \partial_{[i} A_{j]}{}^\alpha + A_i{}^\beta A_j{}^\gamma f_{\beta\gamma}{}^\alpha \right) \Pi^\rho{}_\alpha\,, \qquad\text{and}\label{eq:omegaijrho}\\
    \Omegat_{ijk}          &= 3 \partial_{[i} A_j{}^\alpha A_{k]}{}^\beta \kappa_{\alpha\beta} + A_i{}^\alpha A_j{}^\beta A_k{}^\gamma f_{\alpha\beta\gamma}\,.\label{eq:omegaijk}
\end{align}

Therefore, for the components of~\eqref{eqn:ident} which are relevant to the identification of $A_i{}^\alpha t_\alpha$ and $\Pi^{\alpha\beta} t_\beta$, one finds
\begin{align}
    K: && \cF_{\aL\bR\cR} &= \tfrac{1}{2\sqrt{2}} \left( \Omegat_{abc} - 2 \omega^{(-)}_{abc} \right), &  
    \cF_{\aR\bL\cL} &= \tfrac{1}{2\sqrt{2}} \left( \Omegat_{abc} + 2 \omega^{(+)}_{abc} \right),\nn
    \\ 
    R_1: && \cF_{\aL\bR}{}^{\gammaR} &= \tfrac{1}{2}  \Omegat_{ab}{}^{\gammaR}\,, & 
    \cF_{\aR\bL}{}^{\gammaL} &= \tfrac{1}{2}  \Omegat_{ab}{}^{\gammaL}\,, \nn
    \\ \label{eq:identflatin}
    R_2: && \cF_{\aL}{}^{\betaR\gammaR} &= \tfrac{1}{\sqrt{2}} \Omegat_a{}^{\betaR\gammaR}\,, &
    \cF_{\aR}{}^{\betaL\gammaL} &= \tfrac{1}{\sqrt{2}} \Omegat_a{}^{\betaL\gammaL}\,,
\intertext{and}
    K: && \cF^{\alphaL}{}_{\bR\cR} &= \tfrac12 \Omegat_{bc}{}^{\alphaL}\,, &
    \cF^{\alphaR}{}_{\bL\cL} &= \tfrac12 \Omegat_{bc}{}^{\alphaR}\,, \nn\\
    R_1: && \cF^{\alphaL}{}_{\bR}{}^{\gammaR} &= \tfrac{1}{\sqrt{2}} \Omegat_{b}{}^{\gammaR\alphaL}\,, &
    \cF^{\alphaR}{}_{\bL}{}^{\gammaL} &= \tfrac{1}{\sqrt{2}} \Omegat_{b}{}^{\gammaL\alphaR}\,, \nn \\ \label{eq:identfgreek}
    R_2: && \cF^{\alphaL\betaR\gammaR} &= \Omegat^{\alphaL\betaR\gammaR}\,, &
    \cF^{\alphaR\betaL\gammaL} &= \Omegat^{\alphaR\betaL\gammaL}\,,
\end{align}
where, for convenience, we have introduced the quantities $\omega^{(\pm)}_{iab} = \omega_{iab} \pm \tfrac12 H_{iab}$. Useful relations involving these modified connections are presented in appendix \ref{app:geometry}.

In the same vein, we compute the dilatonic one-index flux
\begin{equation}
    \cF_{\Ac} = 2 \Ec_{\Ac}{}^I \partial_I \Phi - \partial_I \Ec_{\Ac}{}^I\,.
\end{equation}
Using the parametrization \eqref{eqn:mega-frame}, this expression simplifies to
\begin{equation}
    \cF_\cA = \begin{cases} F_A = 2 E_A{}^I \partial_I \Phi - \partial_I E_A{}^I\,,\\
    \cF^\alpha=0\,.
    \end{cases}
\end{equation}
As pointed out before in section~\ref{sec:torsionconst}, the dilatonic flux is not relevant for the identification and will only appear in the action, as we discuss in section~\ref{sec:action}.

\subsection{The identification mechanism}\label{sec:gBdR}
Now that we have the explicit expressions of the fluxes, we will come to the actual identification. As we noticed previously, we have to consider the megaspace torsion \eqref{eq:megaspacetorsion} and impose the torsion and gauge fixing constraints \eqref{eq:gBdRi}. This leads to the the identification
\begin{equation}
    \cA_{\Acl}{}^{\deltaR} (\tau_{\deltaR})_{\Bcr\Ccr} = - \cF_{\Acl\Bcr\Ccr}\,, \quad\text{and}\quad
    \cA_{\Acr}{}^{\deltaL} (\tau_{\deltaL})_{\Bcl\Ccl} = - \cF_{\Acr\Bcl\Ccl}\,,
\end{equation}
where the generators $(\tau_\delta)_{\Bc\Cc}$ perform a similar splitting of indices as described in \eqref{eq:splittingindices}, allowing for the writing of covariant expressions relating the projections of $\cA_{\Ac}{}^{\delta}$ and $\cF_{\Ac\Bc\Cc}$.
However, we know from \eqref{eqn:taualpha} that the difference between the $\tau_\alpha$ and $t_\alpha$ generators is $R_\alpha$ from \eqref{eq:Ralpha}. Therefore, we can rewrite the expressions as
\begin{equation}\label{eq:tRidentification}
    \cA_{\Acl}^{(n)\deltaR} (t_{\deltaR})_{\Bcr\Ccr} + \cA_{\Acl}^{(n)\deltaR} (R_{\deltaR})_{\Bcr\Ccr} = - \cF^{(n)}_{\Acl\Bcr\Ccr}\,, \quad\text{and}\quad
    \cA_{\Acr}^{(n)\deltaL} (t_{\deltaL})_{\Bcl\Ccl} + \cA_{\Acr}^{(n)\deltaL} (R_{\deltaL})_{\Bcl\Ccl} = - \cF^{(n)}_{\Acr\Bcl\Ccl}\,.
\end{equation}
Here we introduced the superscript $(n)$ to label terms which contain $n$ derivatives. This notation will be extremely useful in the following because, as we noticed in the previous sections, the identification procedure is recursive in the order of derivatives.

It was shown in \cite{Gitsis:2024gfb} that
\begin{equation}
    \cF^{(n)}_{\Acl\Bcr\Ccr} = - \cA_{\Acl}^{(n)\deltaR} (R_{\deltaR})_{\Bcr\Ccr} + \cF^{(n)}_{\Acl\Bcr\Ccr}(\mathcal{A}^{(<n)})\,, \quad\text{and}\quad
    \cF^{(n)}_{\Acr\Bcl\Ccl} = - \cA_{\Acr}^{(n)\deltaL} (R_{\deltaL})_{\Bcl\Ccl} + \cF^{(n)}_{\Acr\Bcl\Ccl}(\mathcal{A}^{(<n)})\,,
\end{equation}
where the notation $\cF_{\Ac\Bc\Cc}^{(n)}(\mathcal{A}^{(<n)})$ indicates all the terms with $n$ derivatives appearing in $\cF_{\Ac\Bc\Cc}$ that are built in terms of $\cA$'s up to order $n-1$. Since we want the procedure to be fully recursive, we need to isolate on the left-hand side all the quantities defined at the $n$-th step and on the right-hand side all the ones depending only on the previous steps. Therefore, exploiting \eqref{eq:tRidentification}, we can write the identification as\footnote{To compare with the gBdR procedure, this equation corresponds with equation (1.4) from \cite{Baron:2020xel}.}
\begin{equation}\label{eq:finalidentification}
    \cA_{\Acl}^{(n)\deltaR} (t_{\deltaR})_{\Bcr\Ccr} = - \cF^{(n)}_{\Acl\Bcr\Ccr}(\mathcal{A}^{(<n)})\,, \quad\text{and}\quad
    \cA_{\Acr}^{(n)\deltaL} (t_{\deltaL})_{\Bcl\Ccl} = - \cF^{(n)}_{\Acr\Bcl\Ccl}(\mathcal{A}^{(<n)})\,.
\end{equation}

Here lies an important difference with the previous approach carried on in \cite{Gitsis:2024gfb}: There, the starting point of the analysis was the algebra corresponding to the $\tau_\alpha$ generators. Then, in order to perform the identification in an iterative way, one needed to turn to the $t_\alpha$ generators, and this led to a tricky regularization procedure there dubbed \emph{collapse of towers}. In the present work, we instead started analyzing directly the algebra defined by the $t_\alpha$ generators in section \ref{sec:structgroup}. Therefore, we can directly proceed with the identification exploiting the index-splitting procedure \eqref{eq:splittingindices}.

Employing the parametrization \eqref{eqn:paramcE}, the relevant components coming from the first equation in \eqref{eq:finalidentification} read
\begin{align}
    K: && A_i^{(n)\la \bR\cR\ra } &= - \sqrt{2} e^{\al}{}_i \cF^{(n)}_{\al}{}^{\bR\cR}(A^{(<n)}) \quad &
    \Pi^{(n)\alphaL\la \bR\cR\ra } &= - \cF^{(n)\alphaL\bR\cR}(A^{(<n)}) \,, \nn \\
    R_1: && A_i^{(n)\la \bR\gammaR\ra } &= - \sqrt{2} e^{\al}{}_i \cF_{\al}^{(n)\bR\gammaR}(A^{(<n)})  &
    \Pi^{(n)\alphaL\la \bR\gammaR\ra } &= - \cF^{(n)\alphaL\bR\gammaR}(A^{(<n)})\,, \nn \\
    R_2: && A_i^{(n)\la \betaR\gammaR\ra } &= - \sqrt{2} e^{\al}{}_i \cF_{\al}^{(n)\betaR\gammaR}(A^{(<n)})  &
    \Pi^{(n)\alphaL\la \betaR\gammaR\ra } &= -\cF^{(n)\alphaL\betaR\gammaR}(A^{(<n)}) \,,
\end{align}
while from the second equation we obtain identical relations with opposite chiralities.

Comparing with \eqref{eq:identflatin} and \eqref{eq:identfgreek}, they can be written as
\begin{equation*}\tag{i0a}\label{eq:firstordA}
\boxed{
    \begin{aligned}
        K: && A^{(1)}_i{}^{\alphaR} &= A^{(1)}_i{}^{\la \aR\bR\ra } = \omega^{(-)ab}_i \quad & \quad
     A^{(1)}_i{}^{\alphaL} &= A^{(1)}_i{}^{\la \aL\bL \ra} = -\omega^{(+)ab}_i\,,
    \end{aligned}
}
\end{equation*}
at the leading order, and, for $n>1$, as
\begin{equation*}\tag{i0b}\label{eq:identificationcomponents}
\boxed{\begin{aligned}
    K: && A_i^{(n)\la \bR\cR\ra } &= - \tfrac12 e^{\bR j} e^{\cR k} \Omegat^{(n)}_{ijk}(A^{(<n)}) \quad &
    \Pi^{(n)\alphaL\la \bR\cR\ra } &= - \tfrac{1}{2} e^{\bR i} e^{\cR j} \Omegat_{ij}^{(n) \alphaL}(A^{(<n)}) \,, \\
    R_1: &&  A_i^{(n)\la \bR\gammaR\ra } &=  \tfrac{1}{\sqrt{2}} e^{\bR j} \Omegat_{ij}^{(n) \gammaR}(A^{(<n)}) &
    \Pi^{(n)\alphaL\la \bR\gammaR\ra } &=  \tfrac{1}{\sqrt{2}}e^{\bR i} \Omegat_i^{(n) \gammaR \alphaL}(A^{(<n)})\,, \\
    R_2: && A_i^{(n)\la \betaR\gammaR\ra } &= - \Omegat_{i}^{(n)\betaR\gammaR}(A^{(<n)}) &
    \Pi^{(n)\alphaL\la \betaR\gammaR\ra } &= -\Omegat^{(n)\alphaL\betaR\gammaR}(A^{(<n)}) \,,\\
    K: && A_i^{(n)\la \bL\cL\ra } &= - \tfrac12 e^{\bL j} e^{\cL k} \Omegat^{(n)}_{ijk}(A^{(<n)}) &
    \Pi^{(n)\alphaR\la \bL\cL\ra } &= - \tfrac{1}{2} e^{\bL i} e^{\cL j} \Omegat_{ij}^{(n) \alphaR}(A^{(<n)}) \,, \\
    R_1: && A_i^{(n)\la \bL\gammaL\ra } &=  -\tfrac{1}{\sqrt{2}} e^{\bL j} \Omegat_{ij}^{(n) \gammaL}(A^{(<n)})  &
    \Pi^{(n)\alphaR\la \bL\gammaL\ra } &= - \tfrac{1}{\sqrt{2}}e^{\bL i} \Omegat_i^{(n) \gammaL \alphaR}(A^{(<n)})\,, \\
    R_2: && A_i^{(n)\la \betaL\gammaL\ra } &= - \Omegat_{i}^{(n)\betaL\gammaL}(A^{(<n)})  &
    \Pi^{(n)\alphaR\la \betaL\gammaL\ra } &= -\Omegat^{(n)\alphaR\betaL\gammaL}(A^{(<n)}) \,.
\end{aligned}}
\end{equation*}

\subsubsection{More details on the identification}\label{sec:detailsID}
We now continue with the analysis of the identification on an order-by-order basis. For $n=1$, as we noticed above, we obtain the equations \eqref{eq:firstordA}. These relations are crucial as they seed all the next orders of the identification. An important consequence of \eqref{eq:firstordA} is that nearly all higher order contributions can be written in terms of covariant derivatives \eqref{eq:+-connections} for the modified spin connections $\omega^{(-)ab}_i$ and $\omega^{(+)ab}_i$. To see how this work, we will look in the following at all the terms appearing on the right-hand side of \eqref{eq:identificationcomponents}. In order to do that, we will develop iterative expressions for the quantities \eqref{eq:omegamunurho} -- \eqref{eq:omegaijk}.
These formulas are essential for the implementation of the identification and, therefore, we will label them by \eqref{eq:OmegatijrhoID} -- \eqref{eq:finalOmegatmunurho}.

The most straightforward of them is \eqref{eq:omegaijrho}, which we can write as
\begin{equation*}\tag{i1}\label{eq:OmegatijrhoID}
    \boxed{\Omegat^{(n)}_{ij}{}^\rho = \sum_{m=2}^{n} F^{(m)\beta}_{ij} \kappa_{\beta\gamma} \Pi^{(n-m)\rho\gamma}\,,}
\end{equation*}
after introducing the non-Abelian field strength 
\begin{equation}\label{eq:Fijalpha}
    F_{ij}{}^\alpha = 2 \partial_{[i} A_{j]}^\alpha + A_i^\beta A_j^\gamma f_{\beta\gamma}{}^\alpha\,.
\end{equation}
By expanding $F_{ij}{}^\alpha$ in derivatives, we find that for $n>2$ all the contributions coming from $A^{(1)\alpha}_i$ can be repackaged in terms of the covariant derivatives \eqref{eq:+-connections}. In the end, we are just left with
\begin{align}
    F^{(n)}_{ij}{}^{\alphaR} &= 2 \nabla^{(-)}_{[i} A^{(n-1)\alphaR}_{j]} + H_{ij}{}^k A^{(n-1)\alphaR}_k + \sum_{m=2}^{n-2} A^{(m)\betaR}_i A^{(n-m)\gammaR}_j f_{\betaR\gammaR}{}^{\alphaR}\,, \nn \\ \label{eq:Fijexpanded}
    F^{(n)}_{ij}{}^{\alphaL} &= 2 \nabla^{(+)}_{[i} A^{(n-1)\alphaL}_{j]} - H_{ij}{}^k A^{(n-1)\alphaL}_k + \sum_{m=2}^{n-2} A^{(m)\betaL}_i A^{(n-m)\gammaL}_j f_{\betaL\gammaL}{}^{\alphaL}\,.
\end{align}
For $n=2$ something interesting happens. At this order, we are dealing with the two relevant contributions
\begin{align}
    F^{(2)\gammaR }_{ij} &=  2 \partial_{[i} A_{j]}^{(1)\gammaR} + A_i^{(1)\deltaR} A_j^{(1)\epsilonR} f_{\deltaR\epsilonR}{}^{\gammaR}\,, \nn \\ \label{eq:F2components}
    F^{(2)\gammaL }_{ij} &=  2 \partial_{[i} A_{j]}^{(1)\gammaL} + A_i^{(1)\deltaL} A_j^{(1)\epsilonL} f_{\deltaL\epsilonL}{}^{\gammaL}\,.
\end{align}
To compute them we use the iterative definition of the structure constants \eqref{eq:expandedconstr} and notice that, since $A^{(1)\alpha}_i$ are Lorentz valued and the Lorentz algebra is a closed subalgebra of the structure algebra, the only relevant structure constants for these terms are
\begin{align}
    f_{\la \aR_1 \aR_2 \ra \la \bR_1 \bR_2 \ra}{}^{\la \cR_1 \cR_2 \ra}&=-2 \delta_{[\aR_1}^{[\cR_1} \eta_{\aR_2] [\bR_1} \delta_{\bR_2]}^{\cR_2]}\,,\nn \\
    f_{\la \aL_1 \aL_2 \ra \la \bL_1 \bL_2 \ra}{}^{\la \cL_1 \cL_2 \ra}&=-2 \delta_{[\aL_1}^{[\cL_1} \eta_{\aL_2] [\bL_1} \delta_{\bL_2]}^{\cL_2]}\,.
\end{align}
Then, splitting the $\gamma$ index, we see that
\begin{align}
    F^{(2)\la \cR\dR\ra }_{ij} &= \phantom{+,} \left( 2 \partial_{[i} \omega_{j]}^{(-)cd} + \omega^{(-)ce}_i \omega^{(-)d}_{je}  - \omega^{(-)de}_i \omega^{(-)c}_{je}\right), \nn \\
    F^{(2)\la \cL\dL\ra }_{ij} &= - \left( 2 \partial_{[i} \omega_{j]}^{(+)cd} + \omega^{(+)ce}_i \omega^{(+)d}_{je}  - \omega^{(+)de}_i \omega^{(+)c}_{je}\right).
\end{align}
This result can be simplified introducing modified Riemann tensors for the connections $\omega^{(\pm)}$, as in \eqref{eq:chiralRiemann}, such that
\begin{equation}\label{eq:Fij2}
    F_{i j}^{(2)\alphaR} = F_{i j}^{(2)\la \aR\bR\ra } = R^{(-)}_{i j}{}^{a b}\,, \qquad\text{and}\qquad  F_{i j}^{(2)\alphaL} = F_{i j}^{(2)\la \aL\bL\ra } =- R^{(+)}_{i j}{}^{a b}\,.
\end{equation}

We come now to the second term we need for the identification, namely the expansion of \eqref{eq:omegaijk}.
For $n>3$, it can be written as
\begin{align}
    \Omegat_{ijk}^{(n)} =&\quad 3 \kappa_{\alpha\beta} \left(\sum_{m=2}^{n-3} \partial_{[i} A_j^{(m)\alpha} A_{k]}^{(n-m-1)\beta} + \partial_{[i} A_j^{(1)\alpha} A_{k]}^{(n-2)\beta} + \partial_{[i} A_j^{(n-2)\alpha} A_{k]}^{(1)\beta} \right) \nn \\
    &+f_{\alpha\beta\gamma} \left( \sum_{m=2}^{n-3} \sum_{l=2}^{n-m-2} A^{(m)\alpha}_i A^{(l)\beta}_j A^{(n-m-l)\gamma}_k + 3 \sum_{m=2}^{n-3} A^{(1)\alpha}_i A^{(m)\beta}_j A^{(n-m-1)\gamma}_k \right. \nn \\
    &\left. \phantom{+f_{\alpha\beta\gamma}+}+ 3 A^{(1)\alpha}_i A^{(1)\beta}_j A^{(n-2)\gamma}_k \right)\,.
\end{align}
As before, we collect the terms which do not contain a naked spin connection to obtain
\begin{equation*}\tag{i2}\label{eq:finalOmegatijk}
    \boxed{\begin{aligned}
        \Omegat_{ijk}^{(n)} =& \frac12 \sum_{m=2}^{n-3} \left( 3 F^{(m+1)\alpha}_{[ij} A^{(n-m-1)\beta}_{k]} \kappa_{\alpha\beta} - \sum_{l=2}^{n-m-2} A^{(m)\alpha}_i A^{(l)\beta}_j A^{(n-m-l)\gamma}_k f_{\alpha\beta\gamma}\right) \\
        & + 3 F^{(2)\alpha}_{[ij} A^{(n-2)\beta}_{k]} \kappa_{\alpha\beta} + 3 \partial_{[i} b^{(n-1)}_{jk]}\,, 
    \end{aligned}}
\end{equation*}
where
\begin{equation}\label{eq:bn-1fieldredef}
    b^{(n-1)}_{ij} = A^{(n-2)\alpha}_{[i} A^{(1)\beta}_{j]} \kappa_{\alpha\beta}
\end{equation}
denotes a non-Lorentz-covariant contribution that can be removed through a field redefinition. Note that this result excludes $n=3$, which we will discuss at the end of this subsection.

In order to treat $\Omegat_i{}^{\nu\rho}$ in a similar fashion, we rewrite \eqref{eq:omegainurho} as
\begin{equation}\label{eq:OmegatIndexfree}
    (\Omegat_i)^\nu{}_\rho = \left(\partial_i \Pi \Pi^{-1} + \Pi A_i \Pi^{-1}\right)^\nu{}_\rho\,,
\end{equation}
with $(\Pi)^\alpha{}_\beta = \Pi^\alpha{}_\beta$, and $(A_i)^\alpha{}_\beta = A_i^\gamma f_{\gamma}{}^\alpha{}_\beta$. To obtain this equation, we made use of the relation
\begin{equation}\label{eq:PiTkappaPi}
    \Pi^T \kappa \Pi = \kappa\,,
\end{equation}
which is a direct consequence of \eqref{eqn:constrPI}. It is not too complicated to see that \eqref{eq:OmegatIndexfree} can be rewritten as
\begin{equation}
    (\Omegat_i)^\nu{}_\rho = \left(\left(\partial_i \Pi - [ A_i, \Pi ] \right) \Pi^{-1} + A_i \right)^\nu{}_\rho\,.
\end{equation}
For the aim of the identification we drop the last term, $A_i$, because we want to write iterative expressions of the kind of $(\Omegat^{(n)}_i)^\nu{}_\rho(A^{(<n)})$, where $(\Omegat^{(n)}_i)^\nu{}_\rho$ cannot depend on the $n$-th order $A^{(n)}_i$. Hence, we are left with
\begin{equation*}\tag{i3}\label{eq:OmegatinurhoID}
    \boxed{\Omegat^{(n)}_i{}^{\nu\rho}(A^{(<n)}) = \sum_{m=3}^n G^{(m)\nu\alpha}_i \kappa_{\alpha \beta} \Pi^{(n-m)\rho\beta}\,,}
\end{equation*}
after introducing
\begin{equation}
    G^{(n)\nu\rho}_i = \nabla_i^{(\pm,\pm)} \Pi^{(n-1)\nu\rho} + \sum_{m=2}^{n-2} A^{(m)\beta}_i \left(  f_{\beta\gamma}{}^{\nu} \Pi^{(n-m)\gamma\rho}
        + f_{\beta\gamma}{}^{\rho} \Pi^{(n-m)\nu\gamma} \right)\,.
\end{equation}
Here, we also defined the covariant derivatives $\nabla_i^{(\pm,\pm)} X^{\mu\nu}$ which act differently on the two indices of the tensor $X^{\mu\nu}$. They act as $\nabla_i^{(+)}$ on left chiral-projected (under-barred) indices and as $\nabla_i^{(-)}$ on right chiral-projected (over-barred) ones. For instance $\nabla_i^{(\pm,\pm)} X^{\underline{\mu} \overline{\nu}}=\nabla_i^{(+,-)} X^{\underline{\mu} \overline{\nu}}$, while $\nabla_i^{(\pm,\pm)} X^{\underline{\mu} \underline{\nu}}=\nabla_i^{(+,+)} X^{\underline{\mu} \underline{\nu}}=\nabla_i^{(+)} X^{\underline{\mu} \underline{\nu}}$. At this point one can also easily see the similarity between \eqref{eq:OmegatijrhoID} and \eqref{eq:OmegatinurhoID}.

Finally, we come to equation \eqref{eq:omegamunurho}, for which we obtain the expansion
\begin{equation*}\label{eq:finalOmegatmunurho}\tag{i4}
    \boxed{\Omegat^{(n)\mu\nu\rho} = 3 \sum_{m=2}^{n-2} \Pi^{(m)[\mu|\alpha} \Pi^{(n-m)|\nu|\beta} f_{\alpha\beta}{}^{|\rho]} + \sum_{m=2}^n \sum_{l=2}^{n-m-2} \Pi^{(m)\mu\alpha} \Pi^{(l)\nu\beta} \Pi^{(n-m-l)\rho\gamma} f_{\alpha\beta\gamma}\,.}
\end{equation*}

As we mentioned above, to complete all the equations we need for the identification procedure, we still have to compute $\Omegat^{(3)}_{ijk}$. It depends on the recursive definitions for $\kappa_{\alpha\beta}$ and $f_{\alpha\beta\gamma}$ given in \eqref{eq:kappasplitchiral} and \eqref{eq:fdownsplitchiral}, respectively. Thus, in contrast to the previously computed leading orders, it is model-dependent, meaning that it depends on the choice of $\kappa_{\alpha\beta}$. At this order, we only need
\begin{equation}
\begin{aligned}
    \kappa_{\la \aR\bR\ra \la \cR\dR\ra } &= \phantom{-} \tfrac{a}{2}\eta_{\aR[\cR}\eta_{\dR]\bR}\,, \\
    \kappa_{\la \aL\bL\ra \la \cL\dL\ra } &= -\tfrac{b}{2}\eta_{\aL[\cL}\eta_{\dL]\bL}   \,, 
\end{aligned} \qquad\text{and}\qquad
\begin{aligned}
    f_{\la \aR_1 \aR_2 \ra \la \bR_1 \bR_2 \ra \la \cR_1 \cR_2 \ra} &= \phantom{-} a \eta_{[\aR_1|[\bR_2}\eta_{\bR_1][\cR_1}\eta_{\cR_2]|\aR_2]}\,, \\
    f_{\la \aL_1 \aL_2 \ra \la \bL_1 \bL_2 \ra \la \cL_1 \cL_2 \ra} &= -b \eta_{[\aL_1|[\bL_2}\eta_{\bL_1][\cL_1}\eta_{\cL_2]|\aL_2]}\,,
\end{aligned}
\end{equation}
which, eventually, give rise to
\begin{equation}\label{eq:Htilde3}
    \Omegat^{(3)}_{i j k} = -\tfrac{3}{2}a\Omega^{(-)}_{i j k} + \tfrac{3}{2}b \Omega^{(+)}_{i j k}\,.
\end{equation}
Here, we encounter the gravitational Chern-Simons terms
$\Omega^{(+)}_{i j k}$ and $\Omega^{(-)}_{i j k}$ for the connections $\omega^{(+)}$ and $\omega^{(-)}$. As one can see from their definition in \eqref{eq:chernsimonspm}, they do not transform covariantly under Lorentz transformations. However, employing the relation \eqref{eq:OmegapmintermsofOmega} between $\Omega^{(\pm)}_{ijk}$ and the ordinary Chern-Simons form $\Omega_{iab}$ for the spin connection $\omega_{iab}$, we can rewrite $\Omegat^{(3)}_{ijk}$ as
\begin{align}\label{eq:Omegatijk3}
        \Omegat^{(3)}_{ijk} = 3 \partial_{[i} b^{(2)}_{jk]} &- \tfrac{3}{2} (a - b) \left( \Omega_{ijk} - \tfrac14 \nabla_{[i} H_j{}^{lm} H_{k]lm} \right)\nn\\
        &- \tfrac{3}{4} (a + b) \left( H_{lm[i} R_{jk]}{}^{lm} - \tfrac16 H_{il}{}^m H_{jm}{}^n H_{kn}{}^l \right),
\end{align}
with
\begin{equation}\label{eq:bfieldredef}
    b^{(2)}_{ij} = -\tfrac{1}{4} (a + b) H^{kl}{}_{[i} \omega_{j]kl} \,.
\end{equation}
The reason for singling out the term $b^{(2)}_{ij}$ is that we can again remove the total derivative acting on it by a field redefinition, like we already did in \eqref{eq:finalOmegatijk}. Note that this particular non-Lorentz-covariant field redefinition was already discussed in \cite{Marques:2015vua}. The connection between this term and the Green-Schwarz anomaly cancellation mechanism will be made manifest in section \ref{sec:GStransf}, where we explicitly analyze $B$-field transformations.

At this point, it is natural to distinguish two classes of theories:
\begin{itemize}[topsep=2pt,itemsep=1.5pt,parsep=2pt]
    \item Non-chiral theories, like bosonic or type II string theories: Here $a-b=0$. Therefore, we do not encounter any further Lorentz violating terms after the field redefinition.
    \item Chiral theories, such as heterotic string theory or HSZ theory: Now $a-b\ne 0$ and there is no way to get rid of the Chern-Simons term with respect to the spin connection $\omega_{iab}$.  However, this term will always appear together with the $H$-flux throughout the identification. Thus, we define the combination
    \begin{equation}\label{eq:Hhat}
        \hat{H}_{ijk} = H_{ijk} - \tfrac{3}{2} (a-b) \Omega_{ijk}\,.
    \end{equation}
    This new quantity is invariant under Lorentz transformations and will replace the non-covariant $H_{ijk}$ in the results of the higher-order identifications.
\end{itemize}

\subsection{Action}\label{sec:action}
In analogy with the standard heterotic DFT result \eqref{eq:hetDFTaction}, the expression for the action on the megaspace was computed in~\cite{Gitsis:2024gfb} and found to be
\begin{equation}
    S = \int \dd^d x e^{-2\Phi} \left( 2 \Dc_{\Acr} \cF^{\Acr} - \cF_{\Acr} \cF^{\Acr} + \tfrac12 \cF_{\Acl\Bcr\Ccr} \cF^{\Acl\Bcr\Ccr} + \tfrac16 \cF_{\Acr\Bcr\Ccr} \cF^{\Acr\Bcr\Ccr} - \text{c.c} \right).
\end{equation}
 Because $\cF_{\Ac}$ only receives contributions from $F_A$, the respective terms will combine with the $F_{ABC}$ components of $\cF_{\Ac\Bc\Cc}$ to form the standard two derivative action of the NS-NS sector. To compute the remaining terms, we turn back to the winding basis.
This allows us to rewrite the action as
\begin{align}
        S = \int \dd^d x \sqrt{-g} e^{-2 \phi} \Bigl( & R + 4 \partial_i \phi \partial^i \phi - \tfrac1{12} \widetilde{H}_{ijk} \widetilde{H}^{ijk} - \tfrac14 F_{ij}{}^\alpha F^{ij\beta} h_{\alpha\beta} + \tfrac18 D_i h_{\alpha\beta} D^i h^{\alpha\beta} \nn\\ \label{eqn:action}
        &-\tfrac1{12} f_{\alpha\beta\gamma} f_{\delta\epsilon\lambda} h^{\alpha\delta} h^{\beta\epsilon} h^{\gamma\lambda} - \tfrac1{4} f_{\alpha\gamma}{}^\delta f_{\beta\delta}{}^\gamma h^{\alpha\beta}
        \Bigr)\,,
\end{align}
where we defined the fields
\begin{align}
    \widetilde{H}_{ijk} &= H_{ijk} + 3 \partial_{[i} A_j^\alpha A_{k]}^\beta \kappa_{\alpha\beta} + A_i^\alpha A_j^\beta A_k^\gamma f_{\alpha\beta\gamma}=H_{ijk} +\Omegat_{ijk}\,,\label{eq:Hijk} \\
    h_{\alpha\beta} &= \Pi^\gamma{}_\alpha \widetilde{\kappa}_{\gamma\delta} \Pi^\delta{}_\beta \,,\label{eq:hdef}
\end{align}
in addition to the non-Abelian field strength $F_{ij}{}^\alpha$ which we already encountered in \eqref{eq:Fijalpha} of the last subsection. Furthermore, we defined the covariant derivative
\begin{equation}
    D_i h^{\alpha\beta} = \partial_i h^{\alpha\beta} + 2 A_i^\gamma f_{\gamma\delta}{}^{(\alpha|} h^{\delta|\beta)}\,,
\end{equation}
and the rescaled quantities
\begin{equation}
    \widetilde{\kappa}_{\alphaL\betaR} = \widetilde{\kappa}_{\betaR\alphaL} = 0 \,, 
    \qquad \widetilde{\kappa}_{\alphaR\betaR} = \kappa_{\alphaR\betaR} \,,
    \qquad \widetilde{\kappa}_{\alphaL\betaL} = -\kappa_{\alphaL\betaL}\,.
\end{equation}

This rewriting of the action has a similar form to a gauged DFT \cite{Geissbuhler:2011mx,Aldazabal:2011nj,Grana:2012rr,Geissbuhler:2013uka}. In addition to the physical degrees of freedom, formed by the metric, the $B$-field and the dilaton, it also contains the vector field $A_i^\alpha$ and scalar field $h^{\alpha\beta}$, which furnish the coset $\OO(p, q)/(\OO(p)\times \OO(q))$. Here the parameters $p$ and $q$ capture the range of the indices $\alphaL=1,\dots ,p$ and $\alphaR=1,\dots, q$; Hence, they go to infinity. This is not an issue, since it is still possible to expand the action order-by-order as we show in the next section. Remarkably, all the higher-derivative corrections in this framework arise from the interplay of the modified $H$-flux, and the auxiliary vectors and scalars. They do not introduce any new degrees of freedom because they are all expressed in terms of the fundamental fields and their derivatives through the identification procedure outlined in the last subsections. 

At this point, a few additional comments about the properties of $\Pi^{\alpha\beta}$ and its relation to $h^{\alpha\beta}$ are in order. First, $\Pi^{\alpha\beta}$ splits in two parts: A chiral contribution $\Pi_{\mathrm{C}}^{\alpha\beta}$, and a mixed-chiral one, $\Pi_{\mathrm{M}}^{\alpha\beta}$. The latter is fixed by the identification, and it is antisymmetric under the exchange of its indices, while the former arises from the constraint \eqref{eqn:constrPI}. Solving it explicitly gives rise to
\begin{equation}\label{eq:PiMC}
    \Pi = \Pi_{\mathrm{M}} + \Pi_{\mathrm{C}} = \Pi_{\mathrm{M}} + \sqrt{1 + \Pi_{\mathrm{M}}^2}\,.
\end{equation}
Here, we again suppress the indices of $(\Pi)^\alpha{}_\beta = \Pi^\alpha{}_\beta$ with the convention that the first index is up and the second is down. Moreover, we use the relations
\begin{equation}
    \Pi_{\mathrm{M}}^T \kappa = - \kappa \Pi_{\mathrm{M}}\,,
    \qquad \text{and} \qquad
    \Pi_{\mathrm{M}}^T \widetilde{\kappa} = + \widetilde{\kappa} \Pi_{\mathrm{M}}\,.
\end{equation}
Both are a direct consequence of the antisymmetry under the index exchange mentioned above. Together, these two relations allow us to prove \eqref{eq:PiTkappaPi} directly by computing
\begin{equation}
    \Pi^T \kappa \Pi = \kappa \left( - \Pi_{\mathrm{M}} + \sqrt{1 + \Pi_{\mathrm{M}}^2} \right) \left( \Pi_{\mathrm{M}} + \sqrt{1 + \Pi_{\mathrm{M}}^2} \right) = \kappa\,.
\end{equation}
A similar computation for $h$ gives rise to
\begin{equation}
 h = \Pi^T \widetilde{\kappa} \Pi = \widetilde{\kappa} \left( \Pi_{\mathrm{M}} + \sqrt{1 + \Pi_{\mathrm{M}}^2} \right)^2.
\end{equation}

In the next section, we will perform the identification order-by-order in derivatives and we will make use of the explicit expansions
\begin{align}\label{eq:PiC}
    \Pi_{\mathrm{C}} &= 1 + \tfrac12 \Pi_{\mathrm{M}}^2 - \tfrac18 \Pi_{\mathrm{M}}^4 + \dots\,,  \\
    h &= \widetilde{\kappa} \left( 1 + 2 \Pi_{\mathrm{M}} + 2 \Pi_{\mathrm{M}}^2 + \Pi_{\mathrm{M}}^3 + \dots \right)\,,
\end{align}
which hold up to 8 derivatives. As a consequence and for convenience, we can also split $h$ into chiral- and mixed-chiral contributions as
\begin{equation}\label{eq:hintermsofPi}
    h_{\mathrm{M}} = \widetilde{\kappa} \left( 2 \Pi_{\mathrm{M}} + \Pi_{\mathrm{M}}^3 + \dots \right), \quad \text{and} \quad 
    h_{\mathrm{C}} = \widetilde{\kappa} \left( 1 + 2\Pi^2_{\mathrm{M}}  + \dots \right).
\end{equation}

\subsection{Green-Schwarz transformations}\label{sec:GStransf}
It is well known that, in order to cancel gauge anomalies in chiral string theories, the Lorentz transformations of the $B$-field have to be corrected by additional higher-derivative contributions. In particular, it was shown that this mechanism gives rise to the \emph{Green-Schwarz transformations} for heterotic string theory\cite{Green:1984sg}. In the context of the gBdR identification in DFT, one encounters analogue corrections to double Lorentz transformations whose explicit form at first order in $\alpha'$ was computed in \cite{Marques:2015vua} and then re-analyzed in the context of the biparametric gBdR identification in \cite{Baron:2018lve,Baron:2020xel}. For obvious reasons, these new transformations are called generalized Green-Schwarz transformations. In \cite{Gitsis:2025clo} an all-order expression was derived in the Pol\'a\v{c}ek and Siegel approach and we will now reformulate it in our new framework. 

The starting point of our analysis are general transformations of the megaframe. They are governed by the heterotic generalized Lie derivative and given by
\begin{equation}\label{eq:deltamathcalE}
    \delta \mathcal{E}_{\mathcal{A}}{}^{\mathcal{J}}\mathcal{E}_{\mathcal{B}\mathcal{J}} = \mathcal{L}_{\xi}\mathcal{E}_{\mathcal{A}}{}^{\mathcal{J}}\mathcal{E}_{\mathcal{B}\mathcal{J}}\,,
\end{equation}
where the explicit expression of the right-hand side was computed in \cite{Gitsis:2024gfb}. This can be understood as a lift of the standard definition \eqref{eq:hetgenLie} to the megaspace, and gives rise to
\begin{equation}
    \mathcal{L}_{\xi}\mathcal{E}_{\mathcal{A}}{}^{\mathcal{J}}\mathcal{E}_{\mathcal{B}\mathcal{J}} = - 2 \mathcal{D}_{[\mathcal{A}}\xi_{\mathcal{B}]} + \xi^{\mathcal{C}}\mathcal{F}_{\mathcal{C}\mathcal{A}\mathcal{B}} - \xi_{\langle \mathcal{A}\mathcal{B}\rangle}\, ,
\end{equation}
where $\xi^{\mathcal{A}}$ parametrizes generalized diffeomorphism on the megaspace and
\begin{equation}
    \xi_{\langle \mathcal{A}\mathcal{B}\rangle} = \xi^{\gamma} (t_\gamma)_{\mathcal{A}\mathcal{B}}\, .
\end{equation}
Employing the parametrization \eqref{eqn:mega-frame} of the megaspace frame $\cE_{\Ac}{}^{\Ic}$, the left-hand side of equation \eqref{eq:deltamathcalE} assumes the form
\begin{equation}\label{eqn:master}
    -\delta E^{\mathcal{A}}{}_{\mathcal{I}}E_{\mathcal{A}\mathcal{J}} + \delta \mathcal{A}_{\mathcal{I}}{}^{\mathcal{K}}\mathcal{A}_{\mathcal{J}\mathcal{K}} = -2\partial_{[\mathcal{I}}\widehat{\xi}_{\mathcal{J}]} + \widehat{\xi}^{\alpha}\kappa_{\alpha\beta}\Omegat_{\mathcal{I}\mathcal{J}}{}^{\beta} - \xi_{\langle\mathcal{I}\mathcal{J}\rangle}\,.
\end{equation}
To interpret this formula, we have to make two observations: First of all, we introduced a new generalized diffeomorphism parameter
\begin{equation}\label{eq:widehatxi}
    \widehat{\xi}_{\Ic} =  \xi_{\Ac} E^{\Ac}{}_{\Ic}\,.
\end{equation}
It is very important to keep in mind that this parameter differs from the initial vector $\xi_{\Ic}$, whose flat version is governed by the full megaspace frame
\begin{equation}
    \xi_{\Ic} = \xi_{\Ac} \cE^{\Ac}{}_{\Ic}\,.
\end{equation}
Moreover, since diffeomorphisms do not get explicitly corrected, we have restricted our analysis only to $B$-field gauge transformations, setting effectively the physical diffeomorphism parameter $\xi^j=0$ to zero. Consequentially, $\widehat{\xi}^j=0$ vanishes as well.

To understand the implications of \eqref{eqn:master}, it is useful to expand it in components. By means of \eqref{eqn:paramcE}, we can rewrite the left-hand side as
\begin{align}\label{eqn:lhs}
    &-\delta E^{\mathcal{A}}{}_{\mathcal{I}}E_{\mathcal{A}\mathcal{J}} + \delta \mathcal{A}_{\mathcal{I}}{}^{\mathcal{K}}\mathcal{A}_{\mathcal{J}\mathcal{K}}\nn\\
    &= \begin{pmatrix}
    2 \delta e_a{}^k e^a{}_{[i|}B_{k| j]} + \delta B_{i j} + \delta A_{[i}^{\alpha}A_{j]}^{\beta}\kappa_{\alpha\beta} && -\delta e^a{}_i e_a{}^j && \delta A_i^{\alpha}\Pi^{\underline{\nu}}{}_{\alpha} && \delta A_i^{\alpha}\Pi^{\overline{\nu}}{}_{\alpha}\\
    \delta e^a{}_j e_a{}^i && 0 && 0 && 0\\
    -\delta A_j^{\alpha}\Pi^{\underline{\mu}}{}_{\alpha} && 0 && \delta \Pi^{\underline{\mu}\alpha}\Pi^{\underline{\nu}}{}_{\alpha} && \delta \Pi^{\underline{\mu}\alpha}\Pi^{\overline{\nu}}{}_{\alpha} \\
    -\delta A_j^{\alpha}\Pi^{\overline{\mu}}{}_{\alpha} && 0 && \delta \Pi^{\overline{\mu}\alpha}\Pi^{\underline{\nu}}{}_{\alpha} && \delta \Pi^{\overline{\mu}\alpha}\Pi^{\overline{\nu}}{}_{\alpha}
    \end{pmatrix}.
\end{align}
The expansion of the first two terms on the right-hand side of \eqref{eqn:master} is straightforward. It produces
\begin{align}\label{eq:rhs1}
    -2\partial_{[\mathcal{I}}\widehat{\xi}_{\mathcal{J}]} &= \begin{pmatrix}
        -2\partial_{[i}\widehat{\xi}_{j]} && 0 && -\partial_i \widehat{\xi}^{\underline{\nu}} && -\partial_i\widehat{\xi}^{\overline{\nu}}\\
        0 && 0 && 0&& 0\\
        \partial_j\widehat{\xi}^{\underline{\mu}} && 0 && 0&& 0\\
        \partial_j\widehat{\xi}^{\overline{\mu}} && 0 && 0&& 0
    \end{pmatrix}\,,
    \intertext{and}
    \label{eqn:rhs}
    \widehat{\xi}^{\alpha}\kappa_{\alpha\beta} \Omegat_{\mathcal{I}\mathcal{J}}{}^{\beta} &=\begin{pmatrix}
        \widehat{\xi}^{\alpha}\kappa_{\alpha\beta} \Omegat_{i j}{}^{\beta} && 0 && \widehat{\xi}^{\alpha}\kappa_{\alpha\beta} \Omegat_i{}^{\underline{\nu}\beta} && \widehat{\xi}^{\alpha}\kappa_{\alpha\beta} \Omegat_i{}^{\overline{\nu}\beta}\\
        0 && 0 && 0 && 0\\
        -\widehat{\xi}^{\alpha}\kappa_{\alpha\beta} \Omegat_j{}^{\underline{\mu}\beta} && 0 && \widehat{\xi}^{\alpha}\kappa_{\alpha\beta}\Omegat^{\underline{\mu}\underline{\nu}\beta} && \widehat{\xi}^{\alpha}\kappa_{\alpha\beta}\Omegat^{\underline{\mu}\overline{\nu}\beta}\\
        -\widehat{\xi}^{\alpha}\kappa_{\alpha\beta} \Omegat_j{}^{\overline{\mu}\beta} && 0 && \widehat{\xi}^{\alpha}\kappa_{\alpha\beta}\Omegat^{\overline{\mu}\underline{\nu}\beta} && \widehat{\xi}^{\alpha}\kappa_{\alpha\beta}\Omegat^{\overline{\mu}\overline{\nu}\beta}
    \end{pmatrix}\,,
\end{align}
while the last term on the right-hand side of \eqref{eqn:master} requires more attention.

Let us look at $\xi_{\la\Ic\Jc\ra} = E^{\cA}{}_{\Ic}\xi_{\la \cA \Bc \ra}E^{\Bc}{}_{\Jc}$ component-wise. First of all, we start with
\begin{equation}\label{eq:xiIJ}
    \xi_{\langle I J \rangle} = E^A{}_I \xi_{\langle A B \rangle} E^B{}_J\,.
\end{equation}
Here, we note that the double Lorentz group splits in two fully-chiral components and therefore
\begin{equation}
    \xi_{\langle A B \rangle} = \begin{pmatrix}
    \xi_{\underline{a}\underline{b}} && 0\\
    0 && \xi_{\overline{a}\overline{b}}
    \end{pmatrix}\,.
\end{equation}
Hence, by means of \eqref{eqn:frame}, equation \eqref{eq:xiIJ} can be expanded to
\begin{equation}
    \xi_{\langle I J \rangle} = \begin{pmatrix}
        2\lambda_{a b} e^{b k} e^a{}_{[i|}B_{k| j]} && -e^b{}_i \lambda_{a b}e^{a j}\\
        e^{a i}\lambda_{a b}e^b{}_j && 0
    \end{pmatrix} ,
\end{equation}
where
\begin{equation}
        \xi_{\underline{a}\underline{b}} = -\xi_{\overline{a}\overline{b}} = \lambda_{a b} = \lambda_{[a b]}
\end{equation}
parametrize standard Lorentz transformations. In principle one could also analyze more involved transformations. They would be relevant, for example, for higher-derivative corrections of generalized T-dualities \cite{Hassler:2020tvz,Borsato:2020wwk,Codina:2020yma}, but for simplicity we will not consider them here. In the same vein, we proceed with the remaining components
\begin{equation}
    \xi_{\langle A}{}^{\beta\rangle} = \begin{pmatrix}
        \xi_{\underline{a}}{}^{\underline{\beta}} && 0\\
        0 && \xi_{\overline{a}}{}^{\overline{\beta}}
    \end{pmatrix}
\end{equation}
to find
\begin{equation}
    \xi_{\langle I}{}^{\mu\rangle} = E^A{}_I  \xi_{\langle A}{}^{\mu\rangle} = \begin{pmatrix}
        e^{\underline{a}k}B_{k i}\xi_{\underline{a}}{}^{\underline{\mu}} + e^{\underline{a}}{}_i \xi_{\underline{a}}{}^{\underline{\mu}} && -e^{\overline{a}k}B_{k i}\xi_{\overline{a}}{}^{\overline{\mu}} + e^{\overline{a}}{}_i \xi_{\overline{a}}{}^{\overline{\mu}}\\
        e^{\underline{a}i} \xi_{\underline{a}}{}^{\underline{\mu}} && -e^{\overline{a}i} \xi_{\overline{a}}{}^{\overline{\mu}}
    \end{pmatrix} .
\end{equation}
However, comparing with the relevant components of equations \eqref{eqn:lhs}-\eqref{eqn:rhs}, one can easily check that
\begin{equation}\label{eq:mixedxis}
    \xi_{\underline{a}}{}^{\underline{\beta}} = \xi_{\overline{a}}{}^{\overline{\beta}} = 0\, ,
\end{equation}
leading to the whole $\xi_{\langle I}{}^{\mu\rangle}$ block being trivial. Gathering these results, we eventually find
\begin{equation}\label{eq:xiIcJc}
    \xi_{\langle \mathcal{I}\mathcal{J}\rangle} = \begin{pmatrix}
         2 \lambda_{a b} e^{b k} e^a{}_{[i|}B_{k| j]} && -e^b{}_i \lambda_{a b}e^{a j} && 0 && 0\\
         e^{a i}\lambda_{a b}e^b{}_j && 0 && 0 && 0\\
         0 && 0 && \xi^{\langle \underline{\mu}\underline{\nu}\rangle} && 0\\
         0 && 0 && 0 && \xi^{\langle \overline{\mu}\overline{\nu}\rangle}
    \end{pmatrix} .
\end{equation}

All the restrictions on the parameter $\xi^\alpha$ of the gauge transformations are of course a direct consequence of preserving the initial form of the megaspace frame. Hence, they should be understood as related to the particular \emph{gauge fixing} we use. This gauge fixing is obtained by comparing the left-hand and right-hand sides of \eqref{eqn:master} from which we immediately see that:
\begin{enumerate}[topsep=2pt,itemsep=1.5pt,parsep=2pt]
    \item To have no corrections to the Lorentz transformation of the frame,
    \begin{equation}
        \xi^{(n)}_{\la A B\ra} = 0 \qquad \text{with} \qquad n >0
    \end{equation}
    has to hold;
    \item From \eqref{eq:mixedxis}, all mixed components have to vanish,
    \begin{equation}
        \xi^{(n)\alpha \ra}_{\la A } = 0\,,  \qquad \forall n \,.
    \end{equation}
\end{enumerate}

Still, we need to fix $\xi^{\la\alpha\beta\ra}$. To see how this is done, we first extract the transformations of all the fields by comparing \eqref{eqn:lhs} with \eqref{eq:rhs1}, \eqref{eqn:rhs} and \eqref{eq:xiIcJc}, resulting in
\begin{align}
    \delta e_a{}^i &= -\lambda_a{}^b e_b{}^i\,, \nn \\
    \delta B_{i j} + 2\partial_{[i}\widehat{\xi}_{j]} &= -\delta A_{[i}^{\alpha} A_{j]}^{\beta} \kappa_{\alpha\beta} + \widehat{\xi}^{\alpha} \kappa_{\alpha\beta} \Omegat_{i j}{}^{\beta}\,, \nn \\ 
    \delta A_i^{\alpha}\Pi^{\beta}{}_{\alpha} &= -\partial_i\widehat{\xi}^{\beta} + \widehat{\xi}^{\alpha} \kappa_{\alpha\gamma} \Omegat_i{}^{\beta\gamma}\,, \nn \\ \label{eq:gaugetransf}
    \delta \Pi^{\alpha\gamma}\Pi^{\beta}{}_{\gamma} &= \widehat{\xi}^{\gamma} \kappa_{\gamma\delta} \Omegat^{\alpha\beta\delta} -\xi^{\langle\alpha\beta\rangle}\,,
\end{align}
where $\widehat{\xi}_i$ is the gauge parameter associated to a gauge transformation of the $B$-field ($B \rightarrow B + d\widehat{\xi}$). We know that $\xi^\alpha$ is chiral, therefore the mixed contributions
\begin{equation}\label{eq:mixedchiralxi}
    \xi^{\langle\alphaR\betaL\rangle} = 0\, \quad \text{and} \quad
    \xi^{\langle\alphaL\betaR\rangle} = 0
\end{equation}
have to vanish. This is fine, because the mixed chiral part of the last equation in \eqref{eq:gaugetransf} can be solved order-by-order if the transformation $\delta \Pi_\mathrm{M}$ is chosen appropriately at each order in derivatives. Once this transformation is fixed, we do not have any freedom left for the chiral transformation $\delta \Pi_\mathrm{C}$. The reason is that the chiral part of $\Pi$ is already completely fixed by the mixed part (as shown by \eqref{eq:PiMC}). Fortunately, we have $\xi^{\la\alpha\beta\ra}$, which is gauge-fixed to
\begin{equation}
    \xi^{\langle\alphaR\betaR\rangle} = \widehat{\xi}^{\gamma}\kappa_{\gamma\delta}\Omegat^{\alphaR\betaR\delta} -\delta \Pi^{\alphaR\gamma}\Pi^{\betaR}{}_{\gamma} \, \quad \text{and} \quad
    \xi^{\langle\alphaL\betaL\rangle} = \widehat{\xi}^{\gamma}\kappa_{\gamma\delta}\Omegat^{\alphaL\betaL\delta} -\delta \Pi^{\alphaL\gamma}\Pi^{\betaL}{}_{\gamma} \,,
\end{equation}
and thus ensures that also the chiral components of the last equation in \eqref{eq:gaugetransf} are satisfied. This relation also shows that 
\begin{equation}
    \xi^{(n)\la\alpha\beta\ra} = 0\,,  \qquad n < 2\,.
\end{equation}

To bring the residual gauge transformations into their final form, the third equation of \eqref{eq:gaugetransf} can be solved for $\delta A^\alpha_i$ to find
\begin{equation}
    \delta A^\alpha_i = - \partial_i ( \widehat{\xi}^{\beta} \Pi_\beta{}^\alpha )+ \widehat{\xi}^{\delta} \Pi_\delta{}^\beta A_i^{\gamma} f_{\beta\gamma}{}^{\alpha} \,.
\end{equation}
With this result, the transformation of the $B$-field becomes
\begin{equation}
    \delta B_{ij} = - \partial_{[i}(\widehat{\xi}^\gamma \Pi_\gamma{}^\alpha ) A^\beta_{j]} \kappa_{\alpha\beta}\,,
\end{equation}
where the $B$-field gauge parameter is chosen as
\begin{equation}
    \widehat{\xi}_i = \widehat{\xi}^\alpha \Pi_\alpha{}^\beta A^\gamma_i \kappa_{\beta\gamma}\,
\end{equation}
in order to obtain a simple transformation. The last thing we have to do is to express the parameters $\widehat{\xi}_{\Ic}$ in terms of their unhatted counterparts $\xi_{\Ic}$ through $\widehat{\xi}_{\Ic} = \cA_{\Ic}{}^{\Jc} \xi_{\Jc}$ (which follows directly from \eqref{eq:widehatxi}). In particular, we obtain $\widehat{\xi}_\alpha = \Pi_\alpha{}^\beta \xi_\beta$. This allows us to further simplify the transformations to
\begin{align}\label{eq: dB}
    \delta B_{ij} &= - \partial_{[i} \xi^\alpha A_{j]}^\beta \kappa_{\alpha\beta}\,, \\
    \label{eqn: dA}
    \delta A_i^\alpha &= - \partial_i \xi^\alpha + \xi^\beta A_i^\gamma f_{\beta\gamma}{}^\alpha\,, \\ 
    \label{eq: dPi}
    \delta \Pi^{\alpha\gamma }\Pi^{\beta}{}_{\gamma} &= \xi^{\gamma}f_{\gamma\delta\epsilon}\Pi^{\alpha\delta}\Pi^{\beta\epsilon} - \xi^{\la \alpha \beta \ra}\,,
\end{align}
with the $B$-field gauge parameter expressed as
\begin{equation}\label{eq:Bshift}
    \xi_i = \xi^{\alpha}A_i^{\beta}\kappa_{\alpha\beta}\,.
\end{equation}
Like in \cite{Gitsis:2025clo}, these are all-order results. But here we used a different gauge fixing which further simplified the computation.

\section{Action and symmetries at the leading orders}\label{sec:actionandsym}
After all the preparation carried on in the last section, we have everything we need to compute the action and its gauge symmetries up to arbitrarily high orders in $\alpha'$. To demonstrate this process, and to compare the results with the literature, we now analyze the first two orders.
\subsection{Four derivative action (order \texorpdfstring{$\alpha'$}{α'})}\label{sec:action2}
To compute the corrections at order $\alpha'$, we are interested in terms up to quartic order in derivatives. To write all the possible terms appearing in the action \eqref{eqn:action} at this order we start with the following observations:
\begin{enumerate}[topsep=2pt,itemsep=1.5pt,parsep=2pt]
    \item The expansion of $A_i^{(n)}$ starts from $n=1$, which we already know from \eqref{eq:firstordA}. 
    \item\label{item:Fijalpha2} \eqref{eq:Fijalpha} implies that the expansion of $F_{ij}^{(n)\alpha}$ starts from $n=2$. Again this quantity is known explicitly from \eqref{eq:Fij2}.
    \item From the leading order of \eqref{eq:hintermsofPi}, one finds that $h^{(0)}_{\alpha\beta} = \widetilde{\kappa}_{\alpha\beta}$.
    \item\label{item:Htilde3} Then, from \eqref{eq:Hijk} one sees that the first contribution ($n=1$) in the $\widetilde{H}_{ijk}^{(n)}$ expansion is just the $H$-flux, followed at $n=3$ by the $\Omegat^{(3)}_{ijk}$ in \eqref{eq:Htilde3}.
    \item Lastly, it is easy to check that the terms involving $h^{\alpha\beta}$ in the action drop out because:
    \begin{enumerate}[topsep=2pt,itemsep=1.5pt,parsep=2pt]
        \item One notices from \eqref{eq:hdef}, that $h^{(1)}_{\alpha\beta}=0$, since $\Pi^{(1)\alpha}{}_\beta=0$; The same is true for $h^{(1)\alpha\beta}$.
        \item Then, since $h^{(0)}_{\alpha\beta}$ and $h^{(0)\alpha\beta}$ are covariantly constant, all the terms coming from $\tfrac18 D_i h_{\alpha\beta} D^i h^{\alpha\beta}$ drop out at this order.
        \item Furthermore, since $h^{(2)\alpha\beta}$ and $h^{(3)\alpha\beta}$ are mixed-chiral (as can be seen from \eqref{eq:hintermsofPi}), while $f_{\alpha\beta\gamma}$ and $f_{\alpha\beta}{}^\gamma$ are fully chiral, the combination
        \begin{equation*}
            -\tfrac1{12} f_{\alpha\beta\gamma} f_{\delta\epsilon\lambda} h^{\alpha\delta} h^{\beta\epsilon} h^{\gamma\lambda} - \tfrac1{4} f_{\alpha\gamma}{}^\delta f_{\beta\delta}{}^\gamma h^{\alpha\beta}
        \end{equation*}
        vanishes at this order.
    \end{enumerate}
\end{enumerate}

Combining all these observations with the general expression for the action in \eqref{eqn:action}, it is not difficult to see that the possible terms we can have at order $\alpha'$ in the expansion are
\begin{equation}\label{eqn:act4devs}
    \Lc^{(4)}= - \tfrac1{6} \widetilde{H}^{(3)}_{ijk} H^{ijk}  - \tfrac14 F^{(2)\alpha}_{ij} F^{(2)ij\beta}\widetilde{\kappa}_{\alpha\beta}\,,
\end{equation}
which after taking points \ref{item:Fijalpha2} and \ref{item:Htilde3} from above into account assumes the final form
\begin{equation}\label{eq:fourderaction}
    \Lc^{(4)}=-\tfrac{1}{6}\left(-\tfrac{3}{2}a\Omega^{(-)}_{i j k} + \tfrac{3}{2}b \Omega^{(+)}_{i j k}\right)H^{i j k} + \tfrac{a}{8}R^{(-)}_{i j a b}R^{(-)i j b a} + \tfrac{b}{8}R^{(+)}_{i j a b}R^{(+)i j b a}\,. 
\end{equation}
This result perfectly matches equation (2.7) in \cite{Marques:2015vua} at order $\alpha'$.

To conclude our analysis of the first-order corrections, we will now consider the two subcases of bosonic and heterotic string theories. For the bosonic case, after setting $a=b=-\alpha'$ and removing the $b_{ij}^{(2)}$ in \eqref{eq:Omegatijk3} by the non-Lorentz-covariant field redefinition
\begin{equation}\label{eq:Bredef2}
    B_{ij} \rightarrow B_{ij} - \tfrac{\alpha'}{2} H^{kl}{}_{[i} \omega_{j]kl}\,,
\end{equation}
we obtain the following Lagrangian
\begin{equation}
    \begin{aligned}
        \Lc^{(4)}_{\text{bos}}= \tfrac{\alpha'}{4} \Big(& \tfrac{1}{24} H_{i}{}^{lm} H^{ijk} H_{jl}{}^{n} H_{kmn} + \tfrac{1}{8} H_{ij}{}^{l} H^{ijk} H_{k}{}^{mn} H_{lmn} +  R_{ijkl} R^{ijkl} \\
        & -  H_{i}{}^{lm} H^{ijk} R_{jklm} -  H_{i}{}^{lm} H^{ijk} R_{jlkm} -  \nabla_{[k} H_{l]ij} \nabla^{l} H^{ijk}\Big).
    \end{aligned}
\end{equation}
Rewriting it in terms of modified Riemann tensors, one easily checks that it reproduces equation (1.3) of \cite{Wulff:2024ips}\footnote{To obtain a perfect match, one has to keep in mind the $\alpha'\rightarrow\tfrac1{4}\alpha'$ shift performed in \cite{Wulff:2024ips}, as pointed out right after equation (1.3) there. Notice also that their Riemann tensor $\mathcal{R}_{abcd}$ corresponds to our $R^{(-)}_{abcd}$. Furhtermore, there seems to be a small typo in the third line of that result, where the coefficient of the term $(\nabla H)^2$ should be $+\tfrac1{3}$ instead of $-\tfrac1{3}$.} without the need of any further field redefinition or integration by parts. Then, through field redefinitions\footnote{See appendix \ref{app:fieldredef} for additional details.}, one can recover the classic Metsaev-Tseytlin result from \cite{Metsaev:1987bc}.

For the heterotic case ($a=-\alpha'$, $b=0$), the field redefinition \eqref{eq:Bredef2} is not required. The action \eqref{eq:fourderaction} immediately reduces to
\begin{equation}
    \Lc^{(4)}_{\text{het}}=-\tfrac{\alpha'}{4}\Omega^{(-)}_{ijk} H^{i j k} - \tfrac{\alpha'}{8}R^{(-)}_{i j a b}R^{(-)i j b a} \,,
\end{equation}
and this Lagrangian perfectly matches equation (5.35) from \cite{Hronek:2022dyr} at order $\alpha'$, once one takes into account that the $\Omega_{ijk}$ there is defined as 3 times our $\Omega^{(-)}_{ijk}$.

\subsection{Four derivative symmetries (order \texorpdfstring{$\alpha'$}{α'})}
In order to apply the general transformations rules \eqref{eq: dB} to \eqref{eq: dPi} from the last section, we use the notation $(\delta X)^{(n)}$ to denote the $n$\textsuperscript{th} order correction of the transformation of a field $X$. To avoid cluttering the notation too much, we also neglect the label for the lowest order transformation of each field. 

Starting from the order $n=1$ of \eqref{eqn: dA}, we obtain
\begin{equation}
    \delta A_i^{(1)\alpha} = -\partial_i \xi^{(0)\alpha} + \xi^{(0)\beta}A_i^{(1)\gamma}f_{\beta\gamma}{}^{\alpha}\,.
\end{equation}
Remembering that $A_i^{(1)\alpha}$ is directly related to the spin connection through \eqref{eq:firstordA}, we recover the expected transformation
\begin{equation}
    \delta \omega_i^{(\pm)a b} = \partial_i \lambda^{a b} +\omega_i^{(\pm)a c}\lambda_c{}^b -\lambda^a{}_c\omega_i^{(\pm)c b}\,.
\end{equation}

Next, we would like to recover the Green-Schwarz transformation of the $B$-field. Hence, we expand \eqref{eq: dB} at two derivatives order, along with the gauge parameter \eqref{eq:Bshift}, to find
\begin{align}
    (\delta B_{i j})^{(2)} &= -\partial_{[i}\xi^{(0)\alpha}A_{j]}^{(1)\beta}\kappa_{\alpha\beta} = -\tfrac{a}{2}\partial_{[i}\lambda^{a b}\omega^{(-)}_{j]b a} + \tfrac{b}{2}\partial_{[i}\lambda^{a b}\omega^{(+)}_{j]b a}\,, \nn \\ \label{eqn:GStr2}
    \xi^{(1)}_i &= \xi^{(0)\alpha}A_{i}^{(1)\beta}\kappa_{\alpha\beta} = \tfrac{a}{2}\lambda^{a b}\omega^{(-)}_{i b a} - \tfrac{b}{2}\lambda^{a b}\omega^{(+)}_{i b a}\,.
\end{align}
Indeed, this is the expected transformation as one can see for example by comparing with equation (2.10) in \cite{Marques:2015vua}.
To make the distinction between the bosonic and heterotic cases clearer, we rewrite the previous expressions as
\begin{align}
    (\delta B_{i j})^{(2)} &= -\tfrac{(a-b)}{2}\partial_{[i}\lambda^{a b}\omega_{j]b a} + \tfrac{(a+b)}{4}\partial_{[i}\lambda^{a b}H_{j]b a}\,, \nn \\ \label{eq:GSorder2}
    \xi^{(1)}_i &= \tfrac{(a-b)}{2}\lambda^{a b}\omega_{i b a} - \tfrac{(a+b)}{4}\lambda^{a b}H_{i b a}\,.
\end{align}
In our discussion regarding the action above, we highlighted the field redefinition \eqref{eq:Bredef2} of the $B$-field, needed to make $\widetilde{H}_{ijk}^{(3)}$ covariant for the bosonic case. As mentioned there, this field redefinition is not covariant under Lorentz transformations. Hence, when we perform it, we have to shift the transformation of the $B$-field as
\begin{equation}\label{eq: dBshift}
    (\delta B_{i j})^{(2)} \rightarrow (\delta B_{i j})^{(2)} + \delta b^{(2)}_{i j}\,,
\end{equation}
with $b^{(2)}_{i j}$ defined in \eqref{eq:bfieldredef}. Looking at the new term
\begin{equation}
    \delta b_{ij}^{(2)}=-\tfrac{(a+b)}{4}H_{ab[i}\partial_{j]}\lambda^{ab}\,,
\end{equation}
we see that it exactly removes the term proportional to $(a+b)$ in \eqref{eq:GSorder2}, giving
\begin{equation}\label{eq:deltaB2}
    \begin{aligned}
        (\delta B_{i j})^{(2)} &= -\tfrac{(a-b)}{2}\partial_{[i}\lambda^{a b}\omega_{j]b a}\,.
    \end{aligned}
\end{equation}
Therefore, it is straightforward to check that $B_{ij}$ transforms covariantly for the bosonic case, while for the heterotic one, the transformation is proportional to the spin connection.

As another consistency check,  we compute the transformation of the Riemann tensor. This is done either from
\begin{equation}
    \delta A_i^{(2)\alpha} = \xi^{(0)\beta}A_i^{(2)\gamma}f_{\beta\gamma}{}^{\alpha}\, ,
\end{equation}
or from
\begin{equation}
    \delta \Pi^{(2)\alpha\beta} = 2 \xi^{(0)\gamma}\Pi^{(2)[\alpha|\delta}f_{\gamma\delta}{}^{|\beta]}\, ,
\end{equation}
both yielding the transformation law
\begin{equation}
    \delta R_{i j}^{(\pm)a b} = R_{i j}^{(\pm)a c}\lambda_c{}^b - \lambda^a{}_c R_{i j}^{(\pm) c b}\, ,
\end{equation}
where we have employed the identity
\begin{equation}
    R_{i j}^{(\pm)a b} = e_{d j} R_i^{(\pm) d a b} = e_{c i} e_{d j} R^{(\pm) c d a b}\,,
\end{equation}
along with the transformation law of the frame (first equation in \eqref{eq:gaugetransf}).

Eventually coming to quantities containing three derivatives, we analyze the transformation of the Chern-Simons term. At this point the new notation we introduced at the beginning of this section becomes relevant. Until now, we just focused on the two leading orders ($n=1,2$), where we had
\begin{equation}
    (\delta A_i^{\alpha})^{(n)} = \delta A_i^{(n) \alpha}, \qquad n\le2\,.
\end{equation}
Starting from $n=3$, we rather need
\begin{equation}\label{eq: dA3}
    (\delta A_i^\alpha)^{(3)} = \delta A_i^{(3)\alpha} + (\delta A_i^{(1)\alpha})^{(2)} = \xi^{(0)\beta}A_i^{(3)\gamma}f_{\beta\gamma}{}^{\alpha}\,,
\end{equation}
because $A^{(1)\alpha}_i$ contains the $H$-flux, whose transformation follows directly from \eqref{eqn:GStr2}. First we notice that
\begin{equation}
    (\delta A_i^{(1)\alphaL})^{(2)} = (\delta A_i^{(1)\alphaR})^{(2)} = -(\delta H_i{}^{a b})^{(3)} = -3 e^{a j} e^{b k} \partial_{[i}(\delta B_{j k]})^{(2)}\,.
\end{equation}
To further proceed, we need to consider the term $A_i^{(3)\alpha}$, which we have not computed yet. While the full derivation will be demanded to the next subsection, here we just need to look at the final result for the component we need, namely equation \eqref{eq:A3Kcomp}. Then, since
\begin{equation}
    \Omega_i^{(\pm) a b} = e^{a j} e^{b k}\Omega^{(\pm)}_{i j k}\, ,
\end{equation}
plugging back all the results in \eqref{eq: dA3}, we obtain the expected result
\begin{equation}\label{eq:deltaOmega+-}
    \delta \Omega^{(\pm)}_{i j k} = - \partial_{[i}(\partial_j \lambda^{a b}\omega_{k] b a}^{(\pm)}) = \partial_{[i}\lambda^{ab}\partial_j\omega^{(\pm)}_{k]ba}\,.
\end{equation}

In \eqref{eq:Hhat} we argued that it is possible to combine $H_{ijk}$ with $\Omega_{ijk}$ in order to have covariant objects in the heterotic action. We are now ready to verify this statement explicitly.
First of all, we note that an analogue expression to \eqref{eq:deltaOmega+-} holds for the ordinary $\Omega_{ijk}$:
\begin{equation}
    \delta \Omega_{ijk}=\partial_{[i}\lambda^{ab}\partial_j\omega_{k]ba}\,.
\end{equation}
Then, from \eqref{eq:deltaB2} we can compute the transformation of the $H$-flux as
\begin{equation}
    (\delta H_{ijk})^{(3)}=3\partial_{[i}(\delta B_{jk]})^{(2)}=-\tfrac{3(a-b)}{2}\partial_{[j}\lambda^{ab}\partial_i \omega_{k]ba}\,.
\end{equation}
Therefore, we obtain
\begin{equation}
    \delta \hat{H}^{(3)}_{ijk}=(\delta H_{ijk})^{(3)}-\tfrac{3(a-b)}{2}\delta \Omega_{ijk}=0\,,
\end{equation}
that is, the new quantity $\hat{H}_{ijk}$ transforms covariantly as expected.

\subsection{Six derivative action (order \texorpdfstring{$\alpha'^2$}{α′2})}
To compute the corrections at order $\alpha'^2$, we are interested in terms with up to six derivatives. Exploiting again chirality arguments, and performing an analysis in the same spirit of section~\ref{sec:action2}, the only terms that contribute are
\begin{align}
    \Lc^{(6)}=& - \tfrac1{6} \widetilde{H}^{(5)}_{ijk} H^{ijk}- \tfrac1{12} \widetilde{H}^{(3)}_{ijk} \widetilde{H}^{(3)ijk}  - \tfrac14 F_{ij}{}^{(3)\alpha} F^{(3)ij\beta}\widetilde{\kappa}_{\alpha\beta} \nn \\
    &- \tfrac12 F_{ij}{}^{(2)\alpha} F^{(4)ij\beta}\widetilde{\kappa}_{\alpha\beta}- \tfrac14 F_{ij}{}^{(2)\alpha} F^{(2)ij\beta}h^{(2)}_{\alpha\beta}+\tfrac1{8}\nabla^{(\pm,\pm)}_ih^{(2)}_{\alpha\beta} \nabla^{(\pm,\pm)i} h^{(2)\alpha\beta} \nn \\ \label{eqn:act6devs}
    &-\tfrac1{12}f_{\alpha\beta\gamma}f_{\delta\epsilon\lambda} h^{(2)\alpha\delta} h^{(2)\beta\epsilon} h^{(2)\gamma\lambda}\,.
\end{align}
To compute all these quantities, we now need $A^{(2)\alpha}_i$, $A^{(3)\alpha}_i$, $\Pi^{(2)\alpha\beta}$, and $\Pi^{(3)\alpha\beta}$. 

By taking into account \eqref{eq:identificationcomponents} and \eqref{eq:OmegatijrhoID}, one easily checks that
\begin{equation}
    F^{(2)\alphaR}_{i j} = \sqrt{2} A_i^{(2)\la b\alphaR\ra } e_{b j}\, \qquad \text{and} \qquad F^{(2)\alphaL}_{i j} = -\sqrt{2} A_i^{(2)\la b\alphaL\ra } e_{b j}\,.
\end{equation}
Therefore, splitting the $\alpha$ index, one obtains
\begin{equation}\label{eq:A2identification}
A^{(2)\alphaR}_i = A^{(2)\la \bR\la \cR\dR\ra \ra }_i = \tfrac{1}{\sqrt{2}} R^{(-)}_{i}{}^{bcd}\, \qquad \text{and} \qquad
    A^{(2)\alphaL}_i = A^{(2)\la \bL\la \cL\dL\ra \ra }_i = \tfrac{1}{\sqrt{2}} R^{(+)}_{i}{}^{bcd}\,.
\end{equation}
In the same way, we get from \eqref{eq:identificationcomponents} the first non-vanishing $\Pi^{\alpha\beta}$ terms
\begin{equation}
    \Pi^{(2)\la \aL\bL\ra \la \cR\dR\ra } = \tfrac12 R^{(+)cdab}\, \qquad\text{and}\qquad
    \Pi^{(2)\la \aR\bR\ra \la \cL\dL\ra } = - \tfrac12 R^{(-)cdab}\,.
\end{equation}
From the relation \eqref{eq:relRpm} between the Riemann tensors with torsion, one easily checks that
\begin{equation}
    \Pi^{(2)\la \bR_1 \bR_2\ra \la \aL_1 \aL_2\ra } + \Pi^{(2)\la \aL_1 \aL_2\ra \la \bR_1 \bR_2\ra } = \tfrac12 \left( 
        R^{(+)a_1 a_2 b_1 b_2} - R^{(-)b_1 b_2 a_1 a_2}
    \right) = 0\,,
\end{equation}
and therefore
\begin{equation}
    \Pi^{(2)\alphaR\betaL} + \Pi^{(2)\betaL\alphaR} = 2 \Pi^{(2)(\alphaR\betaL)} = 0\,. 
\end{equation}
This is another consistency check because $\Pi^{\alphaL\betaR}$ has to be antisymmetric.

Next, we look at $A_i^{(3)\alpha}$. According to \eqref{eq:identificationcomponents}, its first components are given by
\begin{equation}\label{eq:A3Kcomp}
    A_i^{(3)\la \aR\bR\ra } = A_i^{(3)\la \aL\bL\ra } = -\tfrac12 \Omegat_i^{(3)ab} = \tfrac{3}{4}a \Omega_i^{(-)a b} - \tfrac{3}{4}b \Omega_i^{(+)a b}\,,
\end{equation}
after taking into account \eqref{eq:Htilde3}. The second set of non-vanishing contributions originate from \eqref{eq:OmegatijrhoID} and takes the form
\begin{align}
    A^{(3)\la \bR\la \cR\la \dR\eR\ra \ra \ra }_i &= \phantom{-} e^{bj} \left(\nabla^{(-)}_{[i}R_{j]}^{(-)c d e} +\tfrac12 H_{ij}{}^k R_{k}^{(-)c d e}\right)\,, \nn \\ \label{eq:A3R1comp}
    A^{(3)\la \bL\la \cL\la \dL\eL\ra \ra \ra }_i &= - e^{bj} \left(\nabla^{(+)}_{[i}R_{j]}^{(+)c d e}-\tfrac12 H_{ij}{}^k R_{k}^{(+)c d e}\right)\,,
\end{align}
after using our previous result for $A^{(2)\alpha}_i$ from above. 
Then, due to the fact that $\Pi^{(2)\alphaL \betaL} = \Pi^{(2)\alphaR \betaR}=0$, we can immediately see that
\begin{equation}
    A_i^{(3)\la \alphaR\betaR \ra} = A_i^{(3)\la \alphaL\betaL \ra} = 0\,.
\end{equation}

Last but not least, we analyze $\Pi^{(3)\alpha\beta}$. Its first non-trivial components read
\begin{align}
    \Pi^{(3)\la \aL \la \bL \cL \ra\ra \la \dR\eR\ra } &=-\tfrac{1}{\sqrt{2}}e^{d i} e^{e j}\left(\nabla^{(+)}_{[i}R_{j]}^{(+)abc}-\tfrac12 H_{ij}{}^k R_{k}^{(+)abc}\right)\,, \nn \\
    \Pi^{(3)\la \aR \la \bR \cR \ra\ra \la \dL\eL\ra } &=-\tfrac{1}{\sqrt{2}}e^{d i} e^{e j}\left(\nabla^{(-)}_{[i}R_{j]}^{(-)abc}+\tfrac12 H_{ij}{}^k R_{k}^{(-)abc}\right)\,,
\end{align}
and arise in exactly the same way as \eqref{eq:A3R1comp} (which is obvious from \eqref{eq:identificationcomponents}). For the second one, we eventually need \eqref{eq:OmegatinurhoID}. It gives rise to
\begin{equation}
    \Pi^{(3)\alphaL \la \bR\gammaR\ra } = \tfrac{1}{\sqrt{2}} e^{bi} \nabla_i^{(+,-)} \Pi^{(2)\gammaR\alphaL}\, \qquad \text{and} \qquad
    \Pi^{(3)\alphaR \la \bL\gammaL\ra } = -\tfrac{1}{\sqrt{2}} e^{bi} \nabla_i^{(-,+)} \Pi^{(2)\gammaL\alphaR}\,.
\end{equation}
Making use of \eqref{eq:pmconnections} and \eqref{eq:relRpm}, one can also rewrite these terms as
\begin{align}
    \Pi^{(3)\la \aL \eL \ra \la \bR\la \cR \dR\ra\ra } &=-\tfrac{1}{2\sqrt{2}} e^{bi} \left( \nabla_{i}^{(-)}R^{(-)aecd}+2H_{if}{}^{[a}R^{(-)e]fcd}\right), \nn \\
    \Pi^{(3)\la \aR \eR \ra \la \bL\la \cL \dL\ra\ra } &=-\tfrac{1}{2\sqrt{2}} e^{bi} \left( \nabla_{i}^{(+)}R^{(+)aecd}-2H_{if}{}^{[a}R^{(+)e]fcd}\right).
\end{align}

These are all the auxiliary fields we need to compute the full six-derivative Lagrangian, that takes the form
\begin{align}
    \Lc^{(6)}=&-\tfrac{a^2}{8}\left[3 H_{ij}{}^{k} R^{(-) ijln} \Omega^{(-)}_{kln}+ 3 R^{(-) ijkl} \nabla^{(-)}_{i} \Omega^{(-)}_{jkl}+ \tfrac{3}{2} \Omega^{(-)}_{i}{}^{jk} \Omega^{(-)i}{}_{jk}\right] \nn \\
    &+\tfrac{b^2}{8}\left[- 3 H_{ij}{}^{k} R^{(+) ijln} \Omega^{(+)}_{kln}+ 3 R^{(+) ijkl} \nabla^{(+)}_{i} \Omega^{(+)}_{jkl}- \tfrac{3}{2} \Omega^{(+)}_{i}{}^{jk} \Omega^{(+)i}{}_{jk} \right] \nn \\
    &+\tfrac{ab}{2} \bigg[ \tfrac{1}{3} R^{(-)}_{m}{}^{sku} R^{(-)j}{}_{s}{}^{l}{}_{u} R^{(-)m}{}_{jkl} - \tfrac{1}{8} R^{(-)rs}{}_{kl} R^{(-)tukl} R^{(+)}_{rstu} \nn \\
    &\phantom{+\tfrac{ab}{2} \bigg[} - \tfrac{1}{8} R^{(-)mnrs} R^{(+)}_{mn}{}^{tu} R^{(+)}_{rstu} + \tfrac{3}{4} H_{ij}{}^{q} R^{(-)rsij} \Omega^{(-)}_{qrs} \nn \\  
    &\phantom{+\tfrac{ab}{2} \bigg[}+ \tfrac{3}{4} H_{ij}{}^{q} R^{(+)rsij} \Omega^{(+)}_{qrs} + \tfrac{3}{4} R^{(-)mnpq} \nabla^{(-)}_{m} \Omega^{(+)}_{npq} \nn \\ 
    &\phantom{+\tfrac{ab}{2} \bigg[}- \tfrac{3}{4} R^{(+)mnpq} \nabla^{(+)}_{m} \Omega^{(-)}_{npq} + \tfrac{3}{4} \Omega^{(-)jln} \Omega^{(+)}_{jln} \nn \\
    &\phantom{+\tfrac{ab}{2} \bigg[}- \tfrac{1}{8} (\nabla^{(+)}_{i} R^{(+)mnpq}-2H_{ir}{}^{[m}R^{(+)n]rpq}) (\nabla^{(+)i} R^{(+)}_{mnpq}-2H^i{}_{r[m}R^{(+)}_{n]}{}^r{}_{pq})\bigg] \nn \\
    &+\Lc^{(6)}_{\text{redef}}\,, \label{eq:sixordL}
\end{align}
which we already presented in the introduction. In order to bring it in a particularly simple form, we note that the additional terms
\begin{align}
    \Lc^{(6)}_{\text{redef}}=-\tfrac{a^2}{8} \Bigg[& R^{(-) ijk}_{l} R^{(-)}_{ik}{}^{mn} R^{(-) l}{}_{jmn}  + 2 R^{(-) i j k}_{l} R^{(-) m}{}_{i}{}^{n}{}_{k} R^{(-) l}{}_{mjn} \nn \\
    &- H_{j}{}^{k}{}_{l} H^{j}{}_{ki} R^{(-) inpq} R^{(-) l}{}_{npq} - \nabla^{(-)i} R^{(-) jklm} \nabla^{(-)}_{i} R^{(-)}_{jklm} \nn \\
    &+ 3 H_{k}{}^{ij} R^{(-) klmp} \nabla^{(-)}_j R^{(-)}_{ilmp} + \nabla^{(-)k} R^{(-) i}{}_{jlp} \nabla^{(-)j} R_{ik}^{(-)lp} \Bigg] \nn \\
    +\tfrac{b^2}{8} \Bigg[& R^{(+) ijk}_{l} R^{(+)}_{ik}{}^{mn} R^{(+) l}{}_{jmn} - 2 R^{(+) i j k}_{l} R^{(+) m}{}_{i}{}^{n}{}_{k} R^{(+) l}{}_{mjn} \nn \\
    &+ H_{j}{}^{k}{}_{l} H^{j}{}_{ki} R^{(+) inpq} R^{(+) l}{}_{npq}+ \nabla^{(+)i} R^{(+) jklm} \nabla^{(+)}_{i} R^{(+)}_{jklm} \nn \\ \label{eq:redefLagr}
    &+ 3 H_{k}{}^{ij} R^{(+) klmp} \nabla^{(+)}_j R^{(+)}_{ilmp} - \nabla^{(+)k} R^{(+) i}{}_{jlp} \nabla^{(+)j} R_{ik}^{(+)lp} \Bigg]\,
\end{align}
that arise from our procedure can be removed by a field redefinition and integration by parts. To find the relevant field redefinitions and boundary terms, we follow the procedure outlined in \cite{Garousi:2019mca} and summarized in appendix~\ref{app:fieldredef}.

After removing $\Lc^{(6)}_{\text{redef}}$ and restricting the action to the heterotic case with $a=-\alpha'$ and $b=0$, we are left with
\begin{equation}
    \Lc^{(6)}_{\text{het}}=-\tfrac{\alpha'^2}{8} \Big[ 3 H_{ij}{}^{k} R^{(-) ijln} \Omega^{(-)}_{kln} + 3 R^{(-) ijkl} \nabla^{(-)}_{i} \Omega^{(-)}_{jkl} + \tfrac{3}{2} \Omega^{(-)}_{i}{}^{jk} \Omega^{(-)i}{}_{jk}
    \Big]\,.
\end{equation}
This action matches the one obtained from the original Bergshoeff-de Roo identification \cite{Bergshoeff:1989de} after the additional field redefinition
\begin{equation}
    B_{ij}\rightarrow B_{ij}+\tfrac{3 a^2}{8} \Omega^{(-)ab}_{[i}\omega^{(-)}_{j]ab}\,.
\end{equation}
This statement can be verified by comparing again with equation (5.35) from \cite{Hronek:2022dyr}, this time at the order $\alpha'^2$, while keeping in mind that the $\Omega_{ijk}$ there is defined as 3 times our $\Omega^{(-)}_{ijk}$.

Similarly, taking $a=b=-\alpha'$ for the bosonic string, we precisely recover the classic Metsaev-Tseytlin result \cite{Gholian:2023kjj}\footnote{Taking into account the $\alpha'\rightarrow\tfrac1{4} \alpha'$ shift.}. This match requires expanding the modified geometrical $(\pm)$ tensors to ordinary ones and performing suitable field redefinitions/integration by parts. Since the expressions of the expanded Lagrangian and the field redefinitions are quite long and not very illuminating, we reported them in appendix \ref{app:fieldredef}, along with the field redefinitions required for the heterotic case and the general procedure we followed to obtain them.

\subsection{Six derivative symmetries (order \texorpdfstring{$\alpha'^2$}{α′2})}
By going to the four-derivatives level, we finally analyze the $\alpha'^2$ corrections to the Green-Schwarz transformations. From \eqref{eq: dB}, they are given by
\begin{equation}
    (\delta B_{i j})^{(4)} = -\partial_{[i}\xi^{(0)\alpha}A_{j]}^{(3)\beta}\kappa_{\alpha\beta}-\partial_{[i}\xi^{(2)\alpha}A_{j]}^{(1)\beta}\kappa_{\alpha\beta}\, .
\end{equation}
Gauge fixing the  fully-chiral components of \eqref{eq: dPi} at order $n=2$ one finds
\begin{equation}
    \xi^{(2)\la \alphaL \betaL\ra} = \xi^{(2)\la \alphaR \betaR\ra} = 0\,,
\end{equation}
and therefore the Green-Schwarz transformations together with the $B$-field shift read
\begin{align}
    (\delta B_{i j})^{(4)} &= -\partial_{[i}\xi^{(0)\alpha}A_{j]}^{(3)\beta}\kappa_{\alpha\beta} = -\tfrac{3 a^2}{8}\partial_{[i}\lambda^{a b}\Omega^{(-)}_{j]b a} + \tfrac{3b^2}{8}\partial_{[i}\lambda^{a b}\Omega^{(+)}_{j]b a} \nn \\
    &\hspace{10.3em}+\tfrac{3ab}{8}(\partial_{[i}\lambda^{a b}\Omega^{(+)}_{j]b a} - \partial_{[i}\lambda^{a b}\Omega^{(-)}_{j]b a})\, \nn \\
    \xi_i^{(3)} &= \xi^{(0)\alpha}A_i^{(3)\beta}\kappa_{\alpha\beta} =  \tfrac{3a^2}{8} \lambda^{a b} \Omega_{i b a}^{(-)} -\tfrac{3b^2}{8}\lambda^{a b} \Omega_{i b a}^{(+)} - \tfrac{3 a b}{8}(\lambda^{a b} \Omega_{i b a}^{(+)}-\lambda^{a b} \Omega_{i b a}^{(-)}) \, .
\end{align}
For the heterotic case, this result reproduces equation (5.32) from \cite{Hronek:2022dyr}\footnote{Notice that there is a small typo in that equation, and the second term should appear with an opposite sign. We thank Linus Wulff for helping us resolving this issue.}.

As a last remark, we show that $B_{i j}$ still transforms covariantly at this order for the bosonic string. As we did for the second order transformations in \eqref{eq: dBshift}, we have to take into account the field redefinitions \eqref{eq:bfieldredef} and \eqref{eq:bn-1fieldredef}. Together, they give rise to
\begin{equation}\label{eq: dB4shift}
    (\delta B_{i j})^{(4)} \rightarrow (\delta B_{i j})^{(4)} + \delta b_{i j}^{(4)} + (\delta b^{(2)}_{i j})^{(2)}\,.
\end{equation}
Computing the last two terms on the right-hand side, we find
\begin{align}
    \delta b^{(4)}_{i j} &= \tfrac{1}{2}\delta(A_i^{(3)\alpha}A_j^{(1)\beta}\kappa_{\alpha\beta}) - \lbrace i \leftrightarrow j \rbrace \nn \\
    &= \tfrac{3 a^2}{16}\partial_{i}\lambda^{a b}\Omega^{(-)}_{j b a} - \tfrac{3b^2}{16}\partial_{i}\lambda^{a b}\Omega^{(+)}_{j b a}-\tfrac{3ab}{16}(\partial_{i}\lambda^{a b}\Omega^{(+)}_{j b a} - \partial_{ i}\lambda^{a b}\Omega^{(-)}_{j b a}) \nn \\
    &\phantom{=,}+\tfrac{3a^2}{16}e_b{}^k e_a{}^l \omega_{i}^{(-)a b}\partial_{[j}\lambda^{c d}\partial_k \omega_{l]d c}^{(-)}-\tfrac{3b^2}{16}e_b{}^k e_a{}^l \omega_{i}^{(+)a b}\partial_{[j}\lambda^{c d}\partial_k \omega_{l]d c}^{(+)} \nn \\
    &\phantom{=,}-\tfrac{3ab}{16}e_b{}^k e_a{}^l (\omega_{i}^{(-)a b}\partial_{[j}\lambda^{c d}\partial_k \omega_{l]d c}^{(+)} - \omega_{i}^{(+)a b}\partial_{[j}\lambda^{c d}\partial_k \omega_{l]d c}^{(-)}) - \lbrace i \leftrightarrow j \rbrace\,, \\
    \intertext{and}
    (\delta b_{i j}^{(2)})^{(2)} &= \tfrac{3(a^2 - b^2)}{16}e_b{}^k e_a{}^l \omega_{i}{}^{a b}\partial_{[j}\lambda^{c d}\partial_k \omega_{l]d c} \nn \\
    &\phantom{=,}+ \tfrac{3(a + b)^2}{16}e_b{}^k e_a{}^l \omega_{i}{}^{a b}\partial_{[j}\lambda^{c d}\partial_k H_{l] d c} - \lbrace i \leftrightarrow j \rbrace\,.
\end{align}
After plugging them back into \eqref{eq: dB4shift}, we are left with the new transformation
\begin{align}
    (\delta B_{i j})^{(4)} + \delta b_{i j}^{(4)} + (\delta  b^{(2)}_{i j}&)^{(2)} \nn \\
    =\tfrac{3}{8}e_b{}^k e_a{}^l \Bigg[&(a^2 - b^2)\left(\omega_{i}{}^{a b}\partial_{[j}\lambda^{c d}\partial_k \omega_{l]d c} -\tfrac{1}{8}H_i{}^{a b}\partial_{[j}\lambda^{c d}\partial_k H_{l]d c} \right) \nn \\
    & - \tfrac{1}{4} (a-b)^2 H_{i}{}^{a b}\partial_{[j}\lambda^{c d}\partial_k \omega_{l]d c}\Bigg] - \lbrace i \leftrightarrow j \rbrace\,.
\end{align}
As expected, this expression clearly vanishes for the bosonic case $a=b$.

\section{Conclusions and future prospects}
In this work, we further expanded the construction discussed in \cite{Hassler:2024yis,Gitsis:2024gfb,Gitsis:2025clo}. After a brief introduction that outlines the challenges of obtaining higher-derivative corrections to the string's low-energy effective action, and a short review of the generalized Bergshoeff-de Roo procedure, we presented the $\alpha'$-bootstrap program. Its main idea is to exploit the known symmetries of string theories in combination with T-duality to constrain the possible terms appearing order-by-order in the $\alpha'$ expansion. To this extent, we turned to an alternative formulation of DFT, extending the original doubled space to a much larger megaspace. In this way we were able to interpret geometrically the original gBdR identification as torsion constraints and a partial gauge fixing. In this framework, the generalized structure group $\GS$ of the theory plays a pivotal role. Thus, in section \ref{sec:structgroup}, we discussed it in full detail. Although we already considered the same group in \cite{Gitsis:2024gfb}, here we found a recursive procedure to construct it and a more adapted basis to express its generators. We noticed in section \ref{sec:gBdR} that in this way we can avoid the cumbersome regularization methods previously required for the identification (called \emph{collapse of towers} in \cite{Gitsis:2024gfb}).
This new approach has two considerable advantages: On the one hand it further clarifies the structures underlying the identification, and, perhaps even more importantly, it simplifies the explicit procedure considerably. As a direct consequence of this paradigm shift, we are able to write simple recursive formulas that allow us to effortlessly compute towers of $\alpha'$-corrections order-by-order in derivatives.

Another important improvement compared to \cite{Gitsis:2024gfb} is that we employ a new gauge fixing for the generalized frame on the megaspace. It was introduced in section \ref{sec:torsionconst}, and considerably simplifies the expressions for the generalized fluxes found in section \ref{sec:genflux}. These fluxes form the basis for the identification and are expanded to all orders in section \ref{sec:detailsID}. Finally we present the all-order action and discuss its symmetries (which include the Green-Schwarz transformations for the heterotic string) in the sections \ref{sec:action} and \ref{sec:GStransf}.

In section \ref{sec:actionandsym}, we explicitly applied the construction, computing the results of the identification up to order $\alpha'^2$ and matching them with known results from the literature for both the bosonic \cite{Gholian:2023kjj}, and the heterotic \cite{Hronek:2022dyr} cases. In order to do so, we had to implement changes of schemes through field redefinitions, whose forms are reported in appendix \ref{app:fieldredef}. Here one witnesses the huge gain in efficiency compared to the original gBdR identification. As all steps are recursive, one is able to reproduce the full $a$ and $b$ towers of corrections of the NS-NS sector of string theories to arbitrary high orders in $\alpha'$.

In a future paper, we plan to push the identification further to $\alpha'^3$, where the well-known issue of $\zeta(3)$ corrections becomes relevant \cite{Hsia:2024kpi}. This problem might, in principle, be solved by an additional deformation of $\kappa_{\alpha\beta}$, similar to the one considered in this work but giving contributions to higher orders. A very powerful feature of our approach, indeed, is the easiness of generalization, depending on the free choice of a symmetric $ad$-invariant bilinear $\kappa_{\alpha\beta}$. In order to recover the $a$ and $b$ towers of corrections, we chose in this article the most obvious form for it, namely a regularized version of the Killing metric, but nothing prevents us to choose another one. To this extent, we plan to investigate all the admissible families of deformations compatible with the algebraic structure developed in section \ref{sec:structgroup}. A natural question to investigate would then be whether \emph{all} the towers of corrections ($\zeta(3), \zeta(5),\dots$) can always be written as higher deformations induced by suitable $\kappa_{\alpha\beta}$'s.

\subsection*{Acknowledgements}
We would like to thank Mohammad Garousi, Diego Marques and Linus Wulff for helpful discussions and comments on the draft. FH additionally would like to thank Daniel Butter for sharing his insights and notes on the new gauge fixing we use in this article. LS thanks in particular Biplab Mahato for useful insights on the coding part. AG, FH and LS are supported by the SONATA BIS grant 2021/42/E/ST2/00304 from the National Science Centre (NCN), Poland. AG and LS acknowledge financial support from the doctoral school of the University of Wrocław.

\appendix

\section{Geometrical quantities and relations}\label{app:geometry}
Throughout all the article, we often employed geometric identities. Here, we will review our conventions and present useful relations for quantities involving the $\omega^{(+)}$ and $\omega^{(-)}$ connections.

First of all, we define the covariant derivative as
\begin{equation}\label{eq:covd}
    \nabla_i e^a{}_j = \partial_i e^a{}_j - \Gamma_{ij}{}^k e^a{}_k - \omega_{i b}{}^a e^b{}_j\,,
\end{equation}
where $\Gamma_{ij}{}^k$ is the ordinary Levi-Civita connection and $\omega_{ia}{}^b$ is the spin connection.

We also introduce the $H$-twisted connections
\begin{equation}\label{eq:pmconnections}
    \omega^{(\pm)}_{i a}{}^b = \omega_{i a}{}^b \pm \tfrac12 H_{i a}{}^b\, \qquad \text{and} \qquad
    \Gamma^{(\pm)k}_{ij} = \Gamma_{ij}{}^k \mp \tfrac12 H_{ij}{}^k\,.
\end{equation}
Based on them, we define two new covariant derivatives as
\begin{align}
        \nabla^{(-)}_i e^{a}{}_j &=\partial_i e^{a}{}_j - \Gamma_{ij}^{(-)k} e^{a}{}_k -\omega^{(-)}_{ib}{}^a e^{b}{}_j\,,\nn \\ \label{eq:+-connections}
        \nabla^{(+)}_i e^{a}{}_j &=\partial_i e^{a}{}_j - \Gamma_{ij}^{(+)k} e^{a}{}_k -\omega^{(+)}_{ib}{}^a e^{b}{}_j\,.
\end{align}
The corresponding Riemann tensors are given by
\begin{equation}\label{eq:chiralRiemann}
    R^{(\pm)}_{ija}{}^b = 2 \partial_{[i} \omega^{(\pm)}_{j]a}{}^b + 2 \omega^{(\pm)}_{[i|a}{}^c \omega^{(\pm)}_{|j]c}{}^b \,,
\end{equation}
while the torsions read
\begin{equation}
    T^{(\pm)k}_{ij} = 2 \Gamma^{(\pm)k}_{[ij]} = \mp H_{ij}{}^k \,.
\end{equation}

Moreover, we have two Bianchi identities for these modified quantities:
\begin{enumerate}
    \item The algebraic identity
    \begin{equation}
        R^{(\pm)}_{[ijk]l} = H_{[ij|}{}^m H_{|k]lm} \pm \nabla^{(\pm)}_l H_{ijk}\,,
    \end{equation}
    \item and the differential identity
    \begin{equation}\label{eq:diffBian}
        \nabla^{(\pm)}_{[i} R^{(\pm)}_{jk]lm} = \mp H_{[ij}{}^n R^{(\pm)}_{k]nlm}\,.
    \end{equation}
\end{enumerate}
From the first Bianchi identity, one can extract a relation between the $(+)$ and $(-)$-Riemann tensors
\begin{equation}\label{eq:relRpm}
    R^{(+)}_{ijkl} - R^{(-)}_{klij} = 2 \nabla_{[i} H_{jkl]} = 0\,,
\end{equation}
and the relation
\begin{equation}
    R^{(\pm)}_{ijkl}-R^{(\pm)}_{klij}= \pm 2\nabla^{(\pm)}_{[i}H_{j]kl} - 3 H_{[ij}{}^m H_{k]lm}
\end{equation}
for the pairwise exchange of indices.

By contracting the second and last index of the Riemann tensors, we obtain the Ricci tensors
\begin{equation}
    R^{(\pm)}_{ij} = R_{ikj}^{(\pm)k} = R_{ij} - \tfrac14 H_{ikl} H_j{}^{kl} \mp \tfrac12 \nabla^k H_{kij}\,. 
\end{equation}
They are proportional to the leading order field equations of the metric and $B$-field for a constant dilaton. Contracting \eqref{eq:diffBian}, one obtains the relation
\begin{equation}
    \nabla^{(\pm)}_i R^{(\mp)i}{}_{jln}=2\nabla^{(\pm)}_{[l|}R^{(\mp)}_{j|n]}\mp 3H_{[l|k}{}^i R^{(\pm)}_{|n]i}{}^k{}_j\,.
\end{equation}

We can also write Chern-Simons forms associated to the connections $\omega^{(\pm)}$, defined as
\begin{equation}\label{eq:chernsimonspm}
    \Omega^{(\pm)}_{i j k} = \omega^{(\pm)a b}_{[i}\partial_j \omega^{(\pm)}_{k]b a}+\tfrac{2}{3}\omega^{(\pm)}_{[i a}{}^b \omega^{(\pm)}_{j b}{}^c \omega_{k]c}^{(\pm)a}\,,
\end{equation}
which are related to the analogue quantities for $\omega$ according to
\begin{equation}\label{eq:OmegapmintermsofOmega}
    \Omega_{i j k}^{(\pm)} = \Omega_{i j k} \pm \tfrac{1}{2}\partial_{[i}(H_{j}{}^{l m} \omega_{k]m l}) \pm\tfrac{1}{2}H_{[i}{}^{l m}R_{j k] m l} +\tfrac{1}{4}\nabla_{[i}H_{j}{}^{l m}H_{k] m l} \pm\tfrac{1}{12}H_{i l}{}^m H_{j m }{}^n H_{k n}{}^l\, .
\end{equation}
In the same vein, $R^{(\pm)a b}_{i j}$ can also be rewritten as
\begin{equation}
    R^{(\pm)a b}_{i j} = R_{i j}{}^{a b} \pm \nabla_{[i}H_{j]}{}^{a b} +\tfrac{1}{2}H_{[i}{}^{a c}H_{j]c}{}^b\, .
\end{equation}

\section{Field redefinitions and relevant Lagrangians}\label{app:fieldredef}
In order to match our results at order $\alpha'$ and $\alpha'^2$ with the literature, we had to take into account the possibility of field redefinitions and integration by parts. We mainly followed the procedure outlined in \cite{Garousi:2019cdn}, implementing it in Wolfram Mathematica and making use of the packages \emph{xAct} \cite{xAct}, \emph{xTras} \cite{Nutma:2013zea} and \emph{xPert} \cite{Brizuela:2008ra}.

We start at order $\alpha'$. The idea is to build a general Lagrangian $\Lc_1$ from all the possible gauge invariant terms allowed at four derivatives, remembering that for the bosonic case only terms that are even under $B$-parity transformations contribute. This can be easily done with \emph{xTras}. Each term in $\Lc_1$ will come with an unspecified parameter $l_i$, where $i$ spans the total number of gauge invariant terms we can build.

We then want to consider all the possible integrations by parts we can have in our Lagrangian. To this end, one notices that a general integration by part is of the form \cite{Garousi:2019cdn}
\begin{equation}
    \int \dd^d x \sqrt{g} e^{-2\phi} \mathcal{J}_1=\int \dd^d x \sqrt{g} \nabla_\alpha(e^{-2\phi} J^\alpha_1)\,.
\end{equation}
Therefore, we build $J^\alpha_1$ from all the possible one-index covariant terms at three derivatives and use it to obtain $\mathcal{J}_1$, parametrizing each term by a free coefficient $j_i$.

After that, we have to consider all the possible field redefinitions we can apply to our Lagrangian:
\begin{align}
    g_{\mu\nu} &\rightarrow g_{\mu\nu}+ \alpha' \delta g_{\mu\nu}^1\,, \nn \\
    B_{\mu\nu} &\rightarrow B_{\mu\nu}+ \alpha' \delta B_{\mu\nu}^1\,, \nn \\
    \phi &\rightarrow\phi+\alpha' \delta \phi^1\,,
\end{align}
where the $\delta\,\cdot\,$ quantities take into account all the possible gauge covariant/invariant tensors we can write with two derivatives. They are parametrized by arbitrary coefficients $k_i$, and will affect the action at order $\alpha'$ according to
\begin{equation}
    \delta S_0 = \frac{\delta S_0}{\delta g_{\alpha\beta}}\delta g_{\alpha\beta}^1+\frac{\delta S_0}{\delta B_{\alpha\beta}}\delta B_{\alpha\beta}^1+\frac{\delta S_0}{\delta \phi}\delta \phi^1 = \int \dd^d x \sqrt{g} e^{-2\phi} \mathcal{K}_1\,.
\end{equation}

At this point, one can tune the coefficients $j_i, k_i$ to reduce the couplings $l_i$ to a subset $l'_i$ in the combination
\begin{equation}\label{eq:check}
    \Lc'_1=\Lc_1+\mathcal{J}_1+\mathcal{K}_1\,.
\end{equation}
However, there is still the possibility to further reduce them through Bianchi identities and reorderings of covariant derivatives. To remove this additional ambiguity, we exploit the trick performed in \cite{Garousi:2019cdn} and evaluate $\Lc'_1$ in a locally inertial frame, where the Bianchi identities are automatically satisfied and derivatives commute.

Finally, we choose as reference Lagrangian   
\begin{align}
    \Lc'_1 = \Lc^{(4)}_{\text{bos,MT}}&=\tfrac{\alpha'}{4}\Big(\tfrac{1}{24} H_{i}{}^{lm} H^{ijk} H_{jl}{}^{n} H_{kmn}-\tfrac{1}{8} H_{ij}{}^{l} H^{ijk} H_{k}{}^{mn} H_{lmn}+  R_{ijkl} R^{ijkl} \nn \\
    &\phantom{\tfrac{\alpha'}{4}\Big(}-\tfrac{1}{2}  H_{i}{}^{lm} H^{ijk} R_{jklm}\Big)
\end{align}
in the Metsaev-Tseytlin scheme \cite{Metsaev:1987bc}, fixing the $l'_i$ parameters. In order to prove that a second Lagrangian $\Lc_1$ is equivalent, we have to find a solution for the free parameters $j_i$, and $k_i$ such that \eqref{eq:check} holds. If such a solution exists, we obtain the required field redefinitions from the coefficients $k_i$ and the boundary terms from $j_i$. Bianchi identities are automatically taken care of.

For the bosonic case, we considered at order $\alpha'$ the following Lagrangians
\begin{align}
    \Lc^{(4)}_{\text{bos,M}}=&\tfrac{\alpha'}{4}\Big( \tfrac{1}{24} H_{i}{}^{lm} H^{ijk} H_{jl}{}^{n} H_{kmn} - \tfrac{1}{8} H_{ij}{}^{l} H^{ijk} H_{k}{}^{mn} H_{lmn} +  R_{ijkl} R^{ijkl} \nn \\
    & \phantom{\alpha'\Big(} - \tfrac{1}{2} H_{i}{}^{lm} H^{ijk} R_{jklm} + \tfrac{1}{144} H_{ijk} H^{ijk} H_{lmn} H^{lmn} + H_{i}{}^{jk} H_{ljk} R^{il} \nn \\
    & \phantom{\alpha'\Big(}-4 R_{ij} R^{ij}-\tfrac{1}{6} H_{ijk} H^{ijk} R+ R^2 -\tfrac{2}{3} H_{ijk} H^{ijk} \nabla^l \nabla_l \phi \nn \\
    & \phantom{\alpha'\Big(}+\tfrac{2}{3} H_{ijk} H^{ijk} \nabla_l \phi \nabla^l \phi + 8 R \nabla_i \phi \nabla^i \phi + 16 \nabla_i \phi \nabla^i \phi \nabla_j \nabla^j \phi \nn \\
    & \phantom{\alpha'\Big(}-16 R_{ij} \nabla^i \phi \nabla^j \phi -16 \nabla_i\phi \nabla^i \phi \nabla_j \phi \nabla^j \phi+2 H_i{}^{jk}H_{ljk} \nabla^i \nabla^l \phi \Big) \\
    \Lc^{(4)}_{\text{bos,W}}=&\alpha' \Big( \tfrac{1}{24} H_{i}{}^{lm} H^{ijk} H_{jl}{}^{n} H_{kmn} + \tfrac{1}{8} H_{ij}{}^{l} H^{ijk} H_{k}{}^{mn} H_{lmn} +  R_{ijkl} R^{ijkl} \nn \\
    & \phantom{\alpha'\Big(} -  H_{i}{}^{lm} H^{ijk} R_{jklm} -  H_{i}{}^{lm} H^{ijk} R_{jlkm} -  \nabla_{[k} H_{l]ij} \nabla^{l} H^{ijk}\Big) \\
    \Lc^{(4)}_{\text{bos,DFT}}=&\tfrac{1}{4} \Lc^{(4)}_{\text{bos,W}}\,, 
\end{align}
namely $\Lc^{(4)}_{\text{bos,M}}$ in the Meissner scheme \cite{Meissner:1996sa}, $\Lc^{(4)}_{\text{bos,W}}$ in the scheme considered in \cite{Wulff:2024ips}, and $\Lc^{(4)}_{\text{bos,DFT}}$ in our scheme. We explicitly checked that they are, of course, all related through Bianchi identities and field redefinitions (and an overall rescaling) to $\Lc^{(4)}_{\text{bos,MT}}$. In particular, the field redefinitions needed to bring $\Lc^{(4)}_{\text{bos,DFT}}$ in the form of $\Lc^{(4)}_{\text{bos,MT}}$ read
\begin{align}
        \delta g^1_{ij}&=-H_{i}{}^{kl}H_{jkl}\,, \nn \\
        \delta B^1_{ij}&=-2\nabla_k H_{ij}{}^{k}+4H_{ijk}\nabla^k\phi\,, \nn \\
        \delta \phi^1&=-\tfrac{1}{4}H_{ijk}H^{ijk}\,.
\end{align}
For the heterotic case, we directly matched equation (5.35) from \cite{Hronek:2022dyr} at order $\alpha'$ without the need of any field redefinition.

The procedure can be easily extended to the next order, $\alpha'^2$. However, there is a small caveat: For the field redefinitions, we now also have to take into account terms coming from the variation of the Lagrangian at order $\alpha'$ that we want to compare. Thanks to the help of the package \emph{xPert} they can be quickly computed by taking into account the previous results. It is then necessary to transform this variation in the reference scheme (Metsaev-Tseytlin for us), before adding it to the final result (see \cite{Garousi:2019cdn} for additional details).

In this case, the field redefinitions will assume the form
\begin{align}
        g_{\mu\nu} &\rightarrow g_{\mu\nu}+ \alpha' \delta g_{\mu\nu}^1+\alpha'^2 \delta g_{\mu\nu}^2\,, \nn \\
        B_{\mu\nu} &\rightarrow B_{\mu\nu}+ \alpha' \delta B_{\mu\nu}^1+\alpha'^2 \delta B_{\mu\nu}^2\,, \nn \\
        \phi &\rightarrow\phi+\alpha' \delta \phi^1+\alpha'^2 \delta \phi^2\,.
\end{align}

Following the procedure for the bosonic case, we related the Lagrangian in the Metsaev-Tseytlin scheme \cite{Gholian:2023kjj}\footnote{Here we restored the $\alpha'\rightarrow\tfrac{1}{4}\alpha'$ shift, for consistency with the expression of $\Lc^{(4)}_{\text{bos,MT}}$ given above, and our rersult.}
\begin{align}
    \Lc^{(6)}_{\text{bos,MT}}=&\tfrac{\alpha'^2}{16}\Big(\nn \\
    &-\tfrac{1}{12} H_{i}{}^{lm} H^{ikn} H_{kl}{}^{p} H_{n}{}^{jq} H_{mj}{}^{r} H_{pqr} + \tfrac{1}{30} H_{ij}{}^{l} H^{ijk} H_{k}{}^{mn} H_{l}{}^{pq} H_{mn}{}^{r} H_{pqr}\nn \\
    &+ \tfrac{3}{10} H_{ij}{}^{l} H^{ijk} H_{k}{}^{mn} H_{lm}{}^{p} H_{n}{}^{qr} H_{pqr} + \tfrac{13}{20} H_{i}{}^{mn} H_{j}{}^{pq} H_{kmn} H_{lpq} R^{ijkl} \nn \\
    &+ \tfrac{2}{5} H_{i}{}^{mn} H_{jm}{}^{p} H_{kn}{}^{q} H_{lpq} R^{ijkl} + \tfrac{18}{5} H_{ik}{}^{m} H_{j}{}^{pn} H_{lp}{}^{q} H_{mnq} R^{ijkl} \nn \\
    &- \tfrac{43}{5} H_{ik}{}^{m} H_{j}{}^{np} H_{lm}{}^{q} H_{npq} R^{ijkl} - \tfrac{16}{5} H_{ik}{}^{m} H_{jl}{}^{n} H_{m}{}^{pq} H_{npq} R^{ijkl} \nn \\
    &- 2 H_{jm}{}^{p} H_{lnp} R_{i}{}^{m}{}_{k}{}^{n} R^{ijkl} - 2 H_{jl}{}^{p} H_{mnp} R_{i}{}^{m}{}_{k}{}^{n} R^{ijkl} - \tfrac{4}{3} R_{i}{}^{m}{}_{k}{}^{n} R^{ijkl} R_{jnlm} \nn \\
    &+ \tfrac{4}{3} R_{ij}{}^{mn} R^{ijkl} R_{kmln} + 3 H_{j}{}^{np} H_{mnp} R^{ijkl} R_{k}{}^{m}{}_{il} + 2 H_{jm}{}^{p} H_{lnp} R^{ijkl} R_{k}{}^{m}{}_{i}{}^{n} \nn \\
    &+ 2 H_{ijm} H_{lnp} R^{ijkl} R_{k}{}^{mnp} + \tfrac{13}{10} H_{i}{}^{kl} H_{jk}{}^{m} H_{l}{}^{np} H_{mnp} \nabla^{j} \nabla^{i} \phi \nn \\
    &+ \tfrac{13}{5} H_{k}{}^{mn} H_{lmn} R_{i}{}^{k}{}_{j}{}^{l} \nabla^{j} \nabla^{i} \phi - \tfrac{52}{5} H_{jl}{}^{n} H_{kmn} R_{i}{}^{klm} \nabla^{j} \nabla^{i} \phi \nn \\
    &- \tfrac{26}{5} H_{ikm} H_{jln} R^{klmn} \nabla^{j} \nabla^{i} \phi + \tfrac{13}{5} \nabla^{j} \nabla^{i} \phi \nabla_{m} H_{jkl} \nabla^{m} H_{i}{}^{kl} \nn \\
    &+ \tfrac{13}{10} H_{jk}{}^{m} H^{jkl} H_{l}{}^{np} \nabla^{i} \phi \nabla_{p} H_{imn} + \tfrac{1}{20} H_{i}{}^{lm} H^{ijk} \nabla_{p} H_{lmn} \nabla^{p} H_{jk}{}^{n} \nn \\
    &- \tfrac{13}{20} H_{i}{}^{jk} H_{lm}{}^{p} H^{lmn} \nabla^{i} \phi \nabla_{p} H_{jkn} + \tfrac{1}{5} H_{i}{}^{lm} H^{ijk} \nabla_{n} H_{kmp} \nabla^{p} H_{jl}{}^{n} \nn \\
    &- \tfrac{6}{5} H_{i}{}^{lm} H^{ijk} \nabla_{p} H_{kmn} \nabla^{p} H_{jl}{}^{n} - \tfrac{6}{5} H_{ij}{}^{l} H^{ijk} \nabla_{n} H_{lmp} \nabla^{p} H_{k}{}^{mn} \nn \\
    &+ \tfrac{17}{10} H_{ij}{}^{l} H^{ijk} \nabla_{p} H_{lmn} \nabla^{p} H_{k}{}^{mn} \Big),
\end{align}
with our result, here expanded in unmodified geometrical quantities:
\begin{align}
    &\Lc^{(6)}_{\text{bos,DFT}}=\alpha'^2\Big( \nn \\
    &-\tfrac{1}{192}H_{i}{}^{lm}H^{ijk}H_{jl}{}^{n}H_{k}{}^{pq}H_{mp}{}^{r}H_{nqr} + \tfrac{1}{32}H_{ij}{}^{l}H^{ijk}H_{k}{}^{mn}H_{l}{}^{pq}H_{mp}{}^{r}H_{nqr} \nn \\
    &+ \tfrac{1}{192}H_{ij}{}^{l}H^{ijk}H_{k}{}^{mn}H_{l}{}^{pq}H_{mn}{}^{r}H_{pqr} + \tfrac{1}{128}H_{ij}{}^{l}H^{ijk}H_{k}{}^{mn}H_{lm}{}^{p}H_{n}{}^{qr}H_{pqr} \nn \\
    &- \tfrac{1}{3}R_{i}{}^{m}{}_{k}{}^{n}R^{ijkl}R_{jmln} + \tfrac{1}{8}R_{ij}{}^{mn}R^{ijkl}R_{klmn} - \tfrac{1}{4}R_{ij}{}^{mn}R^{ijkl}R_{kmln} \nn \\
    &+ \tfrac{3}{4}H^{ijk}H^{lmn}R_{ijl}{}^{p}R_{kmnp} - \tfrac{1}{8}H^{ijk}H^{lmn}R_{ilj}{}^{p}R_{kmnp} - \tfrac{1}{4}H_{i}{}^{lm}H^{ijk}R_{jl}{}^{np}R_{kmnp} \nn \\
    &+ \tfrac{1}{4}H_{i}{}^{lm}H^{ijk}R_{jl}{}^{np}R_{knmp} - \tfrac{1}{8}H_{i}{}^{lm}H^{ijk}R_{j}{}^{n}{}_{l}{}^{p}R_{knmp} - \tfrac{1}{8}H^{ijk}H^{lmn}R_{ijl}{}^{p}R_{kpmn} \nn \\
    &+ \tfrac{3}{8}H_{i}{}^{lm}H^{ijk}R_{j}{}^{n}{}_{l}{}^{p}R_{kpmn} + \tfrac{1}{8}H_{i}{}^{lm}H^{ijk}R_{jk}{}^{np}R_{lmnp} + \tfrac{1}{8}H_{ij}{}^{l}H^{ijk}R_{k}{}^{mnp}R_{lmnp} \nn \\
    &+ \tfrac{1}{4}H_{i}{}^{lm}H^{ijk}R_{jk}{}^{np}R_{lnmp} + \tfrac{1}{4}H_{i}{}^{lm}H^{ijk}R_{j}{}^{n}{}_{k}{}^{p}R_{lnmp} + \tfrac{3}{16}H_{ij}{}^{l}H^{ijk}H_{k}{}^{mn}H_{m}{}^{pq}R_{lnpq} \nn \\
    &+ \tfrac{1}{16}H_{ij}{}^{l}H^{ijk}H_{k}{}^{mn}H_{m}{}^{pq}R_{lpnq} - \tfrac{1}{16}H_{ij}{}^{l}H^{ijk}H_{k}{}^{mn}H_{l}{}^{pq}R_{mnpq} \nn \\
    &- \tfrac{1}{16}H_{i}{}^{lm}H^{ijk}H_{jl}{}^{n}H_{k}{}^{pq}R_{mpnq} - \tfrac{1}{16}H_{ij}{}^{l}H^{ijk}H_{k}{}^{mn}H_{l}{}^{pq}R_{mpnq} \nn \\
    &- \tfrac{1}{16}R_{lmnp}\nabla_{i} H^{klm}\nabla^{i} H_{k}{}^{np} - \tfrac{5}{16}R_{lnmp}\nabla_{i} H^{klm}\nabla^{i} H_{k}{}^{np} + \tfrac{1}{16}H^{klm}H^{npq}\nabla_{i} H_{kln}\nabla^{i} H_{mpq} \nn \\
    &- \tfrac{5}{64}H_{k}{}^{np}H^{klm}\nabla_{i} H_{ln}{}^{q}\nabla^{i} H_{mpq} + \tfrac{5}{64}H_{k}{}^{np}H^{klm}\nabla_{i} H_{lm}{}^{q}\nabla^{i} H_{npq} \nn \\
    &+ \tfrac{1}{16}H_{kl}{}^{n}H^{klm}\nabla_{i} H_{m}{}^{pq}\nabla^{i} H_{npq} + \tfrac{3}{16}\nabla_{i} R^{klmn}\nabla^{i} R_{klmn} - \tfrac{1}{4}H^{klm}\nabla_{i} H_{k}{}^{np}\nabla^{i} R_{lmnp} \nn \\
    &- \tfrac{1}{4}H^{klm}\nabla_{i} H_{k}{}^{np}\nabla^{i} R_{lnmp} + \tfrac{1}{16}H^{ijk}R_{klmn}\nabla_{j} \nabla_{i} H^{lmn} - \tfrac{3}{32}\nabla_{i} \nabla^{n} H^{klm}\nabla^{i} \nabla_{m} H_{kln} \nn \\
    &+ \tfrac{1}{8}H^{ijk}R_{klmn}\nabla_{j} \nabla^{n} H_{i}{}^{lm} - \tfrac{1}{16}H^{ijk}R_{knlm}\nabla_{j} \nabla^{n} H_{i}{}^{lm} + \tfrac{1}{16}H^{ijk}\nabla_{i} H^{lmn}\nabla_{k} R_{jlmn} \nn \\
    &+ \tfrac{1}{64}H^{ijk}H^{lmn}\nabla_{k} H_{mnp}\nabla_{l} H_{ij}{}^{p} + \tfrac{11}{64}H_{i}{}^{lm}H^{ijk}\nabla_{k} H_{mnp}\nabla_{l} H_{j}{}^{np} \nn \\
    &+ \tfrac{1}{64}H_{ij}{}^{l}H^{ijk}\nabla_{k} H^{mnp}\nabla_{l} H_{mnp} + \tfrac{3}{32}\nabla_{k} \nabla_{i} H^{mnp}\nabla^{k} \nabla^{i} H_{mnp} - \tfrac{3}{16}R_{jkmn}\nabla_{i} H_{l}{}^{mn}\nabla^{l} H^{ijk} \nn \\
    &+ \tfrac{1}{16}R_{jmkn}\nabla_{i} H_{l}{}^{mn}\nabla^{l} H^{ijk} - \tfrac{9}{64}H_{i}{}^{lm}H^{ijk}\nabla_{l} H_{j}{}^{np}\nabla_{m} H_{knp} + \tfrac{3}{32}H_{i}{}^{lm}H^{ijk}\nabla_{k} H_{j}{}^{np}\nabla_{m} H_{lnp} \nn \\
    &- \tfrac{1}{4}\nabla_{l} R_{ijkm}\nabla^{m} R^{ijkl} + \tfrac{1}{16}\nabla_{k} \nabla_{j} H_{ilm}\nabla^{m} \nabla^{l} H^{ijk} + \tfrac{1}{32}\nabla_{k} \nabla_{m} H_{ijl}\nabla^{m} \nabla^{l} H^{ijk} \nn \\
    &- \tfrac{1}{8}H^{ijk}H^{lmn}\nabla_{j} H_{il}{}^{p}\nabla_{n} H_{kmp} + \tfrac{5}{32}H^{ijk}H^{lmn}\nabla_{l} H_{ij}{}^{p}\nabla_{n} H_{kmp} + \tfrac{1}{8}H^{ijk}\nabla_{i} H^{lmn}\nabla_{n} R_{jlkm} \nn \\
    &+ \tfrac{1}{4}R_{klmn}\nabla^{l} H^{ijk}\nabla^{n} H_{ij}{}^{m} - \tfrac{5}{16}R_{kmln}\nabla^{l} H^{ijk}\nabla^{n} H_{ij}{}^{m} + \tfrac{1}{16}R_{knlm}\nabla^{l} H^{ijk}\nabla^{n} H_{ij}{}^{m} \nn \\
    &+ \tfrac{1}{2}R_{jkmn}\nabla^{l} H^{ijk}\nabla^{n} H_{il}{}^{m} - \tfrac{1}{2}R_{jmkn}\nabla^{l} H^{ijk}\nabla^{n} H_{il}{}^{m} - \tfrac{1}{8}H^{ijk}\nabla_{k} R_{jlmn}\nabla^{n} H_{i}{}^{lm} \nn \\
    &- \tfrac{1}{16}H^{ijk}\nabla_{k} R_{jnlm}\nabla^{n} H_{i}{}^{lm} - \tfrac{1}{4}H^{ijk}\nabla_{m} R_{jkln}\nabla^{n} H_{i}{}^{lm} + \tfrac{1}{4}H^{ijk}\nabla_{m} R_{jlkn}\nabla^{n} H_{i}{}^{lm} \nn \\
    &+ \tfrac{1}{32}\nabla^{i} \nabla_{m} H_{kln}\nabla^{n} \nabla_{i} H^{klm} - \tfrac{1}{32}\nabla^{i} \nabla_{n} H_{klm}\nabla^{n} \nabla_{i} H^{klm} - \tfrac{3}{32}\nabla_{m} \nabla^{i} H_{kln}\nabla^{n} \nabla_{i} H^{klm} \nn \\
    &- \tfrac{1}{8}H^{ijk}R_{knlm}\nabla^{n} \nabla_{j} H_{i}{}^{lm} - \tfrac{1}{8}H^{ijk}R_{klmn}\nabla^{n} \nabla^{m} H_{ij}{}^{l} - \tfrac{1}{8}H^{ijk}R_{kmln}\nabla^{n} \nabla^{m} H_{ij}{}^{l} \nn \\
    &+ \tfrac{1}{8}H^{ijk}R_{knlm}\nabla^{n} \nabla^{m} H_{ij}{}^{l} + \tfrac{1}{32}H_{i}{}^{lm}H^{ijk}H_{j}{}^{np}\nabla_{p} \nabla_{k} H_{lmn} - \tfrac{1}{16}H_{i}{}^{lm}H^{ijk}H_{j}{}^{np}\nabla_{p} \nabla_{m} H_{kln} \nn \\
    &- \tfrac{1}{32}H_{i}{}^{lm}H^{ijk}H_{j}{}^{np}\nabla_{p} \nabla_{n} H_{klm} - \tfrac{1}{16}H^{ijk}H^{lmn}\nabla_{k} H_{mnp}\nabla^{p} H_{ijl} \nn \\
    &- \tfrac{3}{32}H^{ijk}H^{lmn}\nabla_{n} H_{kmp}\nabla^{p} H_{ijl} + \tfrac{1}{8}H_{i}{}^{lm}H^{ijk}\nabla_{m} H_{lnp}\nabla^{p} H_{jk}{}^{n} \nn \\
    &- \tfrac{1}{64}H_{i}{}^{lm}H^{ijk}\nabla_{n} H_{lmp}\nabla^{p} H_{jk}{}^{n} - \tfrac{5}{8}H_{i}{}^{lm}H^{ijk}\nabla_{m} H_{knp}\nabla^{p} H_{jl}{}^{n} \nn \\
    &+ \tfrac{13}{64}H_{i}{}^{lm}H^{ijk}\nabla_{n} H_{kmp}\nabla^{p} H_{jl}{}^{n} - \tfrac{1}{16}H_{ij}{}^{l}H^{ijk}\nabla_{l} H_{mnp}\nabla^{p} H_{k}{}^{mn} \nn \\
    & - \tfrac{1}{16}H_{ij}{}^{l}H^{ijk}\nabla_{n} H_{lmp}\nabla^{p} H_{k}{}^{mn}\Big) \nn \\
&+\Lc^{(6)\text{odd}}_{\text{bos,DFT}}\,,
\end{align}
where $\Lc^{(6)\text{odd}}_{\text{bos,DFT}}$ contains only odd terms under $B$-parity transformations, and vanishes by Bianchi identities. Thus, the expected $B$-parity of the Lagrangian is preserved. The corresponding field redefinitions at order $\alpha'^2$ are
\begin{align}
    \delta g^2_{mn}=&\,\tfrac{59}{480}H_{klp}H^{klp}H_{m}{}^{ij}H_{nij} - \tfrac{3}{10}H_{j}{}^{lp}H_{klp}H_{m}{}^{ij}H_{ni}{}^{k} - 3H_{ik}{}^{p}H_{jlp}H_{m}{}^{ij}H_{n}{}^{kl} \nn \\
    &- \tfrac{1}{5}H_{ij}{}^{p}H_{klp}H_{m}{}^{ij}H_{n}{}^{kl} - \tfrac{17}{960}H_{ijk}H^{ijk}H_{lpq}H^{lpq} g_{mn} + \tfrac{38}{5}H_{m}{}^{ij}H_{n}{}^{kl}R_{ijkl} \nn \\
    &- \tfrac{37}{4}H_{i}{}^{kl}H_{n}{}^{ij}R_{mjkl} + 6H_{ij}{}^{l}H^{ijk}R_{mknl} - 2R_{m}{}^{ijk}R_{nijk} - \tfrac{26}{5}H_{mi}{}^{k}H_{njk}R^{ij} \nn \\
    &- 4R_{minj}R^{ij} - \tfrac{33}{20}H_{ijk}H_{n}{}^{jk}R_{m}{}^{i} - \tfrac{1}{6}H_{ijk}H^{ijk}R_{mn} + 4R_{m}{}^{i}R_{ni} \nn \\
    &+ \tfrac{11}{48}H_{ijk}H^{ijk} g_{mn}R + 2R_{mn}R - \tfrac{1}{5} g_{mn}{R}^{2} - \tfrac{36}{5}H_{m}{}^{jk}H_{njk}\nabla_{i} \nabla^{i} \phi \nn \\
    &+ \tfrac{17}{20}H_{jpq}H^{jpq} g_{mn}\nabla_{i} \nabla^{i} \phi + 8R_{mn}\nabla_{i} \nabla^{i} \phi - \tfrac{4}{5} g_{mn}R\nabla_{i} \nabla^{i} \phi - 4\nabla_{i} \nabla^{i} R_{mn} \nn \\
    &- \tfrac{13}{30}H^{jpq} g_{mn}\nabla_{i} H_{jpq}\nabla^{i} \phi - \tfrac{199}{10}H_{n}{}^{jk}\nabla_{i} H_{mjk}\nabla^{i} \phi + \tfrac{17}{2}H_{m}{}^{jk}H_{njk}\nabla_{i} \phi\nabla^{i} \phi \nn \\
    &- \tfrac{17}{20}H_{jpq}H^{jpq} g_{mn}\nabla_{i} \phi\nabla^{i} \phi - 8R_{mn}\nabla_{i} \phi\nabla^{i} \phi + \tfrac{4}{5} g_{mn}R\nabla_{i} \phi\nabla^{i} \phi \nn \\
    &+ 16\nabla_{i} R_{mn}\nabla^{i} \phi- 16R_{minj}\nabla^{i} \phi\nabla^{j} \phi - \tfrac{26}{5}H_{mj}{}^{k}H_{nik}\nabla^{j} \nabla^{i} \phi + 8R_{minj}\nabla^{j} \nabla^{i} \phi \nn \\
    & + \tfrac{119}{5}H_{n}{}^{jk}\nabla^{i} \phi\nabla_{k} H_{mij} + \tfrac{119}{10}H_{n}{}^{ij}\nabla_{k} \nabla_{j} H_{mi}{}^{k} + \tfrac{173}{20}H_{n}{}^{ij}\nabla_{k} \nabla^{k} H_{mij} \nn \\
    &+ \tfrac{133}{10}\nabla_{j} H_{nik}\nabla^{k} H_{m}{}^{ij} + \tfrac{87}{20}\nabla_{k} H_{nij}\nabla^{k} H_{m}{}^{ij} + \tfrac{13}{20}H^{ijk} g_{mn}\nabla_{l} \nabla_{k} H_{ij}{}^{l} \nn \\
    &+ \tfrac{13}{20} g_{mn}\nabla_{k} H_{ijl}\nabla^{l} H^{ijk} + \tfrac{13}{5}H^{ijk}\nabla_{k} H_{nij}\nabla_{m} \phi - 16\nabla^{i} \phi\nabla_{m} R_{ni} \nn \\
    &- \tfrac{7}{10}H_{ijk}H_{n}{}^{jk}\nabla_{m} \nabla^{i} \phi - \tfrac{133}{20}H^{ijk}\nabla_{m} \nabla_{k} H_{nij} + 2\nabla_{n} \nabla_{m} R- \tfrac{59}{40}H_{m}{}^{ij}H_{nij}R\,,\\
    \delta B^2_{mn}=&\,40R_{mjnk}\nabla_{i} H^{ijk} - \tfrac{16}{15}H^{jkl}H_{mn}{}^{i}\nabla_{i} H_{jkl} + \tfrac{89}{20}H_{i}{}^{kl}H_{jkl}H_{mn}{}^{j}\nabla^{i} \phi \nn \\
    &- 4H_{jkl}H_{m}{}^{jk}H_{ni}{}^{l}\nabla^{i} \phi + 16H_{ikl}H_{m}{}^{jk}H_{nj}{}^{l}\nabla^{i} \phi + \tfrac{474}{5}H_{n}{}^{jk}R_{mijk}\nabla^{i} \phi \nn \\
    &- 104H_{i}{}^{jk}R_{mjnk}\nabla^{i} \phi + \tfrac{184}{5}H_{nij}R_{m}{}^{j}\nabla^{i} \phi + \tfrac{67}{20}H_{mni}\nabla^{i} R - \tfrac{22}{5}H_{m}{}^{ij}H_{n}{}^{kl}\nabla_{j} H_{ikl} \nn \\
    &- \tfrac{32}{5}R_{n}{}^{i}\nabla_{j} H_{mi}{}^{j} + 5H_{i}{}^{kl}H_{n}{}^{ij}\nabla_{j} H_{mkl} + \tfrac{187}{10}R^{ij}\nabla_{j} H_{mni} - 4\nabla^{i} \phi\nabla_{j} \nabla_{i} H_{mn}{}^{j} \nn \\
    &- 28\nabla^{i} \phi\nabla_{j} \nabla^{j} H_{mni} + 8\nabla_{j} \nabla^{j} \nabla_{i} H_{mn}{}^{i} + 32\nabla^{i} \phi\nabla_{j} H_{mni}\nabla^{j} \phi - \tfrac{308}{5}H_{nij}\nabla^{j} R_{m}{}^{i} \nn \\
    &- \tfrac{26}{5}H_{mnj}\nabla^{i} \phi\nabla^{j} \nabla_{i} \phi + \tfrac{53}{5}\nabla_{i} H_{mnj}\nabla^{j} \nabla^{i} \phi - 48R_{nijk}\nabla^{k} H_{m}{}^{ij} \nn \\
    &- \tfrac{63}{40}H_{i}{}^{jk}H_{mn}{}^{i}\nabla_{l} H_{jk}{}^{l} - 8H_{m}{}^{ij}H_{ni}{}^{k}\nabla_{l} H_{jk}{}^{l} - \tfrac{86}{5}H_{i}{}^{kl}H_{n}{}^{ij}\nabla_{l} H_{mjk} \nn \\
    &+ 2H_{ij}{}^{k}H_{n}{}^{ij}\nabla_{l} H_{mk}{}^{l} - \tfrac{131}{40}H_{ij}{}^{l}H^{ijk}\nabla_{l} H_{mnk} - \tfrac{108}{5}\nabla_{j} H_{ni}{}^{j}\nabla_{m} \nabla^{i} \phi \nn \\
    &+ \tfrac{24}{5}H_{nij}\nabla^{i} \phi\nabla_{m} \nabla^{j} \phi\,,\\
    \delta \phi^2=&-\tfrac{3}{4}H_{i}{}^{lm}H^{ijk}H_{jl}{}^{n}H_{kmn} - \tfrac{139}{320}H_{ij}{}^{l}H^{ijk}H_{k}{}^{mn}H_{lmn} + \tfrac{5}{384}H_{ijk}H^{ijk}H_{lmn}H^{lmn} \nn \\
    &- \tfrac{1}{2}R_{ijkl}R^{ijkl} + \tfrac{337}{80}H_{i}{}^{lm}H^{ijk}R_{jklm} + \tfrac{31}{40}H_{i}{}^{kl}H_{jkl}R^{ij} - R_{ij}R^{ij} \nn \\
    &+ \tfrac{1}{32}H_{ijk}H^{ijk}R + \tfrac{1}{10}{R}^{2} + \tfrac{9}{80}H_{jkl}H^{jkl}\nabla_{i} \nabla^{i} \phi + \tfrac{1}{5}R\nabla_{i} \nabla^{i} \phi - \tfrac{1}{2}\nabla_{i} \nabla^{i} R \nn \\
    &- \tfrac{463}{60}H^{jkl}\nabla_{i} H_{jkl}\nabla^{i} \phi + 2\nabla_{i} \phi\nabla^{i} R - 4R_{ij}\nabla^{i} \phi\nabla^{j} \phi + \tfrac{741}{80}H^{ijk}\nabla_{l} \nabla_{k} H_{ij}{}^{l} \nn \\
    &+ \tfrac{161}{20}\nabla_{k} H_{ijl}\nabla^{l} H^{ijk}\,.
\end{align}

Finally, the field redefinitions employed to remove $\Lc^{(6)}_{\text{redef}}$ for the heterotic string ($a=-\alpha'$, and $b=0$) from \eqref{eq:sixordL} are
\begin{align}
    \delta g^2_{mn}=&\tfrac{1}{192}H_{klp}H^{klp}H_{m}{}^{ij}H_{nij} - \tfrac{1}{32}H_{j}{}^{lp}H_{klp}H_{m}{}^{ij}H_{ni}{}^{k} + \tfrac{1}{64}H_{ik}{}^{p}H_{jlp}H_{m}{}^{ij}H_{n}{}^{kl} \nn \\
    &- \tfrac{1}{64}H_{ij}{}^{p}H_{klp}H_{m}{}^{ij}H_{n}{}^{kl} - \tfrac{1}{1920}H_{ijk}H^{ijk}H_{lpq}H^{lpq}g_{mn} + \tfrac{3}{32}H_{m}{}^{ij}H_{n}{}^{kl}R_{ijkl} \nn \\
    &- \tfrac{5}{32}H_{i}{}^{kl}H_{n}{}^{ij}R_{mjkl} + \tfrac{1}{16}H_{ij}{}^{l}H^{ijk}R_{mknl} - \tfrac{1}{8}R_{m}{}^{ijk}R_{nijk} - \tfrac{1}{8}H_{mi}{}^{k}H_{njk}R^{ij} \nn \\
    &- \tfrac{1}{4}R_{minj}R^{ij} - \tfrac{1}{96}H_{ijk}H^{ijk}R_{mn} + \tfrac{1}{4}R_{m}{}^{i}R_{ni} - \tfrac{1}{16}H_{m}{}^{ij}H_{nij}R \nn \\
    &+ \tfrac{7}{960}H_{ijk}H^{ijk} g_{mn}R + \tfrac{1}{8}R_{mn}R - \tfrac{1}{80}g_{mn}{R}^{2} - \tfrac{1}{4}H_{m}{}^{jk}H_{njk}\nabla_{i} \nabla^{i} \phi \nn \\
    &+ \tfrac{1}{40}H_{jpq}H^{jpq}g_{mn}\nabla_{i} \nabla^{i} \phi + \tfrac{1}{2}R_{mn}\nabla_{i} \nabla^{i} \phi - \tfrac{1}{20}g_{mn}R\nabla_{i} \nabla^{i} \phi - \tfrac{1}{4}\nabla_{i} \nabla^{i} R_{mn} \nn \\
    &+ \tfrac{1}{4}H_{n}{}^{jk}R_{mijk}\nabla^{i} \phi - \tfrac{1}{4}H_{n}{}^{jk}\nabla_{i} H_{mjk}\nabla^{i} \phi + \tfrac{1}{4}H_{m}{}^{jk}H_{njk}\nabla_{i} \phi\nabla^{i} \phi \nn \\
    &- \tfrac{1}{40}H_{jpq}H^{jpq}g_{mn}\nabla_{i} \phi\nabla^{i} \phi - \tfrac{1}{2}R_{mn}\nabla_{i} \phi\nabla^{i} \phi + \tfrac{1}{20} g_{mn}R\nabla_{i} \phi\nabla^{i} \phi + \nabla_{i} R_{mn}\nabla^{i} \phi \nn \\
    &- R_{minj}\nabla^{i} \phi\nabla^{j} \phi - \tfrac{1}{4}H_{nij}\nabla^{j} R_{m}{}^{i} - \tfrac{1}{4}H_{mj}{}^{k}H_{nik}\nabla^{j} \nabla^{i} \phi + \tfrac{1}{2}R_{minj}\nabla^{j} \nabla^{i} \phi \nn \\
    &+ \tfrac{3}{8}H_{n}{}^{jk}\nabla^{i} \phi\nabla_{k} H_{mij} + \tfrac{3}{16}H_{n}{}^{ij}\nabla_{k} \nabla_{j} H_{mi}{}^{k} + \tfrac{1}{8}H_{n}{}^{ij}\nabla_{k} \nabla^{k} H_{mij}  \nn \\
    &+ \tfrac{1}{4}R_{nijk}\nabla^{k} H_{m}{}^{ij} + \tfrac{3}{16}\nabla_{j} H_{nik}\nabla^{k} H_{m}{}^{ij} - \tfrac{1}{16}H_{i}{}^{kl}H_{n}{}^{ij}\nabla_{l} H_{mjk} - \nabla^{i} \phi\nabla_{m} R_{ni}  \nn \\
    &- \tfrac{1}{16}H^{ijk}\nabla_{m} \nabla_{k} H_{nij}+ \tfrac{1}{8}\nabla_{n} \nabla_{m} R\,,\\
    \delta B^2_{mn}=&-\tfrac{1}{16}H_{i}{}^{kl}H_{n}{}^{ij}R_{mjkl} + \tfrac{1}{8}R_{mjnk}\nabla_{i} H^{ijk} - \tfrac{1}{96}H^{jkl}H_{mn}{}^{i}\nabla_{i} H_{jkl} \nn \\
    &+ \tfrac{1}{16}H_{i}{}^{kl}H_{jkl}H_{mn}{}^{j}\nabla^{i} \phi - \tfrac{1}{16}H_{ikl}H_{m}{}^{jk}H_{nj}{}^{l}\nabla^{i} \phi + \tfrac{1}{8}H_{n}{}^{jk}R_{mijk}\nabla^{i} \phi \nn \\
    &+ \tfrac{1}{4}H_{i}{}^{jk}R_{mjnk}\nabla^{i} \phi + \tfrac{1}{16}H_{mni}\nabla^{i} R - \tfrac{1}{32}H_{m}{}^{ij}H_{n}{}^{kl}\nabla_{j} H_{ikl} \nn \\
    &- \tfrac{1}{4}R_{n}{}^{i}\nabla_{j} H_{mi}{}^{j} + \tfrac{1}{32}H_{i}{}^{kl}H_{n}{}^{ij}\nabla_{j} H_{mkl} + \tfrac{1}{4}R^{ij}\nabla_{j} H_{mni} - \tfrac{1}{2}\nabla^{i} \phi\nabla_{j} \nabla_{i} H_{mn}{}^{j} \nn \\
    &+ \tfrac{1}{8}\nabla_{j} \nabla^{j} \nabla_{i} H_{mn}{}^{i} + \tfrac{1}{2}\nabla^{i} \phi\nabla_{j} H_{mni}\nabla^{j} \phi - \tfrac{1}{8}H_{nij}\nabla^{j} R_{m}{}^{i} \nn \\
    &+ \tfrac{1}{4}H_{n}{}^{jk}\nabla^{i} \phi\nabla_{k} H_{mij} + \tfrac{1}{8}H_{n}{}^{ij}\nabla_{k} \nabla_{j} H_{mi}{}^{k} - \tfrac{1}{32}H_{i}{}^{jk}H_{mn}{}^{i}\nabla_{l} H_{jk}{}^{l} \nn \\
    &+ \tfrac{1}{32}H_{m}{}^{ij}H_{ni}{}^{k}\nabla_{l} H_{jk}{}^{l} - \tfrac{1}{16}H_{ij}{}^{l}H^{ijk}\nabla_{l} H_{mnk} + \nabla_{j} H_{ni}{}^{j}\nabla_{m} \nabla^{i} \phi \nn \\
    &+ H_{nij}\nabla^{i} \phi\nabla_{m} \nabla^{j} \phi\,,\\
    \delta \phi^2=&\tfrac{1}{256}H_{i}{}^{lm}H^{ijk}H_{jl}{}^{n}H_{kmn} - \tfrac{5}{256}H_{ij}{}^{l}H^{ijk}H_{k}{}^{mn}H_{lmn} \nn \\
    &+ \tfrac{1}{1280}H_{ijk}H^{ijk}H_{lmn}H^{lmn} - \tfrac{1}{32}R_{ijkl}R^{ijkl} + \tfrac{1}{16}H_{i}{}^{lm}H^{ijk}R_{jklm} \nn \\
    &+ \tfrac{1}{32}H_{i}{}^{kl}H_{jkl}R^{ij} - \tfrac{1}{16}R_{ij}R^{ij} - \tfrac{3}{640}H_{ijk}H^{ijk}R + \tfrac{1}{160}{R}^{2} \nn \\
    &- \tfrac{1}{160}H_{jkl}H^{jkl}\nabla_{i} \nabla^{i} \phi + \tfrac{1}{80}R\nabla_{i} \nabla^{i} \phi - \tfrac{1}{32}\nabla_{i} \nabla^{i} R - \tfrac{3}{32}H^{jkl}\nabla_{i} H_{jkl}\nabla^{i} \phi \nn \\
    &+ \tfrac{1}{8}\nabla_{i} \phi\nabla^{i} R - \tfrac{1}{4}R_{ij}\nabla^{i} \phi\nabla^{j} \phi + \tfrac{1}{8}H^{ijk}\nabla_{l} \nabla_{k}H_{ij}{}^{l} + \tfrac{3}{64}\nabla_{k} H_{ijl}\nabla^{l} H^{ijk}\,.
\end{align}
Note that similar redefinitions apply to $b=-\alpha'$, and $a=0$ after worldsheet parity. Thus, $\Lc^{(6)}_{\text{redef}}$ can be removed for arbitrary values of $a$, and $b$.

\bibliography{ref}
\bibliographystyle{JHEP}

\end{document}